\documentclass[a4paper,11pt]{article}
\pdfoutput=1 

\usepackage{jheppub} 

\usepackage[T1]{fontenc} 
\usepackage{float}
\usepackage{mathrsfs}

\title{\boldmath Monopole deformations of 3d Seiberg-like dualities with adjoint matters}

\author[a]{Chiung Hwang,}
\author[b]{Sungjoon Kim,}
\author[b]{and Jaemo Park}

\affiliation[a]{Department of Applied Mathematics and Theoretical Physics, University of Cambridge, \\ Cambridge CB3 0WA, U.K.}
\affiliation[b]{Department of Physics, Pohang University of Science and Technology (POSTECH), \\ Pohang 37673, Republic of Korea}

\emailAdd{ch911@cam.ac.uk}
\emailAdd{sjkim0305@postech.ac.kr}
\emailAdd{jaemo@postech.ac.kr}

\abstract{We propose new 3d $\mathcal{N}=2$ Seiberg-like dualities by considering various monopole superpotential deformations on 3d $\mathcal{N}=2$ $U(N_c)$ SQCDs with fundamental and adjoint matter fields. We provide nontrivial evidence of these new dualities by comparing the superconformal indices, from which we analyze the change of the moduli space due to the monopole deformation. In addition, we perform the $F$-maximization to check the relevance of the monopole deformation for some examples, one of which is found to exhibit nontrivial symmetry enhancement in the IR. We prove such enhancement of the global symmetry using the superconformal index.}

\begin{document} 
\maketitle
\flushbottom

\section{Introduction}

The infra-red (IR) duality is one of the most interesting phenomena of quantum field theories. It tells us that two high energy theories flow to the same IR fixed point along the renormalization group (RG) flows and describe identical low energy dynamics. While it is a very nontrivial question to answer how such a duality works microscopically, we usually have a better understanding and more control of the duality if the theory is supersymmetric. Indeed, pioneered by the seminal work by Seiberg \cite{Seiberg:1994pq}, various examples of supersymmetric IR dualities have been proposed so far and used to explain the non-perturbative phenomena of supersymmetric gauge theories.

In the last decade, our understanding of the relations of such supersymmetric dualities to each other has been significantly enlarged. For instance, the work of \cite{Aharony:2013dha,Aharony:2013kma} provides a very concrete connection between 4d and 3d Seiberg-like dualities. Also, a new type of 4d IR duality, called mirror-like duality \cite{Pasquetti:2019hxf,Hwang:2020wpd}, has been found and shown to reduce to 3d $\mathcal N = 4$ mirror symmetry \cite{Intriligator:1996ex} upon circle compactification and some real mass deformation. Those 3d and 4d mirror dualities can be derived from a set of two basic duality moves \cite{Hwang:2021ulb}, which originates from the iterative application of particular Seiberg-like dualities \cite{Bottini:2021vms}. In addition, many examples of IR dualities with matter fields in rank-2 tensor representations of the gauge group have been derived from simpler dualities without such tensor representation fields by using a technique called deconfinement. See \cite{Berkooz:1995km,Pouliot:1995me,Luty:1996cg,Pasquetti:2019uop,Pasquetti:2019tix,Benvenuti:2020gvy,Benvenuti:2021nwt,Nii:2016jzi,Sacchi:2020pet,Bajeot:2022kwt,Bottini:2022vpy,Amariti:2022wae} for example. Those nontrivial relations of supersymmetric dualities provide crucial hints for the deeper structures behind those dualities.

Interestingly, in such relations of supersymmetric IR dualities, the monopole superpotential plays an important role. When a 4d duality is reduced to 3d, a nontrivial superpotential involving a monopole operator is generated in the resulting 3d theory \cite{Aharony:2013dha,Aharony:2013kma}, which is crucial to obtain the correct 3d duality. In addition, the deconfinement technique applied to an adjoint field in a 3d $U(N)$ theory is based on the Benini--Benvenuti--Pasquetti duality \cite{Benini:2017dud}, which is a variation of the Aharony duality \cite{Aharony:1997gp} deformed by linear monopole superpotentials. Also, such monopole-deformed theories lead to interesting IR fixed points; e.g., the 3d $\mathcal N = 2$ $U(N)$ theory with $2 N+2$ fundamental flavors and linear monopole superpotentials is identified with the S-duality wall for the 4d $\mathcal N = 2$ SQCD \cite{Benini:2017dud}. Therefore, the study of monopole superpotential is an important problem to understand more aspects of IR dynamics and dualities of 3d supersymmetric gauge theories.

For this reason, in this paper, we discuss the monopole deformation of 3d $\mathcal N = 2$ theories with adjoint matters, which leads to new Seiberg-like dualities in the presence of the monopole superpotential. One of the most powerful tests of such dualities is the matching of the superconformal index \cite{Bhattacharya:2008zy,Bhattacharya:2008bja}, capturing the the spectrum of chiral operators at the superconformal fixed point.\footnote{The 3d superconformal index has been used to test various supersymmetric IR dualities and the AdS/CFT correspondence. Some earlier works can be found in \cite{Cheon:2011th,Imamura:2011uj,Krattenthaler:2011da,Jafferis:2011ns,Kapustin:2011jm,Bashkirov:2011vy,Hwang:2011qt,Gang:2011xp,Hwang:2011ht,Kapustin:2011vz,Gang:2011jj,Dimofte:2012pd,Honda:2012ik}. Especially, via the AdS/CFT correspondence, it can also be used to count the microstates of rotating electric $AdS_4$ black holes \cite{Choi:2019zpz,Bobev:2019zmz,Nian:2019pxj,Benini:2019dyp,Choi:2019dfu}.} One can obtain its exact expression using the supersymmetric localization technique, which is given by a finite-dimensional matrix integral equipped with flux summations \cite{Kim:2009wb,Imamura:2011su}. We will provide nontrivial evidence of the proposed dualities by making explicit comparisons of the superconformal indices.

The remaining part of the paper is organized as follows.
\begin{itemize}
\item In section \ref{sec:one adjoint}, we discuss the monopole deformation of 3d $U(N_c)$ gauge theories with one adjoint and fundamental matters. We first review the known duality of this adjoint SQCD \cite{Kim:2013cma} and propose a new duality in the presence of a linear monopole superpotential. This is a generalization of \cite{Amariti:2018wht,Amariti:2019rhc}, which examined the deformation of the same theory by different monopole superpotentials. We also exhibit the result of the superconformal index computation, which provides nontrivial evidence of the proposed duality.

\item In section \ref{sec:two adjoints}, we move on to the theories with two adjoint matters. Again we first review the known duality for the theory \cite{Hwang:2018uyj} and propose its monopole deformation. Note that the double adjoint theory has two types of monopole operators, carrying one and two units of magnetic flux, respectively. We discuss the monopole deformation by each type of monopole operator.

\item In section \ref{sec:enhancement}, we discuss explicit examples of the single adjoint case. Specifically, we discuss the monopole deformation of $U(2)$ theories with three and four fundamental flavors and a single adjoint field $X$ with the superpotential $\mathrm{Tr} X^3$. We perform the $F$-maximization to check when the monopole deformation is relevant. We also comment on the symmetry enhancement of the $U(2)$ theory with four flavors, in connection with a similar model discussed in \cite{Amariti:2018wht,Benvenuti:2018bav} without the superpotential for the adjoint field.

\item In section \ref{sec:discussions}, we conclude the paper by summarizing the proposed dualities and giving brief discussions on RG flows and conformal manifold, specifically for the theories with two adjoint matters.

\item In appendix \ref{sec:index results}, we list the results of the index computation, which provide strong evidence of the conjectured dualities. All the global symmetry fugacities are omitted for simplicity.
\\
\end{itemize}

\section{3d SQCD with a single adjoint matter and $W = \mathrm{Tr} X^{n+1}$}
\label{sec:one adjoint}
\subsection{Review of the duality without the monopole superpotential}

In this section, we first review the Seiberg-like duality of a 3d $U(N_c)$ gauge theory with one adjoint and fundamental matters \cite{Kim:2013cma}, which we call the Kim--Park duality.
\begin{itemize}
\item Theory A is the 3d $\mathcal N = 2$ $U(N_c)$ gauge theory with $N_f$ pairs of fundamental $Q^a$ and anti-fundamental $\tilde Q^{\tilde a}$, one adjoint chiral multiplet $X$ and the superpotential
\begin{align}
\label{eq:sup_sa}
W_A = \mathrm{Tr} X^{n+1} \,.
\end{align}
\item Theory B is the 3d $\mathcal N = 2$ $U(n N_f-N_c)$ gauge theory with $N_f$ pairs of fundamental $q_{\tilde a}$ and anti-fundemental $\tilde q_a$, one adjoint $\hat X$, and $n N_f{}^2+2 n$ gauge singlet chiral multiplets ${M_i}^{\tilde a a}$ and $V_i^\pm$ for $a, \tilde a = 1,\dots N_f$ and $i = 0,\ldots,n-1$. The superpotential is given by
\begin{align}
\label{eq:single adjoint dual}
W_B = \mathrm{Tr} \hat X^{n+1}+\sum_{i = 0}^{n-1} M_i \tilde q \hat X^{n-1-i} q+\sum_{i = 0}^{n-1} \left(V_i^+ \hat V_{n-1-i}^-+V_i^- \hat V_{n-1-i}^+\right)
\end{align}
where $\hat V_i^\pm$ are the monopole operators of Theory B.
\end{itemize}
The global symmetry and charges are summarized in Table \ref{tab:KP charges}.
\begin{table}[tbp]
\centering
\begin{tabular}{|c|ccccc|}
\hline
 & $U(1)_R$ & $SU(N_f)_t$ & $SU(N_f)_u$ & $U(1)_A$ & $U(1)_T$ \\
\hline
$Q$ & $\Delta_Q$ & $\mathbf N_f$ & $\mathbf 1$ & 1 & 0 \\
$\tilde Q$ & $\Delta_Q$ & $\mathbf 1$ & $\mathbf N_f$ & 1 & 0 \\
$X$ & $\frac{2}{n+1}$ & $\mathbf 1$ & $\mathbf 1$ & 0 & 0 \\
\hline
$q$ & $\frac{2}{n+1}-\Delta_Q$ & $\mathbf 1$ & $\overline{\mathbf N_f}$ & -1 & 0 \\
$\tilde q$ & $\frac{2}{n+1}-\Delta_Q$ & $\overline{\mathbf N_f}$ & $\mathbf 1$ & -1 & 0 \\
$\hat X$ & $\frac{2}{n+1}$ & $\mathbf 1$ & $\mathbf 1$ & 0 & 0 \\
$M_i$ & $2 \Delta_Q+\frac{2 i}{n+1}$ & $\mathbf N_f$ & $\mathbf N_f$ & 2 & 0 \\
$V_i^\pm$ & $\substack{(1-\Delta_Q) N_f\\-\frac{2}{n+1} (N_c-1-i)}$ & $\mathbf 1$ & $\mathbf 1$ & $-N_f$ & $\pm1$ \\
\hline
$\hat V_i^\pm$ & $\substack{(1-\frac{2}{n+1}+\Delta_Q) N_f\\-\frac{2}{n+1} (\tilde N_c-1-i)}$ & $\mathbf 1$ & $\mathbf 1$ & $N_f$ & $\pm1$ \\
\hline
\end{tabular}
\caption{\label{tab:KP charges} The representations of the chiral operators under the global symmetry groups of the theory with one adjoint and fundamental matters. The elementary fields of Theory A are presented in the top box, whereas those of Theory B are presented in the middle box. Both the monopole operators of Theory A and their dual singlets are denoted by $V_i^\pm$, whose charges are shown in the middle box. Similarly, the monopole operators of Theory B are denoted by $\hat V_i^\pm$, whose charges are shown in the bottom box. Note that $\tilde N_c$ is the dual gauge rank, which is defined by $\tilde N_c = n N_f-N_c$.}
\end{table}
One can also generalize this duality to have the different numbers of fundamental and anti-fundamental matters, possibly with a nonzero Chern--Simons term, by giving real mass to some flavors \cite{Hwang:2015wna}. In this paper, we focus on the same number of the fundamental and anti-fundamental matters for simplicity.

The superpotential \eqref{eq:sup_sa} imposes the F-term condition $X^n = 0$ on the adjoint field $X$ such that the traces of powers of $X$ higher than $n$ are truncated in the chiral ring. In 4d, the duality of a theory with such a superpotential, later called the $A_n$ theory \cite{Intriligator:2003mi}, was studied in \cite{Kutasov:1995ve,Kutasov:1995np,Kutasov:1995ss}. It was also discussed that two $A_n$ and $A_{n'}$ theories with $n > n'$ are connected by an RG flow triggered by the extra superpotential \cite{Kutasov:2003iy,Intriligator:2016sgx}
\begin{align}
\sum_{i = n'}^{n-1} \mathrm{Tr} X^{i+1} \,.
\end{align}
This deformation leads to a discrete set of the vacuum expectation values of $X$, breaking the theory as follows:
\begin{align}
A_n \quad \longrightarrow \quad A_{n'}+\left(n-n'\right) A_1 \,,
\end{align}
where the $A_1$ theory is an ordinary SQCD without the adjoint because it contains the mass term for $X$.
Especially, once we turn on $\Delta W_X = \mathrm{Tr} \left[f_n(X)\right]$ where $f_n(x)$ is a generic polynomial of degree $n$ in $x$, the VEV of $X$ is parameterized by
\begin{align}
\left<X\right> \sim \oplus_{k = 1}^n x_k \mathbf{1}_{r_k} \,, \qquad \sum_{k = 1}^n r_k = N_c
\end{align}
where $x_k$'s are the $n$ distinct solutions to the equation
\begin{align}
(n+1) x^n+f_n'(x) = 0 \,.
\end{align}
This breaks the gauge group $U(N_c)$ into
\begin{align}
\prod_{k = 1}^n U(r_k)
\end{align}
where each $U(r_k)$ sector only has $N_f$ pairs of fundamental and anti-fundamental fields. The adjoint field becomes massive because $f_n''(x_k) \neq 0$. We will call this deformed theory the broken theory because the gauge group is partially broken.

For the original Theory A, the moduli space of vacua is parametrized by the VEV of chiral operators $M_i \equiv \tilde Q X^i Q$, $\mathrm{Tr} X^{i+1}$, and $V_i^\pm$ for $i = 0, \dots, n-1$. While the branch described by $\mathrm{Tr} X^{i+1}$ is lifted after turning on the polynomial superpotential $\Delta W_X = \mathrm{Tr} \left[f_n(X)\right]$, the dimensions of the other branches remain unchanged. For instance, both the original theory and the broken theory have $2n$-dimensional Coulomb branches, which are respectively described by $V_i^\pm$ for $i = 0,\dots,n-1$ and by $V^{(k)}{}^\pm$ for $k = 1,\dots,n$ where $V^{(k)}{}^\pm$ are a pair of monopole operators of each $U(r_k)$ sector.

One can consider the same deformation for Theory B. The corresponding deformation on the dual side should be the same polynomial superpotential $\Delta W_{\hat X} = \mathrm{Tr} \left[f_n(\hat X)\right]$ of degree $n$ now in $\hat X$ because the term $\mathrm{Tr} X^k$ is mapped to $\mathrm{Tr} \hat X^k$. The VEV of $\hat X$ is parameterized by
\begin{align}
\left<\hat X\right> \sim \oplus_{k = 1}^n \hat x_k \mathbf{1}_{\tilde r_k} \,, \qquad \sum_{k = 1}^n \tilde r_k = \tilde N_c = n N_f-N_c
\end{align}
where $\hat x_k$'s are the $n$ solutions to the equation
\begin{align}
(n+1) \hat x^n+f_n'(\hat x) = 0 \,.
\end{align}
This VEV breaks the dual gauge group $U(\tilde N_c)$ into
\begin{align}
\prod_{k = 1}^n U(\tilde r_k)
\end{align}
where now one can see that $\tilde r_k$ should satisfy
\begin{align}
\tilde r_k = N_f-r_k
\end{align}
so that it is consistent with the duality of the broken theory.  Each $U(\tilde r_k)$ sector then has $N_f$ pairs of fundamental and anti-fundamental fields $(q^{(k)},\tilde q^{(k)})$ and $N_f{}^2+2$ singlets $M^{(k)}, \, V^{(k)}{}^\pm$ interacting via the superpotential:
\begin{align}
W_B^{(k)} = M^{(k)} \tilde q^{(k)} q^{(k)}+V^{(k)}{}^+ \hat V^{(k)}{}^-+V^{(k)}{}^- \hat V^{(k)}{}^+ \,,
\end{align}
which is the Aharony dual theory \cite{Aharony:1997gp} of the $U(r_k)$ sector of the broken theory. Hence, once we turn on the polynomial superpotential $\Delta W_X = \mathrm{Tr} \left[f_n(X)\right]$, the Kim--Park duality lands on the Aharony duality of each broken sector. In the next subsection, we will see that such polynomial superpotential deformation is also useful for investigating new dualities with monopole superpotentials.

\subsection{Dualities with linear monopole superpotentials: $\Delta W = V_\alpha^++V_\alpha^-$}

In 3d, the Coulomb branch of the vacuum moduli space is described by the VEVs of monopole chiral operators. If such monopole operators are relevant, the theory can also be deformed by them and lead to new fixed points in the IR. For example, it is known that a 3d $U(N)$ SQCD can be deformed by the monopole superpotential $\Delta W = V^+ V^-$ \cite{Aharony:2013dha} and flow to an IR fixed point distinct from the original one without the monopole superpotential. Interestingly, such a new fixed point can be reached from a 4d $U(N)$ SQCD put on a circle, whose effective theory is 3-dimensional. Similarly, one can also consider the compactification of a 4d $USp(2 N)$ SQCD, which flows to the same fixed point as the 3d $USp(2 N)$ theory deformed by the monopole superpotential $\Delta W = V$ \cite{Aharony:2013dha}. Moreover, this theory further flows to the 3d $U(N)$ theory with linear monopole superpotential $\Delta W = V^++V^-$ \cite{Benini:2017dud} once we  turn on some real mass breaking $USp(2 N)$ into $U(N)$. Note that all those fixed points are distinct from those of the original 3d theories without the monopole superpotentials if the deformation is relevant.

In addition, such monopole deformation also leads to new IR dualities. For instance, \cite{Benini:2017dud} shows that the 3d $U(N_c)$ theory with $N_f$ flavors deformed by linear monopole superpotential $\Delta W = V^++V^-$ enjoys the following Seiberg-like duality.
\begin{itemize}
\item Theory A is the 3d $\mathcal N = 2$ $U(N_c)$ gauge theory with $N_f$ flavors $(Q,\tilde Q)$ and the superpotential
\begin{align}
\label{eq:BBP}
W_A = V^++V^-
\end{align}
where $V^\pm$ are the monopole operators of Theory A.
\item Theory B is the 3d $\mathcal N = 2$ $U(N_f-N_c-2)$ gauge theory with $N_f$ flavors $(q,\tilde q)$, $N_f{}^2$ gauge singlets $M$, and the superpotential
\begin{align}
W_B = M \tilde q q+\hat V^++\hat V^-
\end{align}
where $\hat V^\pm$ are the monopole operators of Theory B.
\end{itemize}
There are also studies of the monopole deformation of adjoint SQCDs and their dualities \cite{Amariti:2018wht,Amariti:2019rhc}. As reviewed in the previous subsection, the $U(N_c)$ theory with an adjoint matter and the superpotential $W = \mathrm{Tr} \, X^{n+1}$ has $2 n$ monopole operators $V_i^\pm$ for $i = 0,\dots,n-1$. Therefore, one can consider a large variety of monopole superpotentials and corresponding dualities.

In this subsection, we examine the deformation of the single adjoint theory by the following linear monopole superpotential:
\begin{align}
\label{eq:V_alpha}
\Delta W_A = V_\alpha^+ + V_\alpha^- \,.
\end{align}
For $\alpha = 0$, this type of monopole superpotential has been discussed in \cite{Amariti:2018wht}, which shows that the adjoint theory with such monopole superpotential can be obtained by compactifying a 4d $USp(2 N_c)$ SQCD with an antisymmetric matter and turning on suitable real mass breaking $USp(2 N_c)$ into $U(N_c)$. Most importantly, this 4d $USp(2 N_c)$ theory enjoys the Intriligator--Leigh--Strassler duality \cite{Intriligator:1995ax}, which in turn descends to a duality for the 3d $U(N_c)$ SQCD with an adjoint matter and the superpotential
\begin{align}
\label{eq:W_A with V0}
W_A^{\alpha = 0} = \mathrm{Tr} X^{n+1}+ V_0^+ +V_0^- \,.
\end{align}
The dual theory is given by the $U(n N_f-N_c-2 n)$ theory with the superpotential
\begin{align}
\label{eq:W_B with V0}
W_B^{\alpha = 0} = \mathrm{Tr} \hat X^{n+1}+\sum_{i = 0}^{n-1} M_i \tilde q \hat X^{n-1-i} q+\hat V_0^+ + \hat V_0^- \,,
\end{align}
which also includes the linear superpotential of the dual monopole operators $\hat V_0^\pm$. Note that the monopole superpotential lifts the Coulomb branch of the moduli space.

We attempt to generalize this duality for $\alpha > 0$. Since $V_0^\pm$ have the smallest conformal dimension among the $2 n$ monopole operators of the single adjoint theory, it is often the case that $V_0^\pm$ decouple in the IR and its linear deformation breaks supersymmetry. Nevertheless, even in such cases, there could exist interacting $V_\alpha^\pm$ for higher $\alpha > 0$, which would give rise to some relevant deformation of the theory. Hence, it is natural to ask if one can find new dualities by turning on the monopole superpotential \eqref{eq:V_alpha} for $\alpha > 0$.

Our proposal is as follows.
\begin{itemize}
\item Theory A is the 3d $\mathcal N = 2$ $U(N_c)$ gauge theory with $N_f$ pairs of fundamental $Q^a$ and anti-fundamental $\tilde Q^{\tilde a}$, one adjoint chiral multiplet $X$ and the superpotential
\begin{align}
\label{eq:W_A}
W_A^{mon} = \mathrm{Tr} X^{n+1}+V_\alpha^+ + V_\alpha^- \,.
\end{align}
where $V_\alpha^\pm$ are monopole operators of Theory A.
\item Theory B is the 3d $\mathcal N = 2$ $U(n N_f-N_c-2 n+2 \alpha)$ gauge theory with $N_f$ pairs of fundamental $q_{\tilde a}$ and anti-fundemental $\tilde q_a$, one adjoint $\hat X$, and $n N_f{}^2$ gauge singlet chiral multiplets ${M_i}^{\tilde a a}$ for $a,\tilde a = 1, \dots, N_f$ and $i = 0,\ldots,n-1$. The superpotential is given by
\begin{align}
\label{eq:W_B}
W_B^{mon} = \mathrm{Tr} \hat X^{n+1}+\sum_{i = 0}^{n-1} M_i \tilde q \hat X^{n-1-i} q+\hat V_\alpha^+ + \hat V_\alpha^- \,.
\end{align}
where $\hat V_\alpha^\pm$ are the monopole operators of Theory B.
\end{itemize}
Due to the monopole superpotential, monopole operators $V_j^\pm$ of Theory A and $\hat V_j^\pm$ of Theory B become massive for $j \geq \alpha$. The duality map of the remaining monopole operators is given by
\begin{align}
V_i^\pm \quad \longleftrightarrow \quad \hat V_i^\pm \,, \qquad i = 0,\dots,\alpha-1 \,.
\end{align}
\\

Since the superpotential for $\alpha > 0$ is not obtained from the compactification, we should take an alternative approach to obtain the duality deformed by those dressed monopole operators. For this reason, although our main interest is $\alpha > 0$, let us first revisit the $\alpha = 0$ case using such an altenative approach and then discuss $\alpha > 0$ subsequently. We first note that the extra superpotential \eqref{eq:V_alpha} with $\alpha = 0$:
\begin{align}
\label{eq:V0}
\Delta W_A = V_0^+ + V_0^-
\end{align}
is mapped to the superpotential of the same form on the dual side
\begin{align}
\Delta W_B = V_0^+ + V_0^-
\end{align}
where $V_0^\pm$ are now elementary gauge singlets rather than monopole operators of the theory. Together with the original superpotential of the dual theory \eqref{eq:single adjoint dual}, the total superpotential of the deformed dual theory is given by
\begin{align}
\label{eq:single adjoint dual 2}
W_B' &= W_B+\Delta W_B \nonumber \\
&= \mathrm{Tr} \hat X^{n+1}+\sum_{i = 0}^{n-1} M_i \tilde q \hat X^{n-1-i} q+\sum_{i = 0}^{n-1} \left(V_i^+ \hat V_{n-1-i}^-+V_i^- \hat V_{n-1-i}^+\right)+V_0^+ + V_0^- \,,
\end{align}
which gives rise to non-zero VEVs of the dual monopole operators $\hat V_{n-1}^\pm$.

While the direct analysis of the corresponding Higgs mechanism is complicated, we can take a shortcut using another deformation of the theory: we go back to the original side and turn on the polynomial superpotential of the adjoint field:
\begin{align}
\Delta W_X = \mathrm{Tr} \left[f_n(X)\right] .
\end{align}
As seen in the previous subsection, this leads to the nonzero VEV of $X$, which breaks the theory into the $n$ sectors consisting of the $U(r_k)$ gauge theories without an adjoint matter for $k = 1, \dots, n$. Each sector has a pair of monopole operators $V^{(k)}{}^\pm$ and the extra monopole superpotential \eqref{eq:V0} we turn on is expected to descend to the following linear superpotential for each sector:
\begin{align}
\label{eq:broken V0}
W_A^{(k)} = V^{(k)}{}^++V^{(k)}{}^- \,.
\end{align}
Note that this is exactly the linear monopole superpotential \eqref{eq:BBP} discussed in \cite{Benini:2017dud}. Thus, the dual theory of each sector is given by the $U(N_f-r_k-2)$ theory with the superpotential:
\begin{align}
\label{eq:broken V0 dual}
W_B^{(k)} = M^{(k)} \tilde q^{(k)} q^{(k)}+\hat V^{(k)}{}^++\hat V^{(k)}{}^- \,.
\end{align}
We find that the collection of those dual sectors is nothing but the effective low energy description of the $U(n N_f-N_c-2 n)$ theory with the superpotential:
\begin{align}
\label{eq:deformed B}
W = W_B^{\alpha = 0}+\Delta W_{\hat X} \,, \qquad \Delta W_{\hat X} = \mathrm{Tr} \left[f_n(\hat X)\right]
\end{align}
where $W_B^{\alpha = 0}$ is given in \eqref{eq:W_B with V0}. As argued in the previous subsection, $\Delta W_{\hat X}$ and the first term of $W_B^{\alpha = 0}$ result in $n$ distinct vacuum solutions governed by the $U(N_f-r_k-2)$ theories without the adjoint $\hat X$. Then the remaining terms of $W_B^{\alpha = 0}$ become the superpotential \eqref{eq:broken V0 dual} for each $U(N_f-r_k-2)$ sector. Hence, the $U(n N_f-N_c-2 n)$ theory with the superpotential \eqref{eq:deformed B} flows to the collection of the $U(N_f-r_k-2)$ theories with the superpotential \eqref{eq:broken V0 dual}.

One should note that $\Delta W_{\hat X}$ is dual to $\Delta W_X$, the extra superpotential we introduce on the original side. Thus, once we turn off both of them, we have a duality between the $U(N_c)$ theory with the superpotential \eqref{eq:W_A with V0} and the $U(n N_f-N_c-2 n)$ theory with the superpotential \eqref{eq:W_B with V0}, which should be the one obtained from the theory with \eqref{eq:single adjoint dual 2} after Higgsing. Indeed, this is exactly the duality proposed in \cite{Amariti:2018wht} obtained by compactifying the 4d duality. The duality and its deformation by the adjoint polynomial superpotential can be summarized as follows:
\begin{align}
\begin{array}{ccc}
U(N_c) & \quad \longleftrightarrow \quad & U(n N_f -N_c-2 n) \\
\big\downarrow & & \big\downarrow \\
{\displaystyle \prod_{k=1}^{n} U(r_k)} & \quad \longleftrightarrow \quad & {\displaystyle \prod_{k=1}^{n} U(N_f-r_k-2)}
\end{array}
\end{align}
where a horizontal arrow denotes the duality between the two theories, while a vertical arrow indicates the deformation by the polynomial superpotential of the adjoint field.

Notice that the Coulomb branch of each broken sector on the right hand side is completely lifted by the monopole superpotential \eqref{eq:broken V0}. Therefore, we expect that the monopole superpotential \eqref{eq:V0} also lifts the entire Coulomb branch on the left hand side. This will be confirmed by the superconformal index shortly in section \ref{sec:V}.
\\

Now let us move on to the generalization to $\alpha > 0$. We first recall that the theory with an adjoint matter has the $2 n$-dimensional Coulomb branch described by $V_i^\pm$ for $i = 0,\dots,n-1$, whose dimension is not affected even if we turn on the polynomial superpotential $\Delta W_X = \mathrm{Tr} \left[f_n(X)\right]$. Once the polynomial superpotential is turned on, the theory is decomposed into $n$ sectors, each of which has a 2-dimensional Coulomb branch described by their own monopole operators $V^{(k)}{}^\pm$. Therefore, the total Coulomb branch is still a $2 n$-dimensional one, now described by $V^{(k)}{}^\pm$ for $k = 1,\dots,n$.

We expect that the monopole superpotential \eqref{eq:V0} completely lifts this $2 n$-dimensional Coulomb branch because it descends to the monopole superpotential \eqref{eq:broken V0} of each $U(r_k)$ sector of the broken theory. Now let us imagine we turn on the monopole superpotential only for $m < n$ sectors of the broken theory. In that case, we still have un-lifted Coulomb branch for $n-m$ sectors, which enjoy the Aharony duality \cite{Aharony:1997gp} rather than the Benini--Benvenuti--Pasquetti duality \cite{Benini:2017dud}. Namely, the dual relations of the broken sectors are given by
\begin{align}
\left\{\begin{array}{cccc}
U(r_k) & \quad \longleftrightarrow \quad & U(N_f-r_k) & \qquad \text{without the monopole superpotential,} \\
U(r_k) & \quad \longleftrightarrow \quad & U(N_f-r_k-2) & \qquad \text{with the monopole superpotential,}
\end{array}\right.
\end{align}
depending on the presence of the monopole superpotential. Now if we turn off the adjoint polynomial superpotential so that the left hand side is restored to the $U(N_c)$ gauge theory, the dual sectors also combine into a $U(n N_f-N_c-2 m)$ gauge theory for $m < n$.\footnote{Given the fixed number of the sectors with the monopole superpotential, there is a subtlety in choosing which sector has the monopole superpotential and which sector doesn't. However, this is not significant for our purpose as the full Weyl symmetry of $U(N_c)$ will be restored once the adjoint polynomial superpotential is turned off, which is what we are interested in.} From this relation, one would expect that there exists a monopole deformation of the original theory lifting the Coulomb branch only partially such that the remaining Coulomb branch is $2 (n-m)$-dimensional.

We propose that the deformation \eqref{eq:V_alpha} plays such a role. Namely, if we turn on the monopole superpotential
\begin{align}
\Delta W_A = V_\alpha^+ + V_\alpha^- \,,
\end{align}
we expect that this lifts the components of the Coulomb branch described by $V_j^\pm$ for $j \geq \alpha$. In the next subsection, we will provide nontrivial evidence of this claim using the superconformal index. Furthermore, now the duality of each broken sector combines to predict a new duality between a $U(N)$ theory and a $U(n N_f-N_c-2 n+2 \alpha)$ theory with monopole superpotentials, which is exactly what we proposed at the beginning of this subsection.
\\

We need to check if new monopole terms in the superpotentials \eqref{eq:W_A} and \eqref{eq:W_B} are consistent with the global charges shown in Table \ref{tab:KP charges}. First we note that the monopole terms in \eqref{eq:W_A} break the $U(1)_A$ and $U(1)_T$. In addition, they also demand the $R$-charge of the monopole operators $V_\alpha^\pm$ to be 2; i.e.,
\begin{align}
(1-\Delta_Q) N_f-\frac{2}{n+1} (N_c-1-\alpha) = 2 \,,
\end{align}
which is satisfied only if the $R$-charge $\Delta_Q$ of $Q$ and $\tilde Q$ is given by
\begin{align}
\label{eq:single adjoint Delta_Q}
\Delta_Q= \frac{(n+1) N_f-2 N_c-2 n+2 \alpha}{(n+1) N_f} \,.
\end{align}
The $R$-charge $\Delta_q$ of $q$ and $\tilde q$ is then determined by
\begin{align}
\Delta_q = \frac{2}{n+1}-\Delta_Q = \frac{-(n-1) N_f+2 N_c+2 n-2 \alpha}{(n+1) N_f} \,.
\end{align}
This requires the $R$-charge of the dual monopole operators $\hat V_\alpha^\pm$ to be
\begin{align}
&\quad \left(1-\Delta_q\right) N_f-\frac{2}{n+1} \left(\tilde N_c-1-\alpha\right) \nonumber \\
&= \left(1-\frac{2}{n+1}+\Delta_Q\right) N_f-\frac{2}{n+1} (n N_f-N_c-2 n+\alpha-1) \nonumber \\
&= 2 \,,
\end{align}
which is consistent with the monopole terms in the dual superpotential \eqref{eq:W_B}. The resulting global charges are summarized in Table \ref{tab:V_alpha charges}.
\begin{table}[tbp]
\centering
\begin{tabular}{|c|ccc|}
\hline
 & $U(1)_R$ & $SU(N_f)_t$ & $SU(N_f)_u$ \\
\hline
$Q$ & $\frac{N_f-N_c+\tilde N_c}{(n+1) N_f}$ & $\mathbf N_f$ & $\mathbf 1$ \\
$\tilde Q$ & $\frac{N_f-N_c+\tilde N_c}{(n+1) N_f}$ & $\mathbf 1$ & $\mathbf N_f$ \\
$X$ & $\frac{2}{n+1}$ & $\mathbf 1$ & $\mathbf 1$ \\
$V_i^\pm$ & $2+\frac{2}{n+1} (i-\alpha)$ & $\mathbf 1$ & $\mathbf 1$ \\
\hline
$q$ & $\frac{N_f-\tilde N_c+N_c}{(n+1) N_f}$ & $\mathbf 1$ & $\overline{\mathbf N_f}$ \\
$\tilde q$ & $\frac{N_f-\tilde N_c+N_c}{(n+1) N_f}$ & $\overline{\mathbf N_f}$ & $\mathbf 1$ \\
$\hat X$ & $\frac{2}{n+1}$ & $\mathbf 1$ & $\mathbf 1$ \\
$M_i$ & $\frac{2 N_f-2 N_c+2 \tilde N_c}{(n+1) N_f}+\frac{2 i}{n+1}$ & $\mathbf N_f$ & $\mathbf N_f$ \\
$\hat V_i^\pm$ & $2+\frac{2}{n+1} (i-\alpha)$ & $\mathbf 1$ & $\mathbf 1$ \\
\hline
\end{tabular}
\caption{\label{tab:V_alpha charges} The representations of the chiral operators under the global symmetry groups of the single adjoint theory with the linear monopole superpotential $\Delta W_A = V_\alpha^+ + V_\alpha^-$ and its dual. Now the monopole operators $V_i^\pm$ and $\hat V_i^\pm$ are presented together with the elementary fields in the upper box and the lower box, respectively. Note that $\tilde N_c$ is the dual gauge rank, which is now defined by $\tilde N_c = n N_f-N_c-2 n+2 \alpha$.}
\end{table}

\subsection{The superconformal index and the chiral ring truncation}
\label{sec:V}

In this subsection, we attempt to support our proposal by providing nontrivial evidence using the superconformal index. We will see that the evaluated indices show the perfect agreement under the proposed duality. Furthermore, from the index, we also show that if the monopole superpotential
\begin{align}
\label{eq:V_alpha 2}
\Delta W_A = V_\alpha^++V_\alpha^- \,.
\end{align}
is turned on, the monopole operators $V_j^\pm$ for $j \geq \alpha$ become $Q$-exact and vanish in the chiral ring, captureing the algebraic structure of the moduli space, as we claimed in the previous subsection.

The definition of the 3d superconformal index is given by \cite{Bhattacharya:2008zy,Bhattacharya:2008bja}:
\begin{align}
I = \mathrm{Tr} (-1)^F e^{\beta' \{Q,S\}} x^{E+j} \prod_i t_i^{F_i} \,.
\end{align}
Here $F$ is the fermion number operator. $Q$ and $S = Q^\dagger$ are supercharges chosen such that only the BPS states saturating
\begin{align}
\{Q,S\} = E-j-R \geq 0
\end{align}
contribute to the index, where energy $E$, angular momentum $j$, and $R$-charge $R$ are three Cartans of the bosonic subgroup of the 3d $\mathcal N = 2$ superconformal group $SO(3,2) \times SO(2)$. $F_i$ are the Cartan charges of the global symmetry.

This index can be evaluated using the localization technique, either localization on the Coulomb branch \cite{Kim:2009wb,Imamura:2011su} or that on the Higgs branch \cite{Fujitsuka:2013fga,Benini:2013yva} leading to the factorized form of the index \cite{Hwang:2012jh}.\footnote{The factorized index is also closely related to the notion of holomorphic block \cite{Beem:2012mb}.} Indeed, the index of the single adjoint theory without the monopole superpotential was computed using the Coulomb branch formula in \cite{Kim:2013cma} and using the factorization formula in \cite{Hwang:2015wna}. For the index without the monopole superpotential, the $R$-charges of the monopole operators $V_0^\pm$ are free parameters, whose values at the IR fixed point should be independently determined by the $F$-maximization \cite{Jafferis:2010un} because the $R$-symmetry can vary along the RG-flow as it mixes with other abelian global symmetries. On the other hand, the index with the monopole superpotential can be obtained using the same formulas but with fixed $R$-charges and fugacities because they are constrained by the given monopole superpotential. For examples, since the monopole superpotential \eqref{eq:V_alpha 2} breaks $U(1)_A$ and $U(1)_T$, there is no $U(1)$ global symmetry that can be mixed with $U(1)_R$ along the RG-flow unless there is an emergent $U(1)$ symmetry. In such a case, the superconformal $R$-charge at the IR fixed point is the same as the UV value, which is fixed by the monopole superpotential.

In order to compute the index for the single adjoint theory with the monopole superpotential, we use the factorization formula derived in \cite{Hwang:2015wna} and insert the $R$-charge value $\Delta_Q$ determined by \eqref{eq:single adjoint Delta_Q}, which makes the $R$-charge of $V_\alpha^\pm$ two. A few examples of the results we obtain are shown in Table \ref{tab:V_alpha}, whereas more results can be found in appendix \ref{sec:index results}.
\begin{table}[tbp]
	\centering
	\begin{tabular}{|c|c|c|c|}
		\hline
			n	&		$\Delta W_{A}$		&	$\!\!(N_f,N_c,\tilde N_c)\!\!$	&	SCI	\\
		\hline
			2	&	$V_0^{+}\!+\!V_0^{-}\!\!$	&			$(3,2,0)$			&	$\begin{array}{c}
																		1+9 x^{2/9}+45 x^{4/9}+165 x^{2/3}+504 x^{8/9}+1359 x^{10/9} \\
																		+3327 x^{4/3}+7515 x^{14/9}+15876 x^{16/9} \\
																		+31681 x^2+O\left(x^{19/9}\right)
																		\end{array}$	\\
		\cline{3-4}
		\hline
			3	&	$V_1^{+}\!+\!V_1^{-}\!\!$	&			$(5,5,6)$			&	$\begin{array}{c}
																		1+\sqrt{x}+25 x^{3/5}+2 x+50 x^{11/10}+325 x^{6/5}+4 x^{3/2} \\
																		+100 x^{8/5}+950 x^{17/10}+2925 x^{9/5}-43 x^2+175 x^{21/10} \\
																		+2225 x^{11/5}+11050 x^{23/10}+20475 x^{12/5}-88 x^{5/2} \\
																		-900 x^{13/5}+4125 x^{27/10}+30225 x^{14/5}+93600 x^{29/10} \\
																		+118573 x^3+O\left(x^{31/10}\right)
																		\end{array}$	\\
		\cline{3-4}
				&						&			$(5,11,0)$			&	$\begin{array}{c}
																		1+25 x^{2/5}+325 x^{4/5}+2925 x^{6/5}+20450 x^{8/5} \\
																		+118130 x^2+O\left(x^{21/10}\right)
																		\end{array}$	\\
		\hline  
			4	&	$V_1^{+}\!+\!V_1^{-}\!\!$	&			$(3,4,2)$			&	$\begin{array}{c}
																		1+9 x^{2/15}+45 x^{4/15}+166 x^{2/5}+513 x^{8/15}+1413 x^{2/3} \\
																		+3575 x^{4/5}+8451 x^{14/15}+18909 x^{16/15}+40443 x^{6/5} \\
																		+83259 x^{4/3}+165807 x^{22/15}+320729 x^{8/5}+604629 x^{26/15} \\
																		+1113786 x^{28/15}+2009121 x^2+O\left(x^{31/15}\right)
																		\end{array}$	\\
		\hline
	\end{tabular}
	\caption{\label{tab:V_alpha} The superconformal index results for the single adjoint theories with $\Delta W_A = V_\alpha^+ + V_\alpha^-$. Here we list a specific examples with $(n,\alpha,N_f,N_c) = (2,0,3,2), \, (3,1,5,5), \, (3,1,5,11)$ and $(4,1,3,4)$, whereas more results are given in appendix \ref{sec:index results}. Note that the index for $(n,\alpha,N_f,N_c) = (3,1,5,11)$ is obtained after flipping $\tilde Q Q$ and $\tilde Q X Q$, the operators with negative $R$-charges, by the superpotential \eqref{eq:flip}. The $SU(N_f)_t \times SU(N_f)_u$ flavor fugacities are all omitted for simplicity. The gauge rank of the dual theory is given by $\tilde N_c = n N_f-N_c-2 n+2 \alpha$.}
\end{table}
In all the cases we have examined, the indices of each dual pair show the exact agreement, which is strong evidence of the duality we propose.
\\

Capturing the spectrum of the chiral operators, the superconformal index tells us many important aspects of the theory. For example, the algebraic structure of the vacuum moduli space can be deduced from the index because it is parametrized by the VEV of chiral operators. Especially, we can see how the moduli space is deformed once we turn on the monopole superpotential \eqref{eq:V_alpha 2}. First recall that, in general, the moduli space of the theory \emph{without} the monopole superpotential is parameterized by the following chiral operators:
\begin{align}
\begin{aligned}
\label{eq:generators}
\tilde Q X^i Q \quad &\longleftrightarrow \quad M_i \,, \qquad \qquad i = 0, \dots, n-1 \,, \\
\mathrm{Tr} X^i \quad &\longleftrightarrow \quad \mathrm{Tr} \hat X^i \,, \qquad \quad i = 1, \dots, n-1 \,, \\
V_i^\pm \quad &\longleftrightarrow \quad V_i^\pm \,, \qquad \quad \,\,\,\,\, i = 0, \dots, n-1 \,,
\end{aligned}
\end{align}
where the right hand side shows the corresponding dual operators, which describe the same moduli space. Note that $V_i^\pm$ on the left hand side are monopole operators, while $V_i^\pm$ on the right hand side are elementary gauge singlet fields.

If we turn on the linear monopole superpotential \eqref{eq:V_alpha 2}, as we will show shortly, we find that some of the monopole operators vanish in the chiral ring as follows:
\begin{align}
\label{eq:monopole truncation}
V_j^\pm \quad \sim \quad 0 \,, \qquad \quad j = \alpha, \dots, n-1
\end{align}
because they combine with some fermionic operators and become $Q$-exact. In addition, on the dual side, the singlets $V_i^\pm$ for all $i$, which used to describe the Coulomb branch of the moduli space, are now all massive. Instead, we have dual monopole operators $\hat V_i^\pm$ for $i = 0,\dots,\alpha-1$, which are mapped to the monopole operators of the original theory as follows:
\begin{align}
\label{eq:monopole map}
V_i^\pm \quad &\longleftrightarrow \quad \hat V_i^\pm \,, \qquad \quad \,\,\,\,\, i = 0, \dots, \alpha-1 \,.
\end{align}
Those monopole operators parameterize the unlifted Coulomb branch of the moduli space, which now has dimension $2 \alpha$ rather than $2 n$.

Therefore, in generic cases, the vacuum moduli space is parameterized by the VEVs of the following gauge invariant operators:
\begin{align}
\tilde Q X^i Q \quad &\longleftrightarrow \quad M_i \,, \qquad \qquad i = 0, \dots, n-1 \,, \\
\mathrm{Tr} X^i \quad &\longleftrightarrow \quad \mathrm{Tr} \hat X^i \,, \qquad \quad i = 1, \dots, n-1 \,, \\
V_i^\pm \quad &\longleftrightarrow \quad \hat V_i^\pm \,, \qquad \quad \,\,\,\,\, i = 0, \dots, \alpha-1
\end{align}
where the right hand side shows the corresponding dual operators describing the same moduli space. In addtion, we will also see that if the gauge ranks are bounded in a certain range, there are additional effects of the monopole superpotential as follows:
\begin{align}
\begin{aligned}
\label{eq:extra truncation}
\mathrm{Tr} X^i \quad &\sim \quad 0 \,, \qquad \qquad i = \mathrm{min}(N_c+1,n N_f-N_c-2 n+2 \alpha+1), \dots, n-1 \,, \\
V_i^\pm \quad &\sim \quad 0 \,, \qquad \qquad i = \mathrm{min}(N_c,n N_f-N_c-2 n+2 \alpha), \dots, \alpha-1
\end{aligned}
\end{align}
if $\mathrm{min}(N_c,n N_f-N_c-2 n+2 \alpha)$ is less than either $n-1$ or $\alpha$.
\\

To see the truncation of the monopole operators \eqref{eq:monopole truncation}, let us examine the case with $(n,N_f,N_c) = (3,5,5)$ and $\Delta W_A = V_1^+ + V_1^-$ as an example. Before turning on the monopole superpotential $\Delta W_A = V_1^+ + V_1^-$, the index is given by\footnote{While we have shown the terms up to $x^2$ to avoid clutter, the next terms up to $x^3$ can be easily obtained by taking the plethystic exponential of \eqref{eq:PL0}.}
\begin{align}
\label{eq:saind}
&I = 1+x^\frac12+\mathbf{5}_t \mathbf{5}_u \tau^2 x^\frac35+2 x+2 \, \mathbf{5}_t \mathbf{5}_u \tau^2 x^\frac{11}{10}+\left(\mathbf{15}_t \mathbf{15}_u+\mathbf{10}_t \mathbf{10}_u\right) \tau^4 x^\frac65 \nonumber \\
&\quad +\left(2+\tau^{-5} \left(w+w^{-1}\right)\right) x^\frac32+4 \, \mathbf{5}_t \mathbf{5}_u \tau^2 x^\frac85 \nonumber \\
&\quad +\left((\mathbf{15}_t+\mathbf{10}_t) (\mathbf{15}_u+\mathbf{10}_u)+\mathbf{15}_t \mathbf{15}_u+\mathbf{10}_t \mathbf{10}_u\right) \tau^4 x^\frac{17}{10} \nonumber \\
&\quad +\left(\mathbf{40}_t \mathbf{40}_u+\mathbf{35}_t \mathbf{35}_u+\overline{\mathbf{10}}_t \overline{\mathbf{10}}_u\right) \tau^6 x^\frac95+\left(1+2 \tau^{-5} \left(w+w^{-1}\right)-\mathbf{24}_t-\mathbf{24}_u\right) x^2+\dots
\end{align}
where $\mathbf n_t$ and $\mathbf n_u$ are the characters of the representation $\mathbf n$ of $SU(N_f)_t$ and that of $SU(N_f)_u$ respectively, and $\tau$ and $w$ are the $U(1)_A$ and $U(1)_T$ fugacities. While we have used a trial UV value of $R$-charge $\Delta_Q = \frac{(n+1) N_f-2 N_c-2 n+2 \alpha}{(n+1) N_f}$, which is the same as the one determined in \eqref{eq:single adjoint Delta_Q}, the superconformal IR value can be obtained by shifting $\tau \rightarrow \tau x^\alpha$ where the mixing coefficient $\alpha$ should be determined by the $F$-maximization. To read the chiral ring relations, it is convenient to take the plethystic logarithm \cite{Benvenuti:2006qr}, which gets the contribution from the single trace operators, both bosonic and fermionic, as well as their relations. For the current example, the plethystic log of the index is given by
\begin{align}
\label{eq:PL0}
&\quad (1-x^2) \, \mathrm{PL}\left[I\right] \nonumber \\
&= x^\frac12+\mathbf{5}_t \mathbf{5}_u \tau^2 x^\frac35+x+\mathbf{5}_t \mathbf{5}_u \tau^2 x^\frac{11}{10}+\tau^{-5} \left(w+w^{-1}\right) x^\frac32+\mathbf{5}_t \mathbf{5}_u \tau^2 x^\frac85\nonumber \\
&\quad+\left(\tau^{-5} \left(w+w^{-1}\right)-\mathbf{24}_t-\mathbf{24}_u-2\right) x^2+\left(\tau^{-5} \left(w+w^{-1}\right)-\mathbf{24}_t-\mathbf{24}_u-2\right) x^\frac52 \nonumber \\
&\quad+\left(-\mathbf{24}_t-\mathbf{24}_u-3\right) x^3+\dots \,,
\end{align}
where both sides are multiplied by $(1-x^2)$ to remove the contribution of the descendants derived by the derivative operator. From this plethystic logarithm of the index, we can easily find the contribution of the monopole operators $V_i^\pm$ for $i = 0,1,2$:
\begin{align}
\label{eq:mon}
\tau^{-5} \left(w+w^{-1}\right) x^\frac{3+i}{2} \,,
\end{align}
which are all independent operators describing the Coulomb branch of the moduli space. On the other hand, we are also interested in the fermionic operators giving the following contributions:
\begin{align}
\label{eq:ferm}
\psi_Q^\dagger X^{i-1} Q \, &: \qquad -\left(\mathbf{24}_t+1\right) x^{\frac{3+i}{2}} \,, \\
\tilde Q X^{i-1} \psi_{\tilde Q}^\dagger \, &: \qquad -\left(\mathbf{24}_u+1\right) x^{\frac{3+i}{2}}
\end{align}
for $i = 1,2$, which play important roles when we turn on the monopole superpotential. Indeed, if we turn on the monopole superpotential $\Delta W_A = V_1^+ + V_1^-$,\footnote{Here we assume all the terms in the superpotential are relevant in the IR so that the theory flows to the IR fixed point with the expected symmetry, which is $SU(N_f)^2$ in this case. However, this should be independently checked using, e.g., the $F$-maximization, which we wasn't able to conduct for this and subsequent examples due to limited computing resources. Nevertheless, regardless of the actual relevance of the superpotential for those particular examples, the explanation here can be regarded as a demonstration of the general structure of the indices and the chiral rings of the theories when their superpotentials are relevant.} this breaks $U(1)_A$ and $U(1)_T$, whose fugacities $\tau$ and $w$ thus have to be 1. Once we set $\tau = w = 1$ in \eqref{eq:PL0}, we see that the monopole contribution \eqref{eq:mon} for $i = 1,\, 2$ is exactly canceled by $-2 x^\frac{3+i}{2}$, the contribution of the trace parts of $\psi_Q^\dagger X^{i-1} Q$ and $\tilde Q X^{i-1} \psi_{\tilde Q}^\dagger$. In contrast, the monopole contribution \eqref{eq:mon} for $i = 0$ still remains nontrivial. This means that the lowest monopole operators $V_0^\pm$ remain nontrivial in the chiral ring and still parametrize the Coulomb branch when we turn on the monopole superpotential $\Delta W_A = V_1^++V_1^-$, while the other monopole operators $V_i^\pm$ for $i \geq 1$ combine with the trace parts of the fermionic operators $\psi_Q^\dagger X^{i-1} Q$ and $\tilde Q X^{i-1} \psi_{\tilde Q}^\dagger$ and become a long multiplet along the RG-flow. Therefore, the components of the Coulomb branch described by $V_j^\pm$ for $j \geq 1$ are now lifted quantum mechanically.

In fact, this cancelation between the $V_{\alpha+j}^\pm$ and $\left(\psi_Q^\dagger X^j Q, \, \tilde Q X^j \psi_{\tilde Q}^\dagger\right)$ for $j = 0,\dots n-1-\alpha$ is generic because, with $\tau = w = 1$, $V_{\alpha+j}^\pm$ always give the contribution $2 x^{2+\frac{2 j}{n+1}}$, whose charges are fixed by the monopole superpotential $\Delta W_A = V_\alpha^++V_\alpha^-$, whereas the trace parts of $\psi_Q^\dagger X^j Q$ and $\tilde Q X^j \psi_{\tilde Q}^\dagger$ give exactly the same contribution with the opposite sign. Thus, we conclude that the components of the Coulomb branch parameterized by $V_j^\pm$ for $j \geq \alpha$ are lifted by the monopole superpotential $\Delta W_A = V_\alpha^++V_\alpha^-$. The unlifted components are described by massless monopole operators $V_i^{\pm}$ for $i = 0,\dots,\alpha-1$, which are mapped to the dual monopole operators under the proposed duality as shown in \eqref{eq:monopole map}. Note that the monopole operators $\hat V_j^\pm$ for $j \geq \alpha$ of the dual theory also get lifted due to the dual superpotential $\Delta W_B = \hat V_\alpha^++\hat V_\alpha^-$, which is another simple consistency check for the proposed duality because both theories have the same Coulomb branch quantum mechanically.
\\

Next, let us examine the extra truncation of chiral ring generators \eqref{eq:extra truncation} when the gauge ranks are bounded in a particular range. We first recall the case \emph{without} the monopole superpotential, whose chiral ring generators are given by \eqref{eq:generators}. Note that the operators with higher $i \geq n$ are truncated due to the the F-term condition $X^n = 0$. However, if the gauge rank is smaller than $n$, (or $n-1$ for $\mathrm{Tr} X^i$,) they are also subject to the classical rank conditions as follows:\footnote{Here the equation $\mathcal O \sim 0$ means the operator $\mathcal O$ is written in terms of other chiral generators; e.g., $\mathrm{Tr} X^2$ of a $U(1)$ gauge theory is not an independent generator because it is identified with $(\mathrm{Tr} X)^2$.}
\begin{align}
\mathrm{Tr} X^i \quad &\sim \quad 0 \,, \qquad \qquad i = N_c+1, \dots, n-1 \,, \label{eq:adjoint rank condition} \\
\tilde Q X^i Q \quad &\sim \quad 0 \,,  \qquad \qquad i = N_c, \dots, n-1 \,, \\
V_i^\pm \quad &\sim \quad 0 \,, \qquad \qquad i = N_c, \dots, n-1 \label{eq:monopole rank condition}
\end{align}
due to the characteristic equation of the matrix field $X$.
For example, $V_i^\pm$ is constrained because it is mapped to the following state with magnetic flux on $S^2$ via operator state correspondence of conformal field theories
\begin{align}
\mathrm{Tr}_{N_c-1} X^i \left|\pm1,0^{N_c-1}\right> \quad + \quad \text{permutations by the Weyl group}
\end{align}
where the gauge group is broken to $U(N_c-1) \times U(1)$ due to the magnetic flux, and $\mathrm{Tr}_{N_c-1}$ is taken over unbroken $U(N_c-1)$. Then $\mathrm{Tr}_{N_c-1} X^i$ for $i \geq N_c$ is written in terms of $\mathrm{Tr}_{N_c-1} X^j$ with $j < N_c$, and so is $V_i^\pm$.

Among those three types of operators, $\mathrm{Tr} X^i$ is further subject to quantum mechanical truncation, which are expected from the duality. Recall that $\mathrm{Tr} X^i$ of the $U(N_c)$ theory is mapped to $\mathrm{Tr} \hat X^i$ of the dual $U(n N_f-N_c)$ theory. If the dual gauge rank $\tilde N_c = n N_f-N_c$ is smaller than $n-1$, the dual operator $\mathrm{Tr} \hat X^i$ is also constrained by the classical rank condition:
\begin{align}
\mathrm{Tr} \hat X^i \quad \sim \quad 0 \,, \qquad \qquad i = \tilde N_c+1, \dots, n-1 \,.
\end{align}
Due to the duality, the original operator $\mathrm{Tr} X^i$ must satisfy the same condition, which is the quantum effect in the $U(N_c)$ theory instead. Namely, on top of the classical condition \eqref{eq:adjoint rank condition}, the operator $\mathrm{Tr} X^i$ of the $U(N_c)$ theory is constrained quantum mechanically as follows:
\begin{align}
\mathrm{Tr} X^i \quad \sim \quad 0 \,, \qquad \qquad i = n N_f-N_c+1, \dots, \mathrm{min}(n-1,N_c)
\end{align}
if $n N_f-N_c < \mathrm{min}(n-1,N_c)$. On the other hand, the meson operators $\tilde Q X^i Q$ and the monopole operators $V_i^\pm$ are mapped to the gauge singlet operators $M_i$ and $V_i^\pm$ on the dual side respectively, which are not constrained by any classical rank condition. Therefore, we don't expect any extra quantum truncation for them in general.
\\

Now we move on to the case with the monopole superpotential. The same analysis can be applied to the theory with the monopole superpotential $\Delta W_A = V_\alpha^++V_\alpha^-$, leading to an interesting consequence on chiral ring generators. We have seen that once we turn on the monopole superpotential $\Delta W_A = V_\alpha^++V_\alpha^-$, the gauge rank of the dual theory is reduced from $n N_f-N_c$ to $n N_f-N_c-2 n+2 \alpha$. Therefore, if the reduced dual rank is smaller than both $n-1$ and $N_c$, there is quantum mechanical truncation of $\mathrm{Tr} X^i$ as follows:
\begin{align}
\label{eq:adjoint truncation}
\mathrm{Tr} X^i \quad \sim \quad 0 \,, \qquad \quad i = n N_f-N_c-2 n+2 \alpha+1, \dots, \mathrm{min}(n-1,N_c) \,,
\end{align}
which shows that more operators are truncated than the case without the monopole superpotential.

For example, let us consider the case with $(n,N_f,N_c) = (2,3,2)$ and $\Delta W_A = V_0^+ + V_0^-$, whose dual theory is given by the $U(4)$ theory before turning on the monopole superpotential. Since both $N_c = 2$ and $\tilde N_c = 4$ are larger than $n-1 = 1$, we don't expect any further constraint on $\mathrm{Tr} X^i$ other than the F-term condition $X^n = X^2 = 0$. Therefore, $\mathrm{Tr} X$ should be a nontrivial chiral ring generator and parametrize the moduli space. Indeed, this can be confirmed by the superconformal index, which is given by
\begin{align}
\label{eq:SCI2032}
I &= 1+\mathbf 3_t \mathbf 3_u \tau^2 x^\frac29+\left(\mathbf 6_t \mathbf 6_u+\mathbf{\overline 3}_t \mathbf{\overline 3}_u\right) \tau^4 x^\frac49+\left(\left(\mathbf{10}_t \mathbf{10}_u+\mathbf 8_t \mathbf 8_u\right) \tau^6+1\right) x^\frac23 \nonumber \\
&\quad+\left(\left(\mathbf{15}_t \mathbf{15}_u+\mathbf{15'}_t \mathbf{15'}_u+\mathbf{\overline 6}_t \mathbf{\overline 6}_u\right) \tau^8+2 \, \mathbf 3_t \mathbf 3_u \tau^2\right) x^\frac89+\dots
\end{align}
with the plethystic logarithm
\begin{align}
\label{eq:PL2032}
(1-x^2) \, \mathrm{PL}[I] &= \mathbf 3_t \mathbf 3_u \tau^2 x^\frac29+(1-\tau^6) x^\frac23+\mathbf 3_t \mathbf 3_u \tau^2 x^\frac89-\mathbf{\overline 3}_t \mathbf{\overline 3}_u \tau^4 x^\frac{10}{9}+\mathbf 3_t \mathbf 3_u \left(\tau^8-\tau^2\right) x^\frac{14}{9} \nonumber \\
&\quad -\mathbf{\overline 3}_t \mathbf{\overline 3}_u \tau^{10} x^\frac{16}{9}+\left(\left(\mathbf 8_t+\mathbf 8_u+1\right) \tau^6+\tau^{-3} \left(w+w^{-1}\right)-\mathbf 8_t-\mathbf 8_u-3\right) x^2+\dots
\end{align}
where one can see the nontrivial contribution $x^\frac23$ of $\mathrm{Tr} X$.

However, if we turn on the monopole superpotential $\Delta W_A = V_0^+ + V_0^-$, the dual gauge rank becomes $\tilde N_c = 0$; i.e., the dual theory is a Wess--Zumino theory of $M_i$ and $V_i^\pm$. Since there is no $\mathrm{Tr} \hat X$ on the dual side, we expect some quantum mechanical constraint should be imposed on $\mathrm{Tr} X$ on the original side as well. Indeed, this can be seen from the index, which is obtained from \eqref{eq:SCI2032} by taking $\tau = w = 1$ because the monopole superpotential breaks $U(1)_A$ and $U(1)_T$. In the same manner, the plethystic log of the index is given by \eqref{eq:PL2032} with $\tau = w = 1$, where we find that the contribution $x^\frac23$ of $\mathrm{Tr} X$ is exactly canceled by the negative contribution $-\tau^6 x^\frac23$ since we set $\tau = 1$. We note that this negative contribution originally reflects the classical rank condition for meson operators of the $U(2)$ gauge group
\begin{align}
\epsilon_{\tilde a \tilde b \tilde c} \epsilon_{abc} (M_0)^{\tilde a a} (M_0)^{\tilde b b} (M_0)^{\tilde c c} \quad = \quad \epsilon_{\tilde a \tilde b \tilde c} \epsilon_{abc} \tilde Q^{\tilde a}_\alpha Q^{\alpha a} \tilde Q^{\tilde b}_\beta Q^{\beta b} \tilde Q^{\tilde c}_\gamma Q^{\gamma c} \quad \sim \quad 0 \,,
\end{align}
which is due to the fact that the gauge indices $\alpha, \beta, \gamma$ run over 1, 2 only. Thus, the cancelation indicates that $\mathrm{Tr} X$ is now identified with $\tilde \epsilon \epsilon(M_0)^3$, which is now nontrivial in the chiral ring but not an independent generator once we turn on the monopole superpotential $\Delta W_A = V_0^+ + V_0^-$. 

Let us give you another example with $\alpha \neq 0$. We consider the case with $(n,N_f,N_c) = (4,3,4)$ and $\Delta W_A = V_1^+ + V_1^-$, whose dual theory is given by the $U(8)$ theory before turning on the monopole superpotential. Since both $N_c = 4$ and $\tilde N_c = 8$ are larger than $n-1 = 3$, again we don't expect any further constraint on $\mathrm{Tr} X^i$ other than the F-term condition $X^n = X^4 = 0$, implying that $\mathrm{Tr} X^i$ for $i = 1,2,3$ are nontrivial chiral ring generators parametrizing the moduli space. If we look at the plethystic log of the superconformal index:
\begin{align}
\label{eq:PL4134}
(1-x^2) \, \mathrm{PL}[I] &= \mathbf 3_t \mathbf 3_u \tau^2 x^\frac{2}{15}+x^\frac25+\mathbf 3_t \mathbf 3_u \tau^2 x^\frac{8}{15}+x^\frac45+\mathbf 3_t \mathbf 3_u \tau^2 x^\frac{14}{15}+\left(1-\tau^6\right) x^\frac65+\mathbf 3_t \mathbf 3_u \tau^2 x^\frac43 \nonumber \\
&\quad+\mathbf{\overline 3}_t \mathbf{\overline 3}_u \tau^4 x^\frac{22}{15}+\left(\tau^{-3} \left(w+w^{-1}\right)-\tau^6\right) x^\frac85-\mathbf 3_t \mathbf 3_u \tau^2 x^\frac{26}{15} \nonumber \\
&\quad-\mathbf{\overline 3}_t \mathbf{\overline 3}_u \tau^4 x^\frac{28}{15}+\left(\tau^{-3} \left(w+w^{-1}\right)-\tau^6-\mathbf 8_t-\mathbf 8_u-3\right) x^2+\dots \,,
\end{align}
we find the nontrivial contribution $x^\frac{2 i}{5}$ of $\mathrm{Tr} X^i$ for $i = 1,2,3$.

However, once we turn on the monopole superpotential $\Delta W_A = V_1^+ + V_1^-$, the dual gauge rank becomes $\tilde N_c = 2$. Since there is no $\mathrm{Tr} \hat X^3$ on the dual side, we also expect a quantum mechanical constraint imposed on $\mathrm{Tr} X^3$ on the original side, which can be seen from the index by taking $\tau = w = 1$. Once we take $\tau = w = 1$, the plethystic log of the index shows that the contribution $x^\frac65$ of $\mathrm{Tr} X^3$ is exactly canceled by the negative contribution $-\tau^6 x^\frac65$. On the other hand, the contributions of $\mathrm{Tr} X$ and $\mathrm{Tr} X^2$ remain nontrivial as expected. We expect such cancelation should happen whenever the dual gauge rank becomes smaller than both $N_c$ and $n-1$.
\\

Lastly, we expect similar conditions for the monopole operators $V_i^\pm$ because they are now mapped to the dual monopole operators $\hat V_i^\pm$ rather than the gauge singlet $V_i^\pm$ under the monopole-deformed duality. Indeed, one can use a similar argument to show that, if $n N_f-N_c-2 n+2 \alpha < \mathrm{min}(\alpha-1, N_c-1)$, the theory with the monopole superpotential $\Delta W_A = V_\alpha^+ + V_\alpha^-$ yields extra quantum truncation of the monopole operators $V_i^\pm$:
\begin{align}
\label{eq:extra monopole truncation}
V_i^\pm \quad \sim \quad 0 \,, \qquad \quad i = n N_f-N_c-2 n+2 \alpha, \dots, \mathrm{min}(\alpha-1,N_c-1)
\end{align}
in addition to the condition \eqref{eq:monopole truncation} that we have already found and the classical rank condition \eqref{eq:monopole rank condition}.

Let us consider an example. First we recall the example with $(n,N_f,N_c) = (3,5,5)$ and $\Delta W_A = V_1^+ + V_1^-$, where $V_{1,2}^\pm$ are lifted due to the monopole superpotential, while $V_0^\pm$ remain massless. See \eqref{eq:mon} and \eqref{eq:ferm}. In this case, we don't see any extra constraint on $V_i^\pm$ other than \eqref{eq:monopole truncation} because both $N_c = 5$ and $\tilde N_c = 6$ are larger than $\alpha-1 = 0$. On the other hand, if we consider the $(n,N_f,N_c) = (3,5,11)$ case, now the dual rank $\tilde N_c$ becomes zero, requiring $V_0^\pm$ to vanish as well. Before showing its index, we have to comment that for $(n,N_f,N_c) = (3,5,11)$ with $\Delta W_A = V_1^+ + V_1^-$, $\Delta_Q$ is given by $\Delta_Q = \frac{(n+1) N_f-2 N_c-2 n+2 \alpha}{(n+1) N_f} = -\frac{3}{10}$. Thus, the meson operators $\tilde Q Q$ and $\tilde Q X Q$ have negative $R$-charges, which indicates that those operators are decoupled in the IR. In order to obtain the series expansion of the index, we need to \emph{flip} those decoupled operators, i.e., make them massive by introducing extra singlets $m_1$ and $m_2$ with the superpotential
\begin{align}
\label{eq:flip}
\delta W_\text{flip} = m_1 \tilde Q X Q+m_2 \tilde Q Q
\end{align}
so that they do not contribute to the index. The F-term equations of $m_1$ and $m_2$ require $\tilde Q Q = \tilde Q X Q = 0$. Once we flip $\tilde Q Q$ and $\tilde Q X Q$, the index is given by\footnote{In fact, a gauge invariant operator is decoupled if its $R$-charge is less than or equal to 1/2. Thus, to obtain the index of the interacting sector, we have to flip all the gauge invariant operators whose $R$-charges are no greater than 1/2. On the other hand, here, and for the other examples, we only flip those with negative $R$-charges for simplicity.}
\begin{align}
I &= 1+\mathbf{5}_t \mathbf{5}_u \tau^2 x^\frac25+\left(1-\tau^{10}\right) x^\frac12+\left(\mathbf{15}_t \mathbf{15}_u+\mathbf{10}_t \mathbf{10}_u\right) \tau^2 x^\frac45+\left(\mathbf 5_t \mathbf 5_u \tau^2-\mathbf 5_t \mathbf 5_u \tau^{12}\right) x^\frac{9}{10} \nonumber \\
&\quad +2 \left(1-\tau^{10}\right) x+\dots
\end{align}
with the plethystic logarithm
\begin{align}
 (1-x^2) \, \mathrm{PL}\left[I\right] &= \mathbf{5}_t \mathbf{5}_u \tau^2 x^\frac25+\left(1-\tau^{10}\right)x^\frac12+\left(1-\tau^{10}\right) x \nonumber \\
&\quad +\left(\tau^{-5}-\tau^5\right) \left(w+w^{-1}\right) x^\frac32-\mathbf{\overline 5}_t \mathbf{\overline 5}_u \tau^8 x^\frac85 \nonumber \\
&\quad+\left(\left(\mathbf{24}_t+\mathbf{24}_u+2\right) \tau^{10}+\left(\tau^{-5}-\tau^{5}\right) \left(w+w^{-1}\right)-\mathbf{24}_t-\mathbf{24}_u-2\right) x^2+\dots
\end{align}
where we can see that the contribution $\tau^{-5} \left(w+w^{-1}\right) x^\frac32$ of $V_0^\pm$ is canceled by negative contribution $-\tau^{5} \left(w+w^{-1}\right) x^\frac32$ once we set $\tau = w = 1$. This shows that, if the dual rank $\tilde N_c$ is less than or equal to $\alpha-1$, there are extra quantum constraints on the monopole operators as we expect in \eqref{eq:extra monopole truncation}.
\\

\section{3d SQCD with double adjoint matters and $W = \mathrm{Tr} X^{n+1}+\mathrm{Tr} X Y^2$}
\label{sec:two adjoints}

\subsection{Review of the duality without the monopole superpotential}

Now we move on to the duality for a 3d $U(N_c)$ gauge theory with \emph{two} adjoint matters. The duality without the monopole superpotential was proposed in \cite{Hwang:2018uyj} as follows.
\begin{itemize}
\item Theory A is the 3d $\mathcal N = 2$ $U(N_c)$ gauge theory with $N_f$ pairs of fundamental $Q^a$ and anti-fundamental $\tilde Q^{\tilde a}$, two adjoint chiral multiplets $X, \, Y$ and the superpotential
\begin{align}
W_A = \mathrm{Tr} X^{n+1}+\mathrm{Tr} X Y^2 \,.
\end{align}
\item Theory B is the 3d $\mathcal N = 2$ $U(3 n N_f-N_c)$ gauge theory with $N_f$ pairs of fundamental $q_{\tilde a}$ and anti-fundemental $\tilde q_a$, two adjoint $\hat X, \, \hat Y$, and three sets of gauge singlet chiral multiplets:\footnote{Our $V_{n,0}^\pm$ are denoted by $V_{0,2}^\pm$ in \cite{Hwang:2018uyj}.}
\begin{align}
\begin{array}{ll}
M_{s,t}{}^{\tilde a a} \,, \qquad & s = 0,\dots,n-1 \,, \quad t = 0,1,2 \,, \quad a,\tilde a = 1,\dots, N_f \,, \\
V_{s,t}^\pm \,, \qquad & s = 0,\dots,n \,, \quad t = 0,1 \,, \quad s t = 0 \,, \\
W_u^\pm \,, \qquad & u = 0,\dots,\frac{n-3}{2} \,.
\end{array}
\end{align}
The superpotential is given by
\begin{align}
W_B &= \mathrm{Tr} \hat X^{n+1}+\mathrm{Tr} \hat X \hat Y^2+\sum_{s = 0}^{n-1} \sum_{t = 0}^2 M_{s,t} \tilde q \hat X^{n-1-s} \hat Y^{2-t} q \nonumber \\
&\quad +\sum_{s = 0}^{n} V_{s,0}^\pm \hat V_{n-s,0}^\mp+V_{0,1}^\pm \hat V_{0,1}^\mp+\sum_{u = 0}^{\frac{n-3}{2}} W_u^\pm \hat W_{\frac{n-3}{2}-u}^\mp
\end{align}
where $\hat V_{s,t}^\pm$ and $\hat W_u^\pm$ are the monopole operators of Theory B.
\end{itemize}
The global symmetry and charges are summarized in Table \ref{tab:HKP charges}.
\begin{table}[tbp]
	\centering
	\begin{tabular}{|c|ccccc|}
		\hline
		& $SU(N_f)_t$ & $SU(N_f)_u$ & $U(1)_A$ & $U(1)_T$ & $U(1)_R$ \\
		\hline
		$Q$ & $\mathbf{N_f}$ & $\mathbf 1$ & $1$ & $0$ & $\Delta_Q$ \\
		$\tilde Q$ & $\mathbf 1$ & $\mathbf{N_f}$ & $1$ & $0$ & $\Delta_Q$ \\
		$X$ & $\mathbf 1$ & $\mathbf 1$ & $0$ & $0$ & $\frac{2}{n+1}$ \\
		$Y$ & $\mathbf 1$ & $\mathbf 1$ & $0$ & $0$ & $\frac{n}{n+1}$ \\
		\hline
		$q$ & $\mathbf 1$ & $\overline{\mathbf{N_f}}$ & $-1$ & $0$ & $\frac{2-n}{n+1}-\Delta_Q$ \\
		$\tilde q$ & $\overline{\mathbf{N_f}}$ & $\mathbf 1$ & $-1$ & $0$ & $\frac{2-n}{n+1}-\Delta_Q$ \\
		$\hat X$ & $\mathbf 1$ & $\mathbf 1$ & $0$ & $0$ & $\frac{2}{n+1}$ \\
		$\hat Y$ & $\mathbf 1$ & $\mathbf 1$ & $0$ & $0$ & $\frac{n}{n+1}$ \\
		$M_{s,t}$ & $\mathbf{N_f}$ & $\mathbf{N_f}$ & $2$ & $0$ & $2 \Delta_Q+\frac{2s+nt}{n+1}$ \\
		$V_{s,t}^\pm$ & $\mathbf 1$ & $\mathbf 1$ & $-N_f$ & $\pm 1$ & $(1-\Delta_Q)N_f -\frac{1}{n+1} (N_c -1) +\frac{2 s+n t}{n+1}$ \\
		$W_u^\pm$ & $\mathbf 1$ & $\mathbf 1$ & $-2N_f$ & $\pm 2$ & $2(1-\Delta_Q)N_f -\frac{2}{n+1} (N_c -1) +\frac{2+4 u}{n+1}$ \\
		\hline
		$\hat V_{s,t}^\pm$ & $\mathbf 1$ & $\mathbf 1$ & $N_f$ & $\pm 1$ & $(1-\frac{2-n}{n+1}+\Delta_Q) N_f -\frac{1}{n+1} (\tilde N_c -1) +\frac{2 s+n t}{n+1}$ \\
		$\hat W_u^\pm$ & $\mathbf 1$ & $\mathbf 1$ & $2N_f$ & $\pm 2$ & $2 (1-\frac{2-n}{n+1}+\Delta_Q) N_f -\frac{2}{n+1} (\tilde N_c -1) +\frac{2+4 u}{n+1}$ \\
		\hline
	\end{tabular}
	\caption{\label{tab:HKP charges} The representations of the chiral operators under the global symmetry groups of the double adjoint theory. The top box is devoted to the elementary fields of Theory A, whereas the middle box is devoted to those of Theory B. Note that the monopole operators of Theory A have the same charges as the gauge singlets $V_{s,t}^\pm, \, W_u^\pm$ of Theory B. On the other hand, the charges of the monopole operators of Theory B are given in the bottom box where $\tilde N_c$ is the dual gauge rank given by $\tilde N_c = 3 n N_f-N_c$.  The indices of $V_{s,t}^\pm$ satisfy $s t = 0$.}
\end{table}
For a generic case, the vacuum moduli space is described by
\begin{align}
\begin{array}{cl}
M_{s,t} = \tilde Q X^s Y^t Q \,, \qquad &s = 0, \dots, n-1, \quad t = 0,1,2 \,, \\
\mathrm{Tr} X^s \,, \qquad &s = 1, \dots, n \,, \\
\mathrm{Tr} Y^t \,, \qquad &t = 1, 2 \,, \\
V_{s,t}^\pm \,, \qquad &s = 0, \dots, n, \quad t = 0,1,2, \quad st = 0 \,, \\
W_{u}^\pm \,, \qquad &u = 0, \dots, \frac{n-3}{2} \,,
\end{array}
\end{align}
with the constraint
\begin{align}
\mathrm{Tr} X^n \quad &\sim \quad \mathrm{Tr} Y^2 \,, \\
V_{n,0}^\pm \quad &\sim \quad V_{0,2}^\pm \,,
\end{align}
which is due to the F-term condition $X^n \sim Y^2$. Because of this condition, only one linear combination of $\mathrm{Tr} X^n$ and $\mathrm{Tr} Y^2$ and that of $V_{n,0}^\pm$ and $V_{0,2}^\pm$ are nontrivial chiral ring generators, which will be simply denoted by $\mathrm{Tr} X^n$ and $V_{n,0}^\pm$ respectively.

Again it is helpful to examine the deformation by the superpotential $\Delta W_X = \mathrm{Tr} \left[f_n(X)\right]$ where $f_n(x)$ is a generic polynomial of degree $n$ in $x$. The vacuum solutions for the adjoint fields are parametrized by
\begin{align}
\begin{aligned}
\langle X \rangle \sim \mathrm{diag}\left(x^{1d}_1{\bf 1}_{m^{1d}_1}, \cdots, x^{1d}_n{\bf 1}_{m^{1d}_n}, 0 {\bf 1}_{m^{1d}_{n+1}}, 0 {\bf 1}_{m^{1d}_{n+2}}, x^{2d}_1 {\bf 1}_{m^{2d}_{1}} \otimes \sigma_1, \cdots, x^{2d}_{(n-1)/2} {\bf 1}_{m^{2d}_{(n-1)/2}}\otimes \sigma_1\right)\\
\langle Y \rangle \sim \mathrm{diag}\left(0{\bf 1}_{m^{1d}_1}, \cdots, 0{\bf 1}_{m^{1d}_n}, y^{1d}_{n+1} {\bf 1}_{m^{1d}_{n+1}}, y^{1d}_{n+2} {\bf 1}_{m^{1d}_{n+2}}, y^{2d}_1 {\bf 1}_{m^{2d}_{1}}\otimes \sigma_2, \cdots, y^{2d}_{(n-1)/2} {\bf 1}_{m^{2d}_{(n-1)/2}}\otimes \sigma_2\right)
\end{aligned}
\end{align}
where $x^{1d}_k$'s are the $n$ solutions to the equation
\begin{align}
(n+1) x^n+f_n'(x) = 0 \,,
\end{align}
$y^{1d}_{n+1}$ and $y^{1d}_{n+2}$ are the two solutions to the equation
\begin{align}
y^2+f_n'(0) = 0 \,,
\end{align}
and lastly $(x^{2d}_k)^2, \, (y^{2d}_k)^2$ are the $\frac{n-1}{2}$ solutions to the equation
\begin{gather}
(n+1) x^{n-1}+\frac{f_n'(x)-f_n'(-x)}{2 x} = 0 \,, \\
y^2+\frac{f_n'(x)+f_n'(-x)}{2} = 0 \,.
\end{gather}
Note that the adjoint VEVs are decomposed into several block matrices, either 1-dimensional or 2-dimensional.  At a given vacuum solution, both $X$ and $Y$ become massive, and the gauge group $U(N_c)$ is broken into
\begin{align}
\label{eq:broken double adjoint}
\prod_{i=1}^{n+2} U(m^{1d}_i) \prod_{j=1}^{\frac{n-1}{2}} U(m^{2d}_j)
\end{align}
where $N_c = \sum_{i=1}^{n+2} m^{1d}_i+\sum_{j=1}^{\frac{n-1}{2}} 2 m^{2d}_j$ is satisfied. In particular, $U(m^{2d}_j)$ is embedded in $U(N_c)$ as a diagonal subgroup of $U(m^{2d}_j) \times U(m^{2d}_j) \subset U(2 m^{2d}_j) \subset U(N_c)$. Accordingly, a fundamental or an anti-fundamental field of $U(N_c)$ becomes two copies of a fundamental or an anti-fundamental field of $U(m^{2d}_j)$. Thus, the 2-dimensional VEV sectors, with gauge group $\prod_{j=1}^{\frac{n-1}{2}} U(m^{2d}_j)$, have $2 N_f$ pairs of fundamental and anti-fundamental fields. On the other hand, the 1-dimensional VEV sectors, with gauge group $\prod_{i=1}^{n+2} U(m^{1d}_i)$, have $N_f$ pairs of fundamental and anti-fundamental fields. Please see \cite{Hwang:2018uyj} for more detailed discussions.

The same deformation can be considered in Theory B. In the presence of the extra polynomial superpotential $\Delta W_{\hat X} = \mathrm{Tr} \left[f_n(\hat X)\right]$, the expectation values of the adjoints break the dual gauge group as follows:
\begin{align}
U(3nN_f -N_c) \quad \longrightarrow \quad \prod_{i=1}^{n+2} U(N_f-m^{1d}_i) \prod_{j=1}^{\frac{n-1}{2}} U(2N_f -m^{2d}_j)
\end{align}
with
\begin{align}
3nN_f-N_c = \sum_{i=1}^{n+2}(N_f-m^{1d}_i) + 2 \times \sum_{j=1}^{\frac{n-1}{2}} (2N_f-m^{2d}_j)
\end{align}
where each sector is the Aharony dual of the corresponding broken sector \eqref{eq:broken double adjoint} of deformed Theory A; i.e., we have a duality for each broken sector as follows:
\begin{align}
\begin{matrix}
U(m^{1d}_i) \text{ with $N_f$ flavors} (Q,\tilde Q) \\
W_A^{(1d, i)} = 0
\end{matrix}
\quad &\longleftrightarrow \quad
\begin{matrix}
U(N_f-m^{1d}_i) \text{ with $N_f$ flavors } (q,\tilde q) \\ W_B^{(1d, i)} = M^{(i)} \tilde q^{(i)} q^{(i)}+V^{(i)}{}^+ \hat V^{(i)}{}^-+V^{(i)}{}^- \hat V^{(i)}{}^+ \,,
\end{matrix} \\
\begin{matrix}
U(m^{2d}_j) \text{ with $2 N_f$ flavors} (\mathsf Q,\tilde{\mathsf Q}) \\
W_A^{(2d, j)} = 0
\end{matrix}
\quad &\longleftrightarrow \quad
\begin{matrix}
U(2 N_f-m^{2d}_j) \text{ with $2 N_f$ flavors } (\mathsf q,\tilde{\mathsf q}) \\ W_B^{(2d, j)} = \mathsf M^{(j)} \tilde{\mathsf q}^{(j)} \mathsf q^{(j)}+\mathsf V^{(j)}{}^+ \hat{\mathsf V}^{(j)}{}^-+\mathsf V^{(j)}{}^- \hat{\mathsf V}^{(j)}{}^+ \,.
\end{matrix}
\end{align}
As in the single adjoint case, the adjoint Higgs branch is completely lifted by the polynomial superpotential deformation, whereas the mesonic Higgs branch is described by the $N_f \times N_f$ matrix fields $M^{(i)}$ for $i = 1,\dots,n+2$ and the $2 N_f \times 2 N_f$ matrix fields $\mathsf M^{(j)}$ for $j = 1,\dots, \frac{n-1}{2}$, which are thus $3 n {N_f}^2$-dimensional in total, the same as the mesonic Higgs branch dimension of the undeformed theory described by $M_{s,t}$ for $s = 0,\dots, n-1$ and $t = 0,1,2$. Similarly, the Coulomb branch is described by $V^{(i)}{}^\pm$ and $\mathsf V^{(j)}{}^\pm$, which are $3 n+3$-dimensional, again the same as the Coulomb branch dimension of the undeformed theory described by $V_{s,t}^\pm$ and $W_u^\pm$ for $s = 0,\dots,n, \, t = 0,1$ with $s t = 0$ and $u = 0,\dots,\frac{n-3}{2}$. In the next subsection, we will use this adjoint polynomial deformation to deduce the monopole dualities for the double adjoint theory.

\subsection{Dualities with linear monopole superpotentials}

Recall that the double adjoint theory has the two types of monopole operators: $V_{s,t}^\pm$ carrying the unit topological symmetry charge and $W_u^\pm$ carrying the topological symmetry charge $\pm 2$. Thus, one can also consider two types of deformations by linear monopole superpotentials:
\begin{align}
\label{eq:V00}
\Delta W_A = V_{0,0}^+ + V_{0,0}^-
\end{align}
and
\begin{align}
\Delta W_A = W_0^+ + W_0^-
\end{align}
where we focus on the $s=t=0$ and $u=0$ cases for simplicity and leave the other cases for future works.
\\

Firstly, we consider the deformation \eqref{eq:V00} with the simplest monopole operators $V_{0,0}^\pm$. We conjecture that this linear superpotential leads to a new monopole duality as follows.
\begin{itemize}
\item Theory A is the 3d $\mathcal N = 2$ $U(N_c)$ gauge theory with $N_f$ pairs of fundamental $Q^a$ and anti-fundamental $\tilde Q^{\tilde a}$, two adjoint chiral multiplets $X, \, Y$ and the superpotential
\begin{align}
W_A^{mon}= \mathrm{Tr} X^{n+1}+\mathrm{Tr} X Y^2+V_{0,0}^++V_{0,0}^-
\end{align}
where $V_{0,0}^\pm$ are a pair of monopole operators of Theory A with topological symmetry charge $\pm1$.
\item Theory B is the 3d $\mathcal N = 2$ $U(3 n N_f-N_c-4 n-2)$ gauge theory with $N_f$ pairs of fundamental $q_{\tilde a}$ and anti-fundemental $\tilde q_a$, two adjoint $\hat X, \, \hat Y$, and $3 n {N_f}^2$ gauge singlet chiral multiplets $M_{s,t}{}^{\tilde a a}$ for $a, \tilde a = 1, \dots, N_f$, $s = 0,\dots,n-1$ and $t = 0,1,2$.
The superpotential is given by
\begin{align}
\label{eq:V00 dual}
W_B^{mon} = \mathrm{Tr} \hat X^{n+1}+\mathrm{Tr} \hat X \hat Y^2+\sum_{s = 0}^{n-1} \sum_{t = 0}^2 M_{s,t} \tilde q \hat X^{n-1-s} \hat Y^{2-t} q+\hat V_{0,0}^+ + \hat V_{0,0}^- \,.
\end{align}
where $\hat V_{0,0}^\pm$ are a pair of monopole operators of Theory B with topological symmetry charge $\pm1$.
\end{itemize}
As we will see shortly, the Coulomb branch of the moduli space, which was described by the monopole operators in Theory A and by the dual singlets in Theory B, is now completely lifted due to the monopole superpotential.

As in the single adjoint case, let us deform the theories by the polynomial superpotential $\Delta W_X = \mathrm{Tr} \left[f_n(X)\right]$, which makes both adjoint fields $X$ and $Y$ massive. As we have discussed in the previous subsection, the low energy description consists of several broken sectors
\begin{align}
\prod_{i=1}^{n+2} U(m^{1d}_i) \prod_{j=1}^{\frac{n-1}{2}} U(m^{2d}_j)
\end{align}
without the adjoint fields. Furthermore, the extra superpotential \eqref{eq:V00} would descend to the following linear monopole superpotential:
\begin{align}
\label{eq:1d}
W^{(1d,i)}_A = V^{(i)}{}^++V^{(i)}{}^-
\end{align}
for each $U(m^{1d}_i)$ sector and
\begin{align}
\label{eq:2d}
W^{(2d,i)}_A = \mathsf V^{(j)}{}^++\mathsf V^{(j)}{}^-
\end{align}
for each $U(m^{2d}_j)$ sector where $V^{(i)}{}^\pm$ and $\mathsf V^{(j)}{}^\pm$ are the monopole operators of the $U(m^{1d}_i)$ sector and the $U(m^{2d}_j)$ sector respectively. In the presence of such monopole superpotentials, one can apply the Benini--Benvenuti--Pasquetti duality \cite{Benini:2017dud} on each sector, which gives rise to the following duality relations:
\begin{align}
\begin{aligned}
\label{eq:broken V00 duality}
U(m^{1d}_i) \quad &\longleftrightarrow \quad U(N_f-m^{1d}_i-2) \,, \\
U(m^{2d}_j) \quad &\longleftrightarrow \quad U(2 N_f-m^{2d}_j-2)
\end{aligned}
\end{align}
with the dual superpotentials
\begin{align}
\begin{aligned}
\label{eq:broken V00 dual sup}
W^{(1d,i)}_B &= M^{(i)} \tilde q^{(i)} q^{(i)}+\hat V^{(i)}{}^++\hat V^{(i)}{}^- \,, \\
W^{(2d,j)}_B &= \mathsf M^{(j)} \tilde{\mathsf q}^{(j)} \mathsf q^{(j)}+\hat{\mathsf V}^{(j)}{}^++\hat{\mathsf V}^{(j)}{}^-
\end{aligned}
\end{align}
where we also have the linear monopole superpotentials of the dual monopole operators $\hat V^{(i)}{}^\pm$ and $\hat{\mathsf V}^{(j)}{}^\pm$.

Indeed, this is exactly what we expect from Theory B deformed by the same polynomial superpotential. Once we turn on $\Delta W_{\hat X} = \mathrm{Tr} \left[f_n(\hat X)\right]$ for Theory B, the non-zero VEVs of $\hat X$ and $\hat Y$ lead to the low energy description with the broken gauge group
\begin{align}
\prod_{i=1}^{n+2} U(N_f-m^{1d}_i-2) \prod_{j=1}^{\frac{n-1}{2}} U(2N_f -m^{2d}_j-2)
\end{align}
and the superpotential \eqref{eq:broken V00 dual sup}, which is consistent with the BBP duality \eqref{eq:broken V00 duality} for each broken sector.
The proposed duality and its deformation by the adjoint polynomial superpotential can be depicted as follows:
\begin{align}
\begin{array}{ccc}
U(N_c) & \quad \longleftrightarrow \quad &U(3nN_f -N_c-4 n -2)  \\
\big\downarrow & & \big\downarrow \\
{\displaystyle \prod_{i=1}^{n+2} U(m^{1d}_i) \prod_{j=1}^{\frac{n-1}{2}} U(m^{2d}_j)} & \quad \longleftrightarrow \quad & {\displaystyle \prod_{i=1}^{n+2} U(N_f-m^{1d}_i-2) \prod_{j=1}^{\frac{n-1}{2}} U(2N_f -m^{2d}_j-2)}
\end{array}
\end{align}
where a horizontal arrow denotes the duality between the two theories, while a vertical arrow indicates the deformation by the polynomial superpotential of the adjoint field $X$ or $\hat X$. Note that the Coulomb branch of each broken sector on the right hand side is completely lifted by the monopole superpotentials \eqref{eq:1d} and \eqref{eq:2d}. Therefore, we expect that the monopole superpotential \eqref{eq:V00} also lifts the entire Coulomb branch on the left hand side. This will be confirmed by the superconformal index shortly in section \ref{sec:V00}.

We need to check if new monopole terms in the superpotentials \eqref{eq:V00} and \eqref{eq:V00 dual} are consistent with the global charges shown in Table \ref{tab:HKP charges}. Again the monopole superpotential \eqref{eq:V00} breaks the $U(1)_A$ and $U(1)_T$ and also demands the $R$-charge of the monopole operators $V_{0,0}^\pm$ to be 2; i.e.,
\begin{align}
(1-\Delta_Q) N_f-\frac{1}{n+1} (N_c-1) = 2 \,,
\end{align}
which is satisfied only if the $R$-charge $\Delta_Q$ of $Q$ and $\tilde Q$ is given by
\begin{align}
\label{eq:double adjoint Delta_Q 1}
\Delta_Q= \frac{(n+1) N_f-N_c-2 n-1}{(n+1) N_f} \,.
\end{align}
The $R$-charge $\Delta_q$ of $q$ and $\tilde q$ is then determined by
\begin{align}
\Delta_q = \frac{2-n}{n+1}-\Delta_Q = \frac{-(2 n-1) N_f+N_c+2 n+1}{(n+1) N_f} \,.
\end{align}
This requires the $R$-charge of the dual monopole operators $\hat V_\alpha^\pm$ to be
\begin{align}
&\quad \left(1-\Delta_q\right) N_f-\frac{1}{n+1} \left(\tilde N_c-1\right) \nonumber \\
&= \left(1-\frac{2-n}{n+1}+\Delta_Q\right) N_f-\frac{1}{n+1} (3 n N_f-N_c-4 n-3) \nonumber \\
&= 2 \,,
\end{align}
which is consistent with the monopole terms in the dual superpotential \eqref{eq:V00 dual}. The resulting global charges are summarized in Table \ref{tab:V00 charges}.
\begin{table}[tbp]
	\centering
	\begin{tabular}{|c|ccc|}
		\hline
		& $SU(N_f)_t$ & $SU(N_f)_u$ & $U(1)_R$ \\
		\hline
		$Q$ & $\mathbf{N_f}$ & $\mathbf 1$ & $\frac{(2-n) N_f-N_c+\tilde N_c}{2 (n+1) N_f}$ \\
		$\tilde Q$ & $\mathbf 1$ & $\mathbf{N_f}$ & $\frac{(2-n) N_f-N_c+\tilde N_c}{2 (n+1) N_f}$ \\
		$X$ & $\mathbf 1$ & $\mathbf 1$ & $\frac{2}{n+1}$ \\
		$Y$ & $\mathbf 1$ & $\mathbf 1$ & $\frac{n}{n+1}$ \\
		$V_{s,t}^\pm$ & $\mathbf 1$ & $\mathbf 1$ & $2+\frac{2 s+n t}{n+1}$ \\
		$W_u^\pm$ & $\mathbf 1$ & $\mathbf 1$ & $\frac{4 n+6+4 u}{n+1}$ \\
		\hline
		$q$ & $\mathbf 1$ & $\overline{\mathbf{N_f}}$ & $\frac{(2-n) N_f-\tilde N_c+N_c}{2 (n+1) N_f}$ \\
		$\tilde q$ & $\overline{\mathbf{N_f}}$ & $\mathbf 1$ & $\frac{(2-n) N_f-\tilde N_c+N_c}{2 (n+1) N_f}$ \\
		$\hat X$ & $\mathbf 1$ & $\mathbf 1$ & $\frac{2}{n+1}$ \\
		$\hat Y$ & $\mathbf 1$ & $\mathbf 1$ & $\frac{n}{n+1}$ \\
		$M_{s,t}$ & $\mathbf{N_f}$ & $\mathbf{N_f}$ & $\frac{(2-n) N_f-N_c+\tilde N_c}{(n+1) N_f}+\frac{2s+nt}{n+1}$ \\
		$\hat V_{s,t}^\pm$ & $\mathbf 1$ & $\mathbf 1$ & $2+\frac{2 s+n t}{n+1}$ \\
		$\hat W_u^\pm$ & $\mathbf 1$ & $\mathbf 1$ & $\frac{4 n+6+4 u}{n+1}$ \\
		\hline
	\end{tabular}
	\caption{\label{tab:V00 charges} The representations of the chiral operators under the global symmetry groups of the double adjoint theory with the linear monopole superpotential $\Delta W_A = V_{0,0}^++V_{0,0}^-$ and its dual. Now the monopole operators $(V_{s,t}^\pm,W_u^\pm)$ and $(\hat V_{s,t}^\pm,\hat W_u^\pm)$ are presented together with the elementary fields in the upper box and the lower box respectively. Note that $\tilde N_c$ is the dual gauge rank, which is defined by $\tilde N_c = 3 n N_f-N_c-4 n-2$.}
\end{table}
\\

Next, let us consider the deformation
\begin{align}
\label{eq:W0}
\Delta W_A = W_0^+ + W_0^- \,.
\end{align}
The duality we propose is as follows.
\begin{itemize}
\item Theory A is the 3d $\mathcal N = 2$ $U(N_c)$ gauge theory with $N_f$ pairs of fundamental fields $Q^a$ and anti-fundamental fields $\tilde Q^{\tilde a}$, two adjoint fields $X, \, Y$ and the superpotential
\begin{align}
W_A^{mon}= \mathrm{Tr} X^{n+1}+\mathrm{Tr} X Y^2+W_{0}^++W_{0}^-
\end{align}
where $W_{0}^\pm$ are a pair of monopole operators of Theory A with topological symmetry charge $\pm2$.
\item Theory B is the 3d $\mathcal N = 2$ $U(3 n N_f-N_c-2 n+2)$ gauge theory with $N_f$ pairs of fundamental fields $q_{\tilde a}$ and anti-fundemental fields $\tilde q_a$, two adjoint fields $\hat X, \, \hat Y$, and $3 n {N_f}^2$ gauge singlet fields $M_{s,t}{}^{\tilde a a}$ for $a,\tilde a = 1,\dots, N_f$, $s = 0,\dots,n-1$ and $t = 0,1,2$.
The superpotential is given by
\begin{align}
\label{eq:W0 dual}
W_B^{mon} = \mathrm{Tr} \hat X^{n+1}+\mathrm{Tr} \hat X \hat Y^2+\sum_{s = 0}^{n-1} \sum_{t = 0}^2 M_{s,t} \tilde q \hat X^{n-1-s} \hat Y^{2-t} q+\hat W_{0}^+ + \hat W_{0}^- \,.
\end{align}
where $\hat W_{0}^\pm$ are a pair of monopole operators of Theory B with topological symmetry charge $\pm2$.
\end{itemize}

To check the duality, we again turn on the polynomial superpotential $\Delta W = \mathrm{Tr} \left[f_n(X)\right]$, which breaks the gauge group $U(N_c)$ into
\begin{align}
\prod_{i=1}^{n+2} U(m^{1d}_i) \prod_{j=1}^{\frac{n-1}{2}} U(m^{2d}_j) \,.
\end{align}
Recall that $U(m^{2d}_j)$ is embedded in $U(N_c)$ as a diagonal subgroup of $U(m^{2d}_j) \times U(m^{2d}_j) \subset U(2 m^{2d}_j) \subset U(N_c)$. Thus, the unit charged monopole operators of the $U(m^{2d}_j)$ sector should correspond to the doubly charged monopole operators of the unbroken $U(N_c)$ theory. We therefore expect that the deformation \eqref{eq:W0} by $W_0^\pm$ only leads to the monopole superpotential for the 2-dimensional VEV sectors, in contrast to the deformation \eqref{eq:V00} by $V_{0,0}^\pm$ leading to the monopole superpotentials for both the 1-dimensional VEV sectors and 2-dimensional VEV sectors. As a result, the dual theory of each sector is given by
\begin{align}
U(m^{1d}_i) \quad &\longleftrightarrow \quad U(N_f-m^{1d}_i) \,, \\
U(m^{2d}_i) \quad &\longleftrightarrow \quad U(2 N_f-m^{2d}_i-2)
\end{align}
with the superpotential
\begin{align}
\begin{aligned}
\label{eq:broken W0 dual sup}
W^{(1d,i)}_B &= M^{(1d,i)} \tilde q^{(1d,i)} q^{(1d,i)}+V^{(1d,i)}{}^+ \hat V^{(1d,i)}{}^-+V^{(1d,i)}{}^- \hat V^{(1d,i)}{}^+ \,, \\
W^{(2d,i)}_B &= M^{(2d,i)} \tilde q^{(2d,i)} q^{(2d,i)}+\hat V^{(2d,i)}{}^++\hat V^{(2d,i)}{}^-
\end{aligned}
\end{align}
where we have applied the Aharony duality \cite{Aharony:1997gp} for the 1-dimensional VEV sectors and the BBP duality \cite{Benini:2017dud} for the 2-dimensional VEV sectors.

This should be the low energy description of the entire dual theory deformed by the polynomial superpotential $\Delta W = \mathrm{Tr} \left[f_n(\hat X)\right]$. The expected theory leading to such a low energy description when deformed by $\mathrm{Tr} \left[f_n(\hat X)\right]$ is the $U(3 n N_f-N_c-2 n+2)$ theory with the superpotential \eqref{eq:W0 dual}. One can check that once we turn on the polynomial superpotential $\mathrm{Tr} \left[f_n(\hat X)\right]$, the first and second terms lead to the non-zero VEVs of $\hat X$ and $\hat Y$, which break the gauge group $U(3 n N_f-N_c-2 n+2)$ into
\begin{align}
\prod_{i=1}^{n+2} U(N_f-m^{1d}_i) \prod_{j=1}^{\frac{n-1}{2}} U(2N_f -m^{2d}_j-2) \,,
\end{align}
while the remaining terms descend to the superpotential \eqref{eq:broken W0 dual sup} of the broken theory. Those relations can be summarized as:
\begin{align}
\begin{array}{ccc}
U(N_c) & \quad \longleftrightarrow \quad & U(3nN_f -N_c-2 n+2) \\
\big\downarrow & & \big\downarrow \\
{\displaystyle \prod_{i=1}^{n+2} U(m^{1d}_i) \prod_{j=1}^{\frac{n-1}{2}} U(m^{2d}_j)} & \quad \longleftrightarrow \quad & {\displaystyle \prod_{i=1}^{n+2} U(N_f-m^{1d}_i) \prod_{j=1}^{\frac{n-1}{2}} U(2N_f -m^{2d}_j-2)}
\end{array}
\end{align}
where a horizontal arrow denotes the duality between the two theories, while a vertical arrow indicates the perturbation by the polynomial superpotential of the adjoint field $X$ or $\hat X$.

One can check that the superpotentials \eqref{eq:W0} and \eqref{eq:W0 dual} are consistent with the conjectured duality. First we note that the monopole superpotential \eqref{eq:W0} demands that the $R$-charge $\Delta_Q$ of $Q$ and $\tilde Q$ must be
\begin{align}
\label{eq:double adjoint Delta_Q 2}
\Delta_Q = \frac{(n+1) N_f-N_c-n+1}{(n+1) N_f}
\end{align}
because the $R$-charge of the monopole operator $W_0^\pm$:
\begin{align}
2 (1-\Delta_Q) N_f-\frac{2}{n+1} (N_c-1)+\frac{2}{n+1}
\end{align}
should be 2. The $R$-charge of $q$ and $\tilde q$ is then given by
\begin{align}
\frac{2-n}{n+1}-\Delta_Q = \frac{-(2 n-1) N_f+N_c+n-1}{(n+1) N_f} \,.
\end{align}
This requires the $R$-charge of the dual monopole operators $\hat W_0^\pm$ to be
\begin{align}
&\quad 2 (1-\Delta_q) N_f-\frac{2}{n+1} (\tilde N_c-1)+\frac{2}{n+1} \nonumber \\
&= 2 \left(1-\frac{2-n}{n+1}+\Delta_Q\right) N_f-\frac{2}{n+1} (3 n N_f-N_c-2 n+1)+\frac{2}{n+1} \nonumber \\
&= 2 \,,
\end{align}
which is consistent with the dual monopole superpotential \eqref{eq:W0 dual}. The resulting global charges are summarized in Table \ref{tab:W0 charges}.
\begin{table}[tbp]
	\centering
	\begin{tabular}{|c|ccc|}
		\hline
		& $SU(N_f)_t$ & $SU(N_f)_u$ & $U(1)_R$ \\
		\hline
		$Q$ & $\mathbf{N_f}$ & $\mathbf 1$ & $\frac{(2-n) N_f-N_c+\tilde N_c}{2 (n+1) N_f}$ \\
		$\tilde Q$ & $\mathbf 1$ & $\mathbf{N_f}$ & $\frac{(2-n) N_f-N_c+\tilde N_c}{2 (n+1) N_f}$ \\
		$X$ & $\mathbf 1$ & $\mathbf 1$ & $\frac{2}{n+1}$ \\
		$Y$ & $\mathbf 1$ & $\mathbf 1$ & $\frac{n}{n+1}$ \\
		$V_{s,t}^\pm$ & $\mathbf 1$ & $\mathbf 1$ & $\frac{n+2 s+n t}{n+1}$ \\
		$W_u^\pm$ & $\mathbf 1$ & $\mathbf 1$ & $2+\frac{4 u}{n+1}$ \\
		\hline
		$q$ & $\mathbf 1$ & $\overline{\mathbf{N_f}}$ & $\frac{(2-n) N_f-\tilde N_c+N_c}{2 (n+1) N_f}$ \\
		$\tilde q$ & $\overline{\mathbf{N_f}}$ & $\mathbf 1$ & $\frac{(2-n) N_f-\tilde N_c+N_c}{2 (n+1) N_f}$ \\
		$\hat X$ & $\mathbf 1$ & $\mathbf 1$ & $\frac{2}{n+1}$ \\
		$\hat Y$ & $\mathbf 1$ & $\mathbf 1$ & $\frac{n}{n+1}$ \\
		$M_{s,t}$ & $\mathbf{N_f}$ & $\mathbf{N_f}$ & $\frac{(2-n) N_f-N_c+\tilde N_c}{(n+1) N_f}+\frac{2s+nt}{n+1}$ \\
		$\hat V_{s,t}^\pm$ & $\mathbf 1$ & $\mathbf 1$ & $\frac{n+2 s+n t}{n+1}$ \\
		$\hat W_u^\pm$ & $\mathbf 1$ & $\mathbf 1$ & $2+\frac{4 u}{n+1}$ \\
		\hline
	\end{tabular}
	\caption{\label{tab:W0 charges} The representations of the chiral operators under the global symmetry groups of the double adjoint theory with the linear monopole superpotential $\Delta W_A = W_{0}^++W_{0}^-$ and its dual. The monopole operators $(V_{s,t}^\pm,W_u^\pm)$ and $(\hat V_{s,t}^\pm,\hat W_u^\pm)$ are presented together with the elementary fields in the upper box and the lower box respectively. Note that $\tilde N_c$ is the dual gauge rank, which is defined by $\tilde N_c = 3 n N_f-N_c-2 n+2$.}
\end{table}We will also provide further evidence of the duality using the superconformal index. See section \ref{sec:W0}.

\subsection{The superconformal index for $\Delta W = V_{0,0}^+ + V_{0,0}^-$}
\label{sec:V00}

In this subsection, we provide nontrivial evidence of the proposed duality for the double adjoint theory with monopole superpotential
\begin{align}
\label{eq:V00 2}
\Delta W_A = V_{0,0}^+ + V_{0,0}^-
\end{align}
using the superconformal index. The index of the double adjoint theory without the monopole superpotential can be computed using the factorization formula derived in \cite{Hwang:2018uyj} with some trial $R$-charge. Then, as explained in the single adjoint case, the index with the monopole superpotential can be obtained by taking the $R$-charge of the fundamental matters to be $\Delta_Q$ determined by \eqref{eq:double adjoint Delta_Q 1} and turning off the fugacities of $U(1)_A$ and $U(1)_T$, which are broken by the monopole superpotential. We have computed the indices for several dual pairs, some of which are listed in Table \ref{tab:V00}.
\begin{table}[tbp]
	\centering
	\begin{tabular}{|c|c|c|}
		\hline
			n	&	$(N_f , N_c , \tilde{N}_c)$	&	SCI	\\ 
		\hline
				&	$(4,22,0)$				&	$\begin{array}{c}
											1+16 \sqrt[8]{x}+136 \sqrt[4]{x}+832 x^{3/8}+4132 \sqrt{x}+O\left(x^{5/8}\right)
											\end{array}$	\\
		\cline{2-3}
				&	$(4,21,1)$				&	$\begin{array}{c}
											1+16 \sqrt[4]{x}+153 \sqrt{x}+1089 x^{3/4}+6373 x+O\left(x^{5/4}\right)
											\end{array}$	\\
		\cline{2-3}
			3	&	$(4,20,2)$				&	$\begin{array}{c}
											1+16 \sqrt[8]{x}+136 \sqrt[4]{x}+832 x^{3/8}+4133 \sqrt{x}+17712 x^{5/8}+67849 x^{3/4} \\
											+237760 x^{7/8}+774456 x+2372208 x^{9/8}+6893045 x^{5/4} \\
											+19130640 x^{11/8}+50987445 x^{3/2}+O\left(x^{13/8}\right)
											\end{array}$	\\
		\cline{2-3}
				&	$(4,19,3)$				&	$\begin{array}{c}
											1+16 \sqrt[4]{x}+153 \sqrt{x}+1105 x^{3/4}+6614 x+34457 x^{5/4} \\
											+160992 x^{3/2}+O\left(x^{7/4}\right)
											\end{array}$	\\
		\hline  
	\end{tabular}
	\caption{\label{tab:V00} The superconformal index results for the double adjoint theories with $\Delta W_A = V_{0,0}^+ + V_{0,0}^-$. Here we list a few examples with $(n,N_f) = (3,4)$, whereas more results with $(n,N_f) = (3,3)$ are given in appendix \ref{sec:index results}. The $SU(N_f)_t \times SU(N_f)_u$ flavor fugacities are all omitted for simplicity. The gauge rank of the dual theory is given by $\tilde N_c = 3 n N_f-N_c-4 n-2$.}
\end{table}
In some cases, $\Delta_Q$ determined by \eqref{eq:double adjoint Delta_Q 1} becomes negative, and therefore, some mesonic operators have negative conformal dimensions. Since such operators are decoupled from the interacting theory, we remove their contributions from the index by flipping them; i.e., we introduce extra singlets coupled to those decoupled operators so that both of them become massive and integrated out. The evaluated indices show perfect agreement, which is strong evidence of the duality we propose.
\\

Furthermore, we have found that the linear monopole superpotential \eqref{eq:V00 2} results in the nontrivial truncation of the chiral ring generators. First of all, as argued in the previous subsection, all the monopole operators become massive once the superpotential \eqref{eq:V00 2} is turned on:
\begin{align}
\label{eq:monopole truncation 2}
V_{s,t}^\pm \quad &\sim \quad 0 \,, \qquad \qquad s = 0, \dots, n, \quad t = 0,1, \quad st = 0 \,, \\
W_{u}^\pm \quad &\sim \quad 0 \,, \qquad \qquad u = 0, \dots, \frac{n-3}{2} \,,
\end{align}
which can be explicitly confirmed by the superconformal index as we will show shortly. Therefore, in generic cases, the moduli space is parameterized by the following chiral ring generators:
\begin{align}
\tilde Q X^s Y^t Q \quad &\longleftrightarrow \quad M_{s,t} \,, \qquad \qquad s = 0, \dots, n-1, \quad t = 0,1,2 \,, \\
\mathrm{Tr} X^s \quad &\longleftrightarrow \quad \mathrm{Tr} \hat X^s \,, \qquad \quad \;\; s = 1, \dots, n \,, \\
\mathrm{Tr} Y \quad &\longleftrightarrow \quad \mathrm{Tr} \hat Y
\end{align}
where the right hand side shows the corresponding dual operators, which describe the same moduli space. In addition, as in the single adjoint case, there is the additional truncation of chiral ring generators if the gauge ranks are bounded in a particular range:
\begin{align}
\begin{aligned}
\label{eq:V00 adj truncation 2}
\mathrm{Tr} X^s \quad &\sim \quad 0 \,, \quad \quad s = \mathrm{min}(N_c+1,3 n N_f-N_c-4 n-1), \dots, n \,, \\
\mathrm{Tr} Y \quad &\sim \quad 0 \,, \quad \quad \text{if} \quad \mathrm{min}(N_c,3 n N_f-N_c-4 n-2) = 0 \,,
\end{aligned}
\end{align}
which happens when $\mathrm{min}(N_c,3 n N_f-N_c-4 n-2) < n$.
\\

To see the truncation of the monopole operators \eqref{eq:monopole truncation 2}, let us consider the case $(n,N_f,N_c) = (3,3,5)$ as an example. To read off chiral ring relations, it is convenient to evaluate the plethystic logarithm of the index\footnote{The (unrefined) index of the $(n,N_f,N_c) = (3,3,5)$ case is given in appendix \ref{sec:index results}.}, which is given by
\begin{align}
\label{eq:pl335}
&\quad (1-x^2) \, \mathrm{PL}\left[I\right] \nonumber \\
&=\left(\mathbf 3_t \mathbf 3_u \tau^2+1\right) x^\frac12+\dots+\left(\tau^{-3} \left(w+w^{-1}\right)-2 \tau^6-2+\dots\right) x^2+\left(\dots\right) x^\frac94 \nonumber \\
&\quad +\left(\tau^{-3} \left(w+w^{-1}\right)-3 \tau^6-2+\dots\right) x^\frac52+\left(\tau^{-3} \left(w+w^{-1}\right)-3 \tau^6-3+\dots\right) x^\frac{11}{4} \nonumber \\
&\quad +\left(\left(\tau^{-3}-\tau^3\right) \left(w+w^{-1}\right)-2 \tau^6-3+\dots\right) x^3+\left(\dots\right) x^\frac{13}{4} \nonumber \\
&\quad +\left(\left(\tau^{-3}-2 \tau^3\right) \left(w+w^{-1}\right)+5 \tau^6-4+\dots\right) x^\frac{7}{2}+\dots \nonumber \\
&\quad +\left(\tau^{-6} \left(w^2+w^{-2}\right)+33 \tau^6-\tau^{-3} \left(w+w^{-1}\right)-18 \tau^{12}-2+\dots\right) x^\frac{9}{2}+\dots
\end{align}
where we keep $\tau$ and $w$ nontrivial; thus, this corresponds to the index before turning on the monopole superpotential. Note that both sides are multiplied by $(1-x^2)$ to remove the contribution of descendants derived by the derivative operator. We have used $\Delta_Q$ determined by \eqref{eq:double adjoint Delta_Q 1} as a trial $R$-charge and also suppressed many terms irrelevant to our discussion here. One can see the contributions of the monopole operators $V_{s,t}^\pm$:
\begin{align}
\tau^{-3} \left(w+w^{-1}\right) x^\frac{8+2 s+3 t}{4}
\end{align}
for $s = 0,1,2, 3, \, t = 0,1, \, st = 0$ and that of $W_0^\pm$:
\begin{align}
\tau^{-6} \left(w^2+w^{-2}\right) x^\frac{9}{2} \,.
\end{align}
Now recall that the index after turning on the monopole superpotential can be simply obtained by taking $\tau = w = 1$. Then the contributions of $V_{s,t}^\pm$ and $W_0^\pm$ would be canceled by negative contributions in the trivial representation of the $SU(3)_t \times SU(3)_u$ global symmetry, which are shown in \eqref{eq:pl335}. Especially, motivated by the result of the single adjoint case, we expect that, in general, the contributions of $V_{s,t}^\pm$ are canceled by the negative contributions of the following fermionic operators:
\begin{align}
(\psi_Q^\dagger)^a X^s Y^t Q_a \,, \quad \tilde Q^{\tilde a} X^s Y^t (\psi_{\tilde Q}^\dagger)_{\tilde a} \quad &: \qquad -2 x^\frac{8+2 s+3 t}{4}
\end{align}
for $s = 0,1,2, 3, \, t = 0,1, \, st = 0$. On the other hand, we couldn't identify the general pattern of the cancelation for the monopole operators $W_u^\pm$ because their contributions appear at higher orders of $x$, which are hard to evaluate for higher values of $N_f$ and $N_c$. Nevertheless, we have confirmed for $N_c \leq 3$ and $N_c = 5$ that there are some negative contributions, which may cancel the contribution of $W_u^\pm$. Such cancelations imply that the monopole operators $V_{s,t}^\pm$ and $W_0^\pm$ become Q-exact and vanish in the chiral ring; i.e., the Coulomb branch of the moduli space is completely lifted.
\\

In addition, we also find that the monopole superpotential \eqref{eq:V00 2} can give rise to extra constraints on $\mathrm{Tr} X^s$ and $\mathrm{Tr} Y$ as shown in \eqref{eq:V00 adj truncation 2}. As in the single adjoint case, we first note that the operators $\mathrm{Tr} X^s$ for $s = 1, \dots, n$ are classically truncated as
\begin{align}
\mathrm{Tr} X^s \quad &\sim \quad 0 \,, \qquad \quad s = N_c+1, \dots, n
\end{align}
due to the characteristic equation of the matrix field $X$ for $N_c < n$. Note that this doesn't depend on the existence of the monopole superpotential.

On the other hand, there is also quantum mechanical truncation, whose effect does depend on the presence of the monopole superpotential. Before we turn on the monopole superpotential, the operators $\mathrm{Tr} X^s$ get extra quantum constraints:
\begin{align}
\mathrm{Tr} X^s \quad &\sim \quad 0 \,, \qquad \quad s = 3 n N_f-N_c+1, \dots, \mathrm{min}(n,N_c)
\end{align}
 if $3 n N_f-N_c < \mathrm{min}(n,N_c)$ because their dual operators $\mathrm{Tr} \hat X^s$ are classically constrained as follows:
\begin{align}
\mathrm{Tr} \hat X^s \quad &\sim \quad 0 \,, \qquad \quad s = 3 n N_f-N_c+1, \dots, n \,.
\end{align}
Similarly, the operator $\mathrm{Tr} Y$ also gets quantum constraint
\begin{align}
\mathrm{Tr} Y \quad \sim \quad 0
\end{align}
if $3 n N_f-N_c = 0$ because in this case the dual gauge rank is zero, and the dual operator $\mathrm{Tr} \hat Y$ doesn't exist.

Now we turn on the monopole superpotential \eqref{eq:V00 2}. Then the dual gauge rank decreases from $3 n N_f-N_c$ to $3 n N_f-N_c-4 n+2$, which affects the quantum constraints on $\mathrm{Tr} X^s$ and $\mathrm{Tr} Y$. Indeed, if the dual rank is smaller than $\mathrm{min}(n, N_c)$, we expect quantum constraints on $\mathrm{Tr} X^s$ as follows:
\begin{align}
\mathrm{Tr} X^s \quad &\sim \quad 0 \,, \quad \quad s = 3 n N_f-N_c-4 n-1, \dots, \mathrm{min}(n,N_c) \,.
\end{align}
Furthermore, $\mathrm{Tr} Y$ now vanishes when $\tilde N_c = 3 n N_f-N_c-4 n-2 = 0$. Therefore, combined with the original constraints before turning on the monopole superpotential, the complete truncation of $\mathrm{Tr} X^s$ and $\mathrm{Tr} Y$ in the presence of the monopole superpotential \eqref{eq:V00 2} is given by \eqref{eq:V00 adj truncation 2}.

Let us discuss some examples. We are going to consider the cases: $N_c = 19, \, 20, \, 21, \, 22$ with $(n,N_f) = (3,4)$. In those cases, the dual ranks are given by $\tilde N_c = 3, \, 2, \, 1, \, 0$ respectively. Firstly, for $N_c = 19$, the plethystic log of the index is given by
\begin{align}
(1-x^2) \, \mathrm{PL}[I] &= \mathbf 4_t \mathbf 4_u \tau^2 x^\frac14+\left(\mathbf 4_t \mathbf 4_u \tau^2+1\right) x^\frac12+\left(\mathbf 4_t \mathbf 4_u \tau^2+1\right) x^\frac34+x+\mathbf 4_t \mathbf 4_u \tau^2 x^\frac54+x^\frac32 \nonumber \\
&\quad +\dots
\end{align}
where $\tau$ has to be 1 if the monopole superpotential is turned on. One can see that the contributions of $\mathrm{Tr} Y$ and $\mathrm{Tr} X^s$ for $s = 1, \, 2, \, 3$,
\begin{align}
\label{eq:adj cont}
x^\frac34 \,, \qquad x^\frac{s}{2} \,, \quad s = 1, 2, 3 \,,
\end{align}
are all nontrivial regardless of $\tau$. Therefore, those operators are not affected by the monopole superpotential, which is consistent with the fact that the extra quantum truncation happens only when $\mathrm{min}(N_c,3 n N_f-N_c-4 n-2) < n$.

Next, for $N_c = 20$, the plethystic log of the index is given by
\begin{align}
(1-x^2) \, \mathrm{PL}[I] &= \mathbf 4_t \mathbf 4_u \tau^2 x^\frac18+\mathbf 4_t \mathbf 4_u \tau^2 x^\frac38+x^\frac12+\mathbf 4_t \mathbf 4_u \tau^2 x^\frac58+x^\frac34+x+\mathbf 4_t \mathbf 4_u \tau^2 x^\frac98 \nonumber \\
&\quad +\left(1-\tau^{16}\right) x^\frac32+\dots \,.
\end{align}
For $\tau \neq 1$, i.e., before turning on the monopole superpotential, one can find the nontrivial contributions \eqref{eq:adj cont} of $\mathrm{Tr} Y$ and $\mathrm{Tr} X^s$ for $s = 1, \, 2, \, 3$. On the other hand, once we set $\tau = 1$, i.e., after turning on the monopole superpotential, the contribution $x^\frac32$ of $\mathrm{Tr} X^3$ is canceled by $-\tau^{16} x^\frac32$. This shows that $\mathrm{Tr} X^3$ is quantum mechanically truncated, which is consistent with the fact that its dual operator $\mathrm{Tr} \hat X^3$ is classically truncated because the dual gauge rank is 2.

Similarly, for $N_c = 21$, the plethystic log of the index is given by
\begin{align}
(1-x^2) \, \mathrm{PL}[I] &= \mathbf 4_t \mathbf 4_u \tau^2 x^\frac14+\left(\mathbf 4_t \mathbf 4_u \tau^2+1\right) x^\frac12+x^\frac34+\left(\mathbf 4_t \mathbf 4_u \tau^2+1-\tau^{16}\right) x+\dots \,.
\end{align}
where we have only evaluated the index up to $x$ due to the limited computing power. Nevertheless, up to this order, we find all the expected contributions of $\mathrm{Tr} Y$ and $\mathrm{Tr} X^s$ for $s = 1, \, 2$ when $\tau \neq 1$. Once we set $\tau = 1$, the contribution $x$ of $\mathrm{Tr} X^2$ is canceled by $-\tau^{16} x$, which is expected because the dual operator of $\mathrm{Tr} X^2$ is classically truncated in the dual $U(1)$ theory.

Lastly, for $N_c = 22$, the plethystic log of the index is given by
\begin{align}
(1-x^2) \, \mathrm{PL}[I] &= \mathbf 4_t \mathbf 4_u \tau^2 x^\frac18+\mathbf 4_t \mathbf 4_u \tau^2 x^\frac38+\left(1-\tau^{16}\right) x^\frac12+\dots \,.
\end{align}
which is evaluated only up to $x^\frac12$ due to the limited computing power. One can see the contribution of $\mathrm{Tr} X$ when $\tau \neq 1$, which is canceled by $-\tau^{16} x^\frac12$ when $\tau = 1$, which shows the expected quantum truncation \eqref{eq:V00 adj truncation 2} of $\mathrm{Tr} X$.

\subsection{The superconformal index for $\Delta W = W_0^+ + W_0^-$}
\label{sec:W0}

Lastly, we test our proposal for the duality with monopole superpotentail
\begin{align}
\label{eq:W0 2}
\Delta W_A = W_{0}^+ + W_{0}^-
\end{align}
using the superconformal index. Again the index is computed using the factorization formula in \cite{Hwang:2018uyj} with the $R$-charge fixed by the formula \eqref{eq:double adjoint Delta_Q 2}. We have computed the indices for several examples, some of which are shown in Table \ref{tab:W0}, where the operators with negative conformal dimensions are all flipped. We observe the exact agreement of the indices for each dual pair.
\begin{table}[tbp]
	\centering
	\begin{tabular}{|c|c|c|}
		\hline
			n	&	$(N_f , N_c , \tilde{N}_c)$	&	SCI	\\ 
		\hline
				&	$(2,22,0)$				&	$\begin{array}{c}
											1+4 \sqrt[3]{x}+14 x^{2/3}+36 x+81 x^{4/3}+156 x^{5/3}+272 x^2 \\
											+428 x^{7/3}+628 x^{8/3}+O\left(x^{17/6}\right)
											\end{array}$	\\
		\cline{2-3}
				&	$(2,21,1)$				&	$\begin{array}{c}
											1+4 \sqrt[6]{x}+11 \sqrt[3]{x}+28 \sqrt{x}+62 x^{2/3}+131 x^{5/6}+264 x+500 x^{7/6} \\
											+917 x^{4/3}+1619 x^{3/2}+2771 x^{5/3}+4630 x^{11/6}+7510 x^2 \\
											+11915 x^{13/6}+18502 x^{7/3}+28116 x^{5/2}+41987 x^{8/3}+O\left(x^{17/6}\right)
											\end{array}$	\\
		\cline{2-3}
				&	$(2,20,2)$				&	$\begin{array}{c}
											1+4 \sqrt[6]{x}+15 \sqrt[3]{x}+40 \sqrt{x}+105 x^{2/3}+239 x^{5/6}+535 x+1103 x^{7/6} \\
											+2233 x^{4/3}+4284 x^{3/2}+8075 x^{5/3}+14652 x^{11/6}+26146 x^2 \\
											+45332 x^{13/6}+77380 x^{7/3}+129092 x^{5/2}+212259 x^{8/3}+O\left(x^{17/6}\right)
											\end{array}$	\\
		\cline{2-3}
			5	&	$(2,19,3)$				&	$\begin{array}{c}
											1+4 \sqrt[6]{x}+15 \sqrt[3]{x}+44 \sqrt{x}+117 x^{2/3}+287 x^{5/6}+658 x+1439 x^{7/6} \\
											+3008 x^{4/3}+6071 x^{3/2}+11870 x^{5/3}+22569 x^{11/6}+41879 x^2 \\
											+75983 x^{13/6}+135121 x^{7/3}+235897 x^{5/2}+404861 x^{8/3}+O\left(x^{17/6}\right)
											\end{array}$	\\
		\cline{2-3}
				&	$(2,18,4)$				&	$\begin{array}{c}
											1+4 \sqrt[6]{x}+15 \sqrt[3]{x}+44 \sqrt{x}+121 x^{2/3}+299 x^{5/6}+706 x+1567 x^{7/6} \\
											+3359 x^{4/3}+6911 x^{3/2}+13829 x^{5/3}+26856 x^{11/6}+50982 x^2 \\
											+94560 x^{13/6}+172085 x^{7/3}+307324 x^{5/2}+540035 x^{8/3}+O\left(x^{17/6}\right)
											\end{array}$	\\
		\cline{2-3}
				&	$(2,17,5)$				&	$\begin{array}{c}
											1+4 \sqrt[6]{x}+15 \sqrt[3]{x}+44 \sqrt{x}+121 x^{2/3}+303 x^{5/6}+718 x+1615 x^{7/6} \\
											+3487 x^{4/3}+7267 x^{3/2}+14684 x^{5/3}+28880 x^{11/6}+55441 x^2 \\
											+104155 x^{13/6}+191864 x^{7/3}+347173 x^{5/2}+617955 x^{8/3}+O\left(x^{17/6}\right)
											\end{array}$	\\
													\cline{2-3}
				&	$(2,16,6)$				&	$\begin{array}{c}
											1+4 \sqrt[6]{x}+15 \sqrt[3]{x}+44 \sqrt{x}+121 x^{2/3}+303 x^{5/6}+722 x+1627 x^{7/6} \\
											+3535 x^{4/3}+7395 x^{3/2}+15040 x^{5/3}+29740 x^{11/6}+57480 x^2 \\
											+108679 x^{13/6}+201631 x^{7/3}+367444 x^{5/2}+659006 x^{8/3}+O\left(x^{19/6}\right)
											\end{array}$	\\
		\hline  
	\end{tabular}
	\caption{\label{tab:W0} The superconformal index results for the double adjoint theories with $\Delta W_A = W_{0}^+ + W_{0}^-$. Here we list a few examples with $(n,N_f) = (5,2)$, whereas the other results with $(n,N_f) = (5,2)$ and those with $(n,N_f) = (3,1), \, (3,2)$ are given in appendix \ref{sec:index results}. The $SU(N_f)_t \times SU(N_f)_u$ flavor fugacities are all omitted for simplicity. The gauge rank of the dual theory is given by $\tilde N_c = 3 n N_f-N_c-2 n+2$.}
\end{table}
\\

In addition, from the index, we also observe that some chiral ring generators are truncated due to the monopole superpotential \eqref{eq:W0 2}. Firstly, while the monopole operators $V_{s,t}^\pm$ remain massless, the other monopole operators $W_u^\pm$ all become massive once the superpotential \eqref{eq:W0 2} is turned on:
\begin{align}
\label{eq:monopole truncation 3}
W_{u}^\pm \quad &\sim \quad 0 \,, \qquad \qquad u = 0, \dots, \frac{n-3}{2} \,.
\end{align}
As a result, in generic cases, the moduli space are parametrized by
\begin{align}
\tilde Q X^s Y^t Q \quad &\longleftrightarrow \quad M_{s,t} \,, \qquad \quad \;\; s = 0, \dots, n-1, \quad t = 0,1,2 \,, \\
\mathrm{Tr} X^s \quad &\longleftrightarrow \quad \mathrm{Tr} \hat X^s \,, \qquad \quad s = 1, \dots, n \,, \\
\mathrm{Tr} Y \quad &\longleftrightarrow \quad \mathrm{Tr} \hat Y \,, \\
V_{s,t}^\pm \quad &\longleftrightarrow \quad \hat V_{s,t}^\pm \,, \qquad \qquad s = 0, \dots, n, \quad t = 0,1, \quad st = 0
\end{align}
where the right hand side shows the corresponding dual operators, which describe the same moduli space. Again, there is the extra truncation of chiral ring generators if the gauge ranks are in a certain range:
\begin{align}
\begin{aligned}
\label{eq:W0 adj truncation 2}
V_{s,0}^\pm \quad &\sim \quad 0 \,, \qquad \quad s = \mathrm{min}(N_c,3 n N_f-N_c-2 n+2), \dots, n \,, \\
V_{0,t}^\pm \quad &\sim \quad 0 \,, \qquad \quad t = \mathrm{min}(N_c,3 n N_f-N_c-2 n+2), \dots, 1 \,, \\
\mathrm{Tr} X^s \quad &\sim \quad 0 \,, \qquad \quad s = \mathrm{min}(N_c+1,3 n N_f-N_c-2 n+3), \dots, n \,, \\
\mathrm{Tr} Y \quad &\sim \quad 0 \,, \qquad \quad \text{if} \quad \mathrm{min}(N_c,3 n N_f-N_c-2 n+2) = 0 \,.
\end{aligned}
\end{align}
which happens when $\mathrm{min}(N_c,3 n N_f-N_c-2 n+2) \leq n$.
\\

To see the truncation of the monopole operators \eqref{eq:monopole truncation 3}, let us consider the case $(n,N_f,N_c) = (5,2,14)$, whose gauge rank and dual rank are large enough to avoid any accidental truncation of the monopole operators. The plethystic log of the index\footnote{The (unrefined) index of the $(n,N_f,N_c) = (5,2,14)$ case is given in appendix \ref{sec:index results}.} before turning on the monopole superpotential \eqref{eq:W0 2} is given by
\begin{align}
&\quad (1-x^2) \, \mathrm{PL}\left[I\right] \nonumber \\
&= \left(\mathbf 2_t \mathbf 2_u \tau^2\right) x^\frac16+\left(1+\mathbf 2_t \mathbf 2_u \tau^2\right) x^\frac13+\left(\mathbf 2_t \mathbf 2_u \tau^2\right) x^\frac12+\left(1+\mathbf 2_t \mathbf 2_u \tau^2\right) x^\frac23 \nonumber \\
&\quad +\left(1+\tau^{-2} \left(w+w^{-1}\right)+\mathbf 2_t \mathbf 2_u \tau^2\right) x^\frac56+\left(\tau^{-2} \left(w+w^{-1}\right)+\mathbf 2_t \mathbf 2_u \tau^2\right) x^\frac76+\left(1+\mathbf 2_t \mathbf 2_u \tau^2\right) x^\frac43 \nonumber \\
&\quad +\tau^{-2} \left(w+w^{-1}\right) x^\frac32+\left(1+\tau^{-2} \left(w+w^{-1}\right)+\mathbf 2_t \mathbf 2_u \tau^2\right) x^\frac53+\tau^{-2} \left(w+w^{-1}\right) x^\frac{11}{6} \nonumber \\
&\quad +\left(-2+\tau^{-4} \left(w^2+w^{-2}\right)-\mathbf 3_t-\mathbf 3_u+\mathbf 2_t \mathbf 2_u \tau^{-2}+\mathbf 2_t \mathbf 2_u \tau^2\right) x^2+\left(\tau^{-2} \left(w+w^{-1}\right)+\mathbf 2_t \mathbf 2_u \tau^{-2}\right) x^\frac{13}{6} \nonumber \\
&\quad +\left(-2-\tau^4-\mathbf 3_t-\mathbf 3_u+\mathbf 2_t \mathbf 2_u \tau^{-2}\right) x^\frac73+\tau^{-2} \left(w+w^{-1}\right) x^\frac52+ \nonumber \\
&\quad +\left(-2+\tau^{-4} \left(w^2+w^{-2}\right)-\tau^4-\mathbf 3_t-\mathbf 3_u+\mathbf 2_t \mathbf 2_u \left(\tau^{-2}-\tau^6\right)\right) x^\frac83+\dots
\end{align}
where we have used the $R$-charge $\Delta_Q$ determined by \eqref{eq:double adjoint Delta_Q 2}. Again, both sides are multiplied by $(1-x^2)$ to remove the contribution of descendants derived by the derivative operator. We find the contributions of the monopole operators $V_{s,t}^\pm$ for $s = 0,1,2,3,4,5, \, t = 0,1, \, st = 0$:
\begin{align}
\tau^{-2} \left(w+w^{-1}\right) x^\frac{5+2 s+5 t}{6}
\end{align}
and those of $W_u^\pm$ for $u = 0,1$:
\begin{align}
\tau^{-4} \left(w^2+w^{-2}\right) x^\frac{6+2 u}{3} \,.
\end{align}
Once we turn on the monopole superpotential \eqref{eq:W0 2}, we have to set $\tau = w = 1$. Then one can see that the contributions of $W_u^\pm$ are canceled by
\begin{align}
-2 x^\frac{6+2 u}{3} \,,
\end{align}
the contributions of fermionic operators
\begin{align}
(\psi_Q^\dagger)^a X^{2 u} Q_a
\end{align}
for $u = 0,1$, which is consistent with what we expect in \eqref{eq:monopole truncation 2}. On the other hand, the contributions of $V_{s,t}^\pm$ remain nontrivial regardless of $\tau$ and $w$, which also agrees with the expectation.
\\

The second effect of the monopole superpotential \eqref{eq:W0 2} is the quantum truncation of the operators $V_{s,t}^\pm$, $\mathrm{Tr} X^s$, and $\mathrm{Tr} Y$. Recall that, before turning on the monopole superpotential, $V_{s,t}^\pm \, (s = 0,\dots,n, \, t = 0,1, \, st = 0)$, $\mathrm{Tr} X^s \, (s = 0,\dots,n)$, and $\mathrm{Tr} Y$ are all nontrivial in the chiral ring only when both the original gauge rank and dual gauge rank are large enough. Otherwise, there are either classical or quantum constraints:
\begin{align}
\begin{aligned}
\label{eq:W0 adj truncation 1}
V_{s,0}^\pm \quad &\sim \quad 0 \,, \qquad \quad s = \mathrm{min}(N_c,3 n N_f-N_c), \dots, n \,, \\
V_{0,t}^\pm \quad &\sim \quad 0 \,, \qquad \quad t = \mathrm{min}(N_c,3 n N_f-N_c), \dots, 1 \,, \\
\mathrm{Tr} X^s \quad &\sim \quad 0 \,, \qquad \quad s = \mathrm{min}(N_c+1,3 n N_f-N_c+1), \dots, n \,, \\
\mathrm{Tr} Y \quad &\sim \quad 0 \,, \qquad \quad \text{if} \quad \mathrm{min}(N_c,3 n N_f-N_c) = 0 \,.
\end{aligned}
\end{align}
On the other hand, once we turn on the monopole superpotential \eqref{eq:W0 2}, the dual gauge rank decreases from $3 n N_f-N_c$ to $3 n N_f-N_c-2 n+2$, in which case we expect quantum constraints on $V_{s,t}^\pm$ or $\mathrm{Tr} X^s$ as follows:
\begin{align}
V_{s,0}^\pm \quad &\sim \quad 0 \,, \qquad \quad s = 3 n N_f-N_c-2 n+2, \dots, \mathrm{min}(n,N_c-1) \,, \\
V_{0,t}^\pm \quad &\sim \quad 0 \,, \qquad \quad t = 3 n N_f-N_c-2 n+2, \dots, \min(1,N_c-1) \,, \\
\mathrm{Tr} X^s \quad &\sim \quad 0 \,, \qquad \quad s = 3 n N_f-N_c-2 n+3, \dots, \mathrm{min}(n,N_c) \,, \\
\mathrm{Tr} Y \quad &\sim \quad 0 \,, \qquad \quad \text{if} \quad 3 n N_f-N_c-2 n+2 = 0 \,.
\end{align}
Therefore, combined with the original constraints in \eqref{eq:W0 adj truncation 1}, the complete truncation of $\mathrm{Tr} X^s$ and $\mathrm{Tr} Y$ in the presence of the monopole superpotential \eqref{eq:W0 2} is given by \eqref{eq:W0 adj truncation 2}.

Let us discuss some examples. We are going to consider the cases: $16 \leq N_c \leq 22$ with $(n,N_f) = (5,2)$, whose dual ranks are $6 \geq \tilde N_c \geq 0$ respectively. Firstly, for $N_c = 16$, the plethystic log of the index is given by
\begin{align}
&\quad (1-x^2) \, \mathrm{PL}\left[I^{N_c = 16}\right] \nonumber \\
&= \mathbf 2_t \mathbf 2_u \tau^2 x^\frac16+\left(1+\mathbf 2_t \mathbf 2_u \tau^2\right) x^\frac13+\mathbf 2_t \mathbf 2_u \tau^2 x^\frac12+\left(1+\mathbf 2_t \mathbf 2_u \tau^2\right) x^\frac23 \nonumber \\
&\quad +\left(1+\tau^{-2} \left(w+w^{-1}\right)+\mathbf 2_t \mathbf 2_u \tau^2\right) x^\frac56+\left(1+\mathbf 2_t \mathbf 2_u \tau^2\right) x+\tau^{-2} \left(w+w^{-1}\right) x^\frac76 \nonumber \\
&\quad +\left(1+\mathbf 2_t \mathbf 2_u \tau^2\right) x^\frac43+\tau^{-2} \left(w+w^{-1}\right) x^\frac32+\left(1+\tau^{-2} \left(w+w^{-1}\right)+\mathbf 2_t \mathbf 2_u \tau^2\right) x^\frac53 \nonumber \\
&\quad +\tau^{-2} \left(w+w^{-1}\right) x^\frac{11}{6}+\left(-2+\tau^{-4} \left(w^2+w^{-2}\right)-\mathbf 3_t-\mathbf 3_u+\mathbf 2_t \mathbf 2_u \tau^{-2}\right) x^2 \nonumber \\
&\quad +\left(\tau^{-2} \left(w+w^{-1}\right)+\mathbf 2_t \mathbf 2_u \tau^{-2}\right) x^\frac{13}{6}+\left(-2-\tau^4-\tau^8-\mathbf 3_t-\mathbf 3_u+\mathbf 2_t \mathbf 2_u \left(\tau^{-2}-\tau^6\right)\right) x^\frac{7}{3} \nonumber \\
&\quad +\left(\tau^{-2} \left(w+w^{-1}\right)+\mathbf 2_t \mathbf 2_u \tau^{-2}\right) x^\frac52+\dots
\end{align}
where $\tau$ and $w$ have to be 1 if the monopole superpotential is turned on. One can see that the contributions of $V_{s,0}^\pm$ and $V_{0,t}^\pm$ for $s = 1,\dots,5, \, t = 0,1$,
\begin{align}
\label{eq:mon cont}
\tau^{-2} \left(w+w^{-1}\right) x^\frac{5+2 s}{6} \,, \qquad \tau^{-2} \left(w+w^{-1}\right) x^\frac{5 (t+1)}{6} \,, \qquad \quad s = 1,\dots,5, \quad t = 0,1
\end{align}
and those of $\mathrm{Tr} Y$ and $\mathrm{Tr} X^s$ for $s = 1, \dots, 5$,
\begin{align}
\label{eq:adj cont}
x^\frac56 \,, \qquad x^\frac{s}{3} \,, \qquad \quad s = 1, \dots, 5 \,,
\end{align}
are all nontrivial regardless of $\tau$ and $w$. Therefore, those operators are not affected by the monopole superpotential, which is consistent with the fact that the extra quantum truncation happens only when $\mathrm{min}(N_c,3 n N_f-N_c-2 n+2) \leq n$.

Next, for $N_c = 17$, the plethystic log of the index is given by
\begin{align}
&\quad (1-x^2) \, \mathrm{PL}\left[I^{N_c = 17}\right] \nonumber \\
&= \mathbf 2_t \mathbf 2_u \tau^2 x^\frac16+\left(1+\mathbf 2_t \mathbf 2_u \tau^2\right) x^\frac13+\mathbf 2_t \mathbf 2_u \tau^2 x^\frac12+\left(1+\mathbf 2_t \mathbf 2_u \tau^2\right) x^\frac23 \nonumber \\
&\quad +\left(1+\tau^{-2} \left(w+w^{-1}\right)+\mathbf 2_t \mathbf 2_u \tau^2\right) x^\frac56+x+\left(\tau^{-2} \left(w+w^{-1}\right)+\mathbf 2_t \mathbf 2_u \tau^2\right) x^\frac76 \nonumber \\
&\quad +x^\frac43+\left(\tau^{-2} \left(w+w^{-1}\right)+\mathbf 2_t \mathbf 2_u \tau^2\right) x^\frac32+\left(1+\tau^{-2} \left(w+w^{-1}\right)\right) x^\frac53 \nonumber \\
&\quad +\tau^{-2} \left(w+w^{-1}\right) x^\frac{11}{6}+\left(-2+\tau^{-4} \left(w^2+w^{-2}\right)-\tau^8-\mathbf 3_t-\mathbf 3_u+\mathbf 2_t \mathbf 2_u \tau^{-2}\right) x^2 \nonumber \\
&\quad +\left(\tau^{-2} \left(w+w^{-1}\right)+\mathbf 2_t \mathbf 2_u \left(\tau^{-2}-\tau^6\right)\right) x^\frac{13}{6}+\left(-2-\tau^4-\tau^8-\mathbf 3_t-\mathbf 3_u+\mathbf 2_t \mathbf 2_u \tau^{-2} \right) x^\frac{7}{3} \nonumber \\
&\quad +\left(\left(\tau^{-2}-\tau^6\right) \left(w+w^{-1}\right)-\tau^8+\mathbf 2_t \mathbf 2_u \left(\tau^{-2}-\tau^6\right)\right) x^\frac52+\dots
\end{align}
For $\tau, w \neq 1$, i.e., before turning on the monopole superpotential, one can find the nontrivial contributions \eqref{eq:mon cont} and \eqref{eq:adj cont} of $V_{s,t}^\pm$, $\mathrm{Tr} Y$ and $\mathrm{Tr} X^s$, while once we set $\tau = w = 1$, i.e., after turning on the monopole superpotential, the contribution $\tau^{-2} \left(w+w^{-1}\right) x^\frac52$ of $V_{5,0}^\pm$ is canceled by $-\tau^6 \left(w+w^{-1}\right) x^\frac52$. This shows that $V_{5,0}^\pm$ are quantum mechanically truncated, which is consistent with the fact that its dual operators $\hat V_{5,0}^\pm$ are classically truncated because the dual gauge rank is 5.

Similarly, for $18 \leq N_c \leq 22$, the plethystic log of the indices are given by
\begin{align}
&\quad (1-x^2) \, \mathrm{PL}\left[I^{N_c = 18}\right] \nonumber \\
&= \mathbf 2_t \mathbf 2_u x^\frac16+\left(1+\mathbf 2_t \mathbf 2_u\right) x^\frac13+\mathbf 2_t \mathbf 2_u x^\frac12+\left(1+\mathbf 2_t \mathbf 2_u\right) x^\frac23+\left(1+w+w^{-1}\right) x^\frac56+\left(1+\mathbf 2_t \mathbf 2_u\right) x \nonumber \\
&\quad +\left(w+w^{-1}\right) x^\frac76+\left(1+\mathbf 2_t \mathbf 2_u\right) x^\frac43+\left(w+w^{-1}\right) x^\frac32+\left(w+w^{-1}\right) x^\frac53+\left(w+w^{-1}\right) x^\frac{11}{6} \nonumber \\
&\quad +\left(-3+w^2+w^{-2}-\mathbf 3_t-\mathbf 3_u\right) x^2+\left(-1+\mathbf 2_t \mathbf 2_u\right) x^\frac{13}{6}+\left(-4-\mathbf 3_t-\mathbf 3_u\right) x^\frac{7}{3} \nonumber \\
&\quad -\mathbf 2_t \mathbf 2_u \left(w+w^{-1}\right) x^\frac52+\dots \,,
\end{align}
\begin{align}
&\quad (1-x^2) \, \mathrm{PL}\left[I^{N_c = 19}\right] \nonumber \\
&= \mathbf 2_t \mathbf 2_u x^\frac16+\left(1+\mathbf 2_t \mathbf 2_u\right) x^\frac13+\mathbf 2_t \mathbf 2_u x^\frac12+x^\frac23+\left(1+w+w^{-1}+\mathbf 2_t \mathbf 2_u\right) x^\frac56+x \nonumber \\
&\quad +\left(w+w^{-1}+\mathbf 2_t \mathbf 2_u\right) x^\frac76+\left(w+w^{-1}\right) x^\frac32+\left(w+w^{-1}\right) x^\frac53-\left(1+\mathbf 2_t \mathbf 2_u\right) x^\frac{11}{6} \nonumber \\
&\quad +\left(-3+w^2+w^{-2}-\mathbf 3_t-\mathbf 3_u+\mathbf 2_t \mathbf 2_u\right) x^2+\left(-6-\left(1+\mathbf 2_t \mathbf 2_u\right) \left(w+w^{-1}\right)-\mathbf 3_t-\mathbf 3_u\right) x^\frac{7}{3} \nonumber \\
&\quad +\dots \,,
\end{align}
\begin{align}
&\quad (1-x^2) \, \mathrm{PL}\left[I^{N_c = 20}\right] \nonumber \\
&= \mathbf 2_t \mathbf 2_u x^\frac16+\left(1+\mathbf 2_t \mathbf 2_u\right) x^\frac13+\left(1+\mathbf 2_t \mathbf 2_u\right) x^\frac23+\left(1+w+w^{-1}\right) x^\frac56+\mathbf 2_t \mathbf 2_u x+\left(w+w^{-1}\right) x^\frac76 \nonumber \\
&\quad -x^\frac32+\left(w+w^{-1}-\mathbf 2_t \mathbf 2_u\right) x^\frac53+\left(-5+w^2+w^{-2}-w-w^{-1}-\mathbf 3_t-\mathbf 3_u\right) x^2 \nonumber \\
&\quad -\mathbf 2_t \mathbf 2_u \left(w+w^{-1}\right) x^\frac{13}{6}+\left(-5-w^2-w^{-2}-w-w^{-1}-\mathbf 3_t-\mathbf 3_u\right) x^\frac{7}{3} \nonumber \\
&\quad +\left(-2-\left(1+\mathbf 2_t \mathbf 2_u\right) \left(w+w^{-1}\right)-\mathbf 2_t \mathbf 2_u\right) x^\frac52+\dots \,,
\end{align}
\begin{align}
&\quad (1-x^2) \, \mathrm{PL}\left[I^{N_c = 21}\right] \nonumber \\
&= \mathbf 2_t \mathbf 2_u x^\frac16+x^\frac13+\mathbf 2_t \mathbf 2_u x^\frac12+\left(1+w+w^{-1}+\mathbf 2_t \mathbf 2_u\right) x^\frac56-x^\frac76-\mathbf 2_t \mathbf 2_u x^\frac32-2 x^\frac53-\mathbf 2_t \mathbf 2_u x^\frac{11}{6} \nonumber \\
&\quad +\left(-2-\mathbf 2_t \mathbf 2_u \left(w+w^{-1}\right)-\mathbf 3_t-\mathbf 3_u\right) x^2-x^\frac{7}{3}+\dots \,,
\end{align}
\begin{align}
&\quad (1-x^2) \, \mathrm{PL}\left[I^{N_c = 22}\right] = \mathbf 2_t \mathbf 2_u x^\frac13+\mathbf 2_t \mathbf 2_u x^\frac23-\mathbf 2_t \mathbf 2_u x^\frac43-\mathbf 2_t \mathbf 2_u x^\frac53 \,,
\end{align}
where we set $\tau = 1$ for simplicity. One can see that the contributions of $\mathrm{Tr} X^s$ and $\mathrm{Tr} Y$ appearing in the above indices satisfy the condition \eqref{eq:W0 adj truncation 2}. In addition, once we set $w = 1$, the contributions of $V_{s,t}^\pm$ also satisfy the condition \eqref{eq:W0 adj truncation 2}.
\\

\section{The $F$-maximization with $W = \mathrm{Tr} X^3$ and the symmetry enhancement}
\label{sec:enhancement}

Note that our duality assumes the monopole superpotential is relevant, or at least there is a sequence of RG-flows reaching the expected monopole-deformed fixed point in the IR.\footnote{We will elaborate what we mean by this in section \ref{sec:RG}.} Such relevance of monopole superpotential should be independently checked using, e.g., the $F$-maximization \cite{Jafferis:2010un}, which determines the superconformal $R$-charges, and hence the conformal dimensions, of the chiral fields in the IR. An operator having an IR dimension less than two would then trigger an RG flow to a new IR fixed point.

In this section, we examine some explicit examples of the monopole-deformed adjoint SQCD discussed in section \ref{sec:one adjoint} with relevant monopole superpotentials. Specifically, we consider $U(2)$ theories with $N_f = 3, \, 4$ and $W_A = \mathrm{Tr} X^3$. Since this cubic superpotential of the adjoint field $X$ is relevant, we fix $\Delta_X = 2/3$ and perform the $F$-maximization to determine the $R$-charge $\Delta_Q$ of the fundamental field $Q$ and those the monopole operators $V_0^\pm, \, V_1^\pm$, which are given in terms of $\Delta_Q$ as follows:
\begin{align}
\Delta_{V_0} &= N_f (1-\Delta_Q)-\frac23 \,, \\
\Delta_{V_1} &= N_f (1-\Delta_Q) \,,
\end{align}
respectively. One can carry out similar analysis for other values of $N_c$ and $N_f$, which we don't do here due to computational simplicity.
\\

The result of the $F$-maximization is summarized in Table \ref{tab:Delta with one adjoint}.
\begin{table}[tbp]
\centering
\begin{tabular}{|c|c|c|c|}
\hline
 & $\Delta_Q$ & $\Delta_{V_0}$ & $\Delta_{V_1}$ \\
\hline
$N_f = 3$ & $0.3673$ & $1.231$ & $1.898$ \\
\hline
$N_f = 4$ & $0.4005$ & $1.731$ & $2.398$ \\
\hline
\end{tabular}
\caption{\label{tab:Delta with one adjoint} The $R$-charges determined by the $F$-maximization for $U(2)$ theories with $W = \mathrm{Tr} X^3$.}
\end{table}
For $N_f = 3$, both $V_0^\pm$ and $V_1^\pm$ are relevant operators because their dimensions are less than 2. For example, we can turn on
\begin{align}
\label{eq:W=V1}
\Delta W_A = V_1^++V_1^- \,,
\end{align}
which triggers an RG flow to a new IR fixed point having another dual UV description, the $U(2)$ theory with extra matrix fields $M_0, \, M_1$ and the superpotential
\begin{align}
W_B^{mon} = \mathrm{Tr} X^3 + M_1 \tilde q q + M_0 \tilde q \hat X q+\hat V_1^++\hat V_1^-
\end{align}
where the contracted gauge and flavor indices are omitted for simplicity. As seen in \eqref{eq:single adjoint Delta_Q}, the monopole superpotential $\Delta W_A = V_1^++V_1^-$ fixes the $R$-charge of the fundamental field $Q$ to $\Delta_Q = 1/3$ so that $\Delta_{V_1}$ becomes two.

We can compute the superconformal index with this value of $\Delta_Q$, which is given by
\begin{align}
I &= 1+\left(\mathbf{3}_t \mathbf{3}_u+1\right) x^\frac23+\left(\mathbf{6}_t \mathbf{6}_u+2 \, \mathbf{3}_t \mathbf{3}_u+\overline{\mathbf{3}}_t \overline{\mathbf{3}}_u+3\right) x^\frac43 \nonumber \\
&\quad+\left(\mathbf{10}_t \mathbf{10}_u+\mathbf{8}_t \mathbf{8}_u+2 \, \mathbf{6}_t \mathbf{6}_u+\mathbf{6}_t \overline{\mathbf{3}}_u+\overline{\mathbf{3}}_t \mathbf{6}_u+3 \, \mathbf{3}_t \mathbf{3}_u+\overline{\mathbf{3}}_t \overline{\mathbf{3}}_u+2-\mathbf{8}_t-\mathbf{8}_u\right) x^2+\dots
\end{align}
where $\mathbf n_t$ is the character of the representation $\mathbf n$ of the $SU(3)_t$ global symmetry, and $\mathbf n_u$ is that of $SU(3)_u$. Note that $U(1)_A \times U(1)_T$ is broken by the monopole superpotential. We observe negative terms of order $x^2$, which are the contributions of the current multiplet \cite{Razamat:2016gzx}. In this case, it is in the adjoint representation $(\mathbf{8},\mathbf{8})$ of $SU(3)_t \times SU(3)_u$, as expected.

Furthermore, we have also computed the index of the dual theory, giving exactly the same index, which is strong evidence of the proposed duality. Thus, we expect that the monopole superpotential \eqref{eq:W=V1} indeed leads to a new fixed point in the IR, to which both Theory A and Theory B flow.
\\

On the other hand, for $N_f = 4$, only $V_0^\pm$ are relevant operators, while $V_1^\pm$ are not. Thus, we can now turn on $\Delta W_A = V_0^+ +V_0^-$, leading to the following superpotential of Theory A:
\begin{align}
\label{eq:sup 1}
W_A^{mon} = \mathrm{Tr} X^3+V_0^+ + V_0^- \,.
\end{align}
The conjectured dual theory is the same $U(2)$ theory with extra matrix fields $M_0, \, M_1$ and the superpotential
\begin{align}
W_B^{mon} = \mathrm{Tr} \hat X^3+M_1 \tilde q q+M_0 \tilde q \hat X q+\hat V_0^+ + \hat V_0^- \,.
\end{align}
The monopole terms in the superpotential fix the $R$-charge of the fundamental field $Q$ to $\Delta_Q = 1/3$, with which we can compute the superconformal index at the monopole-deformed fixed point as follow:
\begin{align}
\label{eq:ind1}
I &= 1+\left(\mathbf{4}_t \mathbf{4}_u+1\right) x^\frac23+\left(\mathbf{10}_t \mathbf{10}_u+\mathbf{6}_t \mathbf{6}_u+2 \, \mathbf{4}_t \mathbf{4}_u+1\right) x^\frac43 \nonumber \\
&\quad +\left(\mathbf{20}_t \mathbf{20}_u+\mathbf{20}'_t \mathbf{20}'_u+2 \, \mathbf{10}_t \mathbf{10}_u+\mathbf{10}_t \mathbf{6}_u+\mathbf{6}_t \mathbf{10}_u+\mathbf{6}_t \mathbf{6}_u+\mathbf{4}_t \mathbf{4}_u-\mathbf{15}_t-\mathbf{15}_u\right) x^2+\dots \,.
\end{align}
The contribution of the conserved current multiplet is given by the negative contribution of order $x^2$, which is $-\left(\mathbf{15}_t+\mathbf{15}_u\right) x^2$ in this case. This shows that the theory preserves the global symmetry
\begin{align}
SU(4)_t \times SU(4)_u
\end{align}
without any abelian symmetries as expected because they are broken by the monopole superpotential. We have also computed the index of the dual theory, which gives exactly the same index.
\\

Note that those examples are \emph{almost} self-dual because Theory A and Theory B are identical up to extra gauge singlets flipping the meson operators in Theory B. Interestingly, one can make those dual pairs \emph{exactly} self-dual by re-flipping part of those extra singlets in Theory B, and equivalently, flipping part of the mesons in Theory A. In general, such self-duality implies an emergent $\mathbb Z_2$ symmetry of the theory in the IR. If the theory enjoys more self-dualities, more emergent discrete symmetries would appear. Surprisingly, such emergent discrete symmetries sometimes lead to the enhancement of the continuous global symmetry \cite{Razamat:2017hda,Razamat:2018gbu,Hwang:2020ddr}, whose Weyl group is constructed from that of the manifest symmetry and the emergent discrete symmetries induced by the self-dualities.

We will see that the $U(2)$ theory with $N_f = 4$ examined above is one such example. Using the superconformal index, we will show that the $SU(4)_t \times SU(4)_u$ UV symmetry of this theory is enhanced to $SO(12)$ in the IR if we flip part of its mesons. This example is closely related to a model proposed in \cite{Amariti:2018wht,Benvenuti:2018bav}, where a similar adjoint QCD was discussed but without the cubic superpotential for the adjoint field $X$.\footnote{This 3d model can be obtained from a 4d $USp(2 N)$ theory with one antisymmetric and eight fundamental chirals. With even $N$ and some extra gauge singlets, this 4d model exhibits the enhanced $E_7 \times U(1)$ global symmetry in the IR \cite{Razamat:2017hda}.} This model without the cubic superpotential was shown to have the $S^3_b$ partition function invariant under the discrete symmetries belonging to the $SO(12)$ Weyl group \cite{Bult}, which strongly signals that the model has the enhanced $SO(12)$ symmetry in the IR. We will also prove this claim by computing its superconformal index.

As explained, one can make the duality exactly self-dual by flipping part of the extra singlets in Theory B, either $M_0$ or $M_1$ in this case, which corresponds to $\tilde Q Q$ or $\tilde Q X Q$ in Theory A, respectively. Here we choose to flip $\tilde Q Q$ in Theory A, and accordingly $M_0$ in Theory B. Thus, we introduce new gauge singlets $m_1$ in the representation $(\overline{\mathbf{4}},\overline{\mathbf{4}})$ of $SU(4)_t \times SU(4)_u$ and an extra superpotential interaction
\begin{align}
\Delta W_A = m_1 \tilde Q Q \,,
\end{align}
which leads to the total superpotential
\begin{align}
\label{eq:self-dual}
W_A^\text{self-dual} = \mathrm{Tr} X^3+m_1 \tilde Q Q+V_0^+ + V_0^- \,.
\end{align}
On the dual side, the extra superpotential corresponds to
\begin{align}
\Delta W_B = m_1 M_0 \,.
\end{align}
Once we solve the F-term equations for $m_1$ and $M_0$, $M_0$ becomes massive, and $m_1$ is identified with $\tilde q \hat X q$. The resulting superpotential of the dual theory is given by
\begin{align}
W_B^\text{self-dual} = \mathrm{Tr} \hat X^3+M_1 \tilde q q+\hat V_0^+ + \hat V_0^- \,.
\end{align}
This is exactly the same as the superpotential \eqref{eq:self-dual}, while the duality still nontrivially maps the meson operators as follows:
\begin{align}
\tilde Q X Q \quad &\longleftrightarrow \quad M_1 \,, \\
m_1 \quad &\longleftrightarrow \quad \tilde q \hat X q \,.
\end{align}
In other words, this is a self-duality of the theory exchanging two chiral ring generators:
\begin{align}
\tilde Q X Q \quad &\longleftrightarrow \quad m_1 \,.
\end{align}

Now we compute the index of the self-dual theory after the flipping. The corresponding index is given by
\begin{align}
\label{eq:ind2}
I^\text{self-dual} = 1+x^\frac23+\left(\mathbf{4}_t \mathbf{4}_u+\overline{\mathbf{4}}_t \overline{\mathbf{4}}_u+1\right) x^\frac43-\left(\mathbf{15}_t+\mathbf{15}_u+\mathbf{6}_t \mathbf{6}_u\right) x^2+\dots \,.
\end{align}
Note that the negative $x^2$ terms, which are supposed to capture the conserved currents, fit the adjoint representation $\mathbf{66}$ of $SO(12)$ because $\mathbf{66}$ is decomposed under $SU(4) \times SU(4) \subset SO(12)$ as follows:
\begin{align}
\mathbf{66} \quad \longrightarrow \quad (\mathbf{15},\mathbf{1})\oplus(\mathbf{1},\mathbf{15})\oplus(\mathbf{6},\mathbf{6}) \,.
\end{align}
Similarly, $\mathbf{32}$ of $SO(12)$ is decomposed as
\begin{align}
\mathbf{32} \quad \longrightarrow \quad (\mathbf{4},\mathbf{4})\oplus(\overline{\mathbf{4}},\overline{\mathbf{4}}) \,.
\end{align}
Therefore, the expanded index can be written in terms of the $SO(12)$ characters as follows:
\begin{align}
I^\text{self-dual} = 1+x^\frac23+\left(\mathbf{32}_{t,u}+1\right) x^\frac43-\mathbf{66}_{t,u} \, x^2+\dots \,,
\end{align}
where $\mathbf{n}_{t,u}$ is the character of the $SO(12)$ representation $\mathbf{n}$ written in terms of the $SU(4)_t \times SU(4)_u$ fugacities $t, \, u$. This shows that the theory exhibits the following enhancement of the symmetry in the IR:
\begin{align}
SU(4)_t \times SU(4)_u \quad \longrightarrow \quad SO(12) \,.
\end{align}

Notice that this theory is the same as the $U(N)$ adjoint SQCD with four flavors discussed in \cite{Amariti:2018wht,Benvenuti:2018bav} up to the cubic superpotential for the adjoint field $X$. This model without the cubic superpotential was shown to have the $S^3_b$ partition function invariant under the $SO(12)$ Weyl group actions \cite{Bult}. To see the connection between the two models, let us introduce an extra singlet $x_1$ with a superpotential term flipping the operator $\mathrm{Tr} X$:
\begin{align}
\Delta W' = x_1 \, \mathrm{Tr} X \,.
\end{align}
Note that such a flip of $\mathrm{Tr} X$ does not spoil the self-duality.
The total superpotential is now written as
\begin{align}
\label{eq:self-dual 2}
W^{\text{self-dual}'} = \mathrm{Tr} X^3+x_1 \, \mathrm{Tr} X+m_1 \tilde Q Q+V_0^+ + V_0^- \,.
\end{align}
One can compute the corresponding index, which is naively given by
\begin{align}
I^{\text{self-dual}'} = 1+\left(\mathbf{4}_t \mathbf{4}_u+\overline{\mathbf{4}}_t \overline{\mathbf{4}}_u+1\right) x^\frac43-\left(\mathbf{15}_t+\mathbf{15}_u+\mathbf{6}_t \mathbf{6}_u+\mathbf{4}_t \mathbf{4}_u+\overline{\mathbf{4}}_t \overline{\mathbf{4}}_u+1\right) x^2+\dots \,.
\end{align}
However, the negative $x^2$ terms capturing the conserved current do not fit the adjoint representation of any Lie group, which is inconsistent. It turns out this is because we are missing some abelian symmetries whose contribution to the superconformal $R$-symmetry in the IR is nontrivial. To find such abelian symmetries, let us reexamine the superpotential \eqref{eq:self-dual 2}. We note that the first term can actually be written as
\begin{align}
\mathrm{Tr} X^3 \sim \left(\mathrm{Tr} X\right) \left(\mathrm{Tr} X^2\right)+(\mathrm{Tr} X)^3
\end{align}
up to suitable coefficients because $X$, an adjoint field of the $U(2)$ gauge group, is a $2 \times 2$ matrix field. According to \cite{Benvenuti:2017lle}, those two terms do not satisfy the condition called the \emph{chiral ring stability}, because the F-term equation of $x_1$ set them zero, and has to be dropped. Once we drop those terms, we have an extra $U(1)_X$ symmetry, which rotates the adjoint field $X$. More precisely, now the monopole operators are also charged under such $U(1)_X$ with charge $-N_c+1 = -1$, and the monopole superpotential terms in \eqref{eq:self-dual 2} break $U(1)_A \times U(1)_T \times U(1)_X$ into a single $U(1)_v$ symmetry in such a way that the monopole operators $V_0^\pm$ are neutral under $U(1)_v$. Since $V_0^\pm$ are $U(1)_v$ neutral, their $R$-charges are independent of the mixing between the $R$-symmetry and $U(1)_v$, which requires that the $R$-charges of $X$ and $Q$ must satisfy
\begin{align}
\Delta_X = 2-4 \Delta_Q
\end{align}
to ensure the $R$-charge of $V_0^\pm$: $\Delta_{V_0} = 4-4 \Delta_Q-\Delta_X$ to be always two.
With this condition, we conduct the $F$-maximization allowing the mixing between the $R$-symmetry and $U(1)_v$, which results in the following $R$-charge of $X$:
\begin{align}
\Delta_X \approx 0.1323 \,.
\end{align}
This indicates that the dimension of the operator $\mathrm{Tr} X^2$ also falls below the unitarity bound $\Delta \geq 1/2$ and decouples from the interacting sector. Hence, we need to flip $\mathrm{Tr} X^2$ by another singlet $x_2$, which results in the $U(2)$ adjoint SQCD with four flavors and the superpotential
\begin{align}
\label{eq:sup_AB}
W^{\text{self-dual}''} = x_1  \mathrm{Tr} X+x_2 \, \mathrm{Tr} X^2+m_1 \tilde Q Q+V_0^+ + V_0^- \,.
\end{align}
This model is identical to the one discussed in \cite{Amariti:2018wht,Benvenuti:2018bav} up to the flip of $\mathrm{Tr} X^i$.
The $R$-charges obtained from the $F$-maximization are
\begin{align}
\Delta_Q = 0.625 \,, \qquad \Delta_X = -0.5 \,,
\end{align}
with which no more operators hit the unitarity bound. Using those $R$-charges, we obtain the index
\begin{align}
I^{\text{self-dual}''} = 1+\mathbf{32} _{t,u} \sqrt{v} x^\frac34+\mathbf{462}_{t,u} v x^\frac32-\left(\mathbf{66}_{t,u}+1\right) x^2+\dots \,,
\end{align}
where $v$ is the fugacity of $U(1)_v$ normalized such that the adjoint field $X$ has charge 1. The fugacities $t$ and $u$ of the manifest $SU(4)_t \times SU(4)_u$ symmetry are neatly organized into the characters of $SO(12)$ representations, which are decomposed under $SU(4)_t \times SU(4)_u \subset SO(12)$ as follows:
\begin{align}
\mathbf{32} &= (\mathbf{4},\mathbf{4})+(\overline{\mathbf{4}},\overline{\mathbf{4}}) \,, \\
\mathbf{66} &= (\mathbf{6},\mathbf{6})+(\mathbf{15},\mathbf{1})+(\mathbf{1},\mathbf{15}) \,, \\
\mathbf{462} &= (\mathbf{1},\mathbf{1})+(\mathbf{6},\mathbf{6})+(\mathbf{10},\mathbf{10})+(\overline{\mathbf{10}},\overline{\mathbf{10}})+(\mathbf{15},\mathbf{15}) \,.
\end{align}
In particular, the contribution of the conserved currents, $-\left(\mathbf{66}_{t,u}+1\right) x^2$, is in the adjoint representation of $SO(12) \times U(1)_v$. This proves that the model with the superpotential \eqref{eq:sup_AB} exhibits the following enhancement of the symmetry in the IR:
\begin{align}
SU(4)_t \times SU(4)_u \times U(1)_v \quad \longrightarrow \quad SO(12) \times U(1)_v \,.
\end{align}
Note that this model has an additional $U(1)_v$ symmetry compared to the previous model with the superpotential \eqref{eq:self-dual}.
\\

\section{Discussions}
\label{sec:discussions}

\subsection{Summary of the dualities}

In this paper, we have examined the monopole deformation of the 3d $\mathcal N = 2$ $U(N)$ gauge theories with adjoint matters and fundamental flavors and their new Seiberg-like dualities. The first duality we propose is for the $U(N)$ gauge theory with one adjoint matter deformed by a linear monopole superpotential. We have proposed that the following pair of theories are dual to each other.
\begin{itemize}
\item Theory A is the 3d $\mathcal N = 2$ $U(N_c)$ gauge theory with $N_f$ pairs of fundamental $Q^a$ and anti-fundamental $\tilde Q^{\tilde a}$, one adjoint chiral multiplet $X$ and the superpotential
\begin{align}
W_A^{mon} = \mathrm{Tr} X^{n+1}+V_\alpha^+ + V_\alpha^- \,.
\end{align}
where $V_\alpha^\pm$ are monopole operators of Theory A.
\item Theory B is the 3d $\mathcal N = 2$ $U(n N_f-N_c-2 n+2 \alpha)$ gauge theory with $N_f$ pairs of fundamental $q_{\tilde a}$ and anti-fundemental $\tilde q_a$, one adjoint $\hat X$, and $n N_f{}^2+2 n$ gauge singlet chiral multiplets ${M_i}^{\tilde a a}$ and $V_i^\pm$ for $a,\tilde a = 1,\dots,N_f$ and $i = 0,\ldots,n-1$. The superpotential is given by
\begin{align}
W_B^{mon} = \mathrm{Tr} \hat X^{n+1}+\sum_{i = 0}^{n-1} M_i \tilde q \hat X^{n-1-i} q+\hat V_\alpha^+ + \hat V_\alpha^- \,.
\end{align}
where $\hat V_\alpha^\pm$ are the monopole operators of Theory B.
\end{itemize}

The second and third dualities we propose are for the $U(N)$ theory with two adjoint matters. The theory with two adjoint matters has two types of monopole operators, which we call $V_{s,t}^\pm$ and $W_u^\pm$. Thus, we have considered two monopole dualities including $V_{0,0}^\pm$ and $W_0^\pm$ in the superpotential, respectively. For the deformation by $V_{0,0}^\pm$, two dual theories are as follows.
\begin{itemize}
\item Theory A is the 3d $\mathcal N = 2$ $U(N_c)$ gauge theory with $N_f$ pairs of fundamental fields $Q^a$ and anti-fundamental fields $\tilde Q^{\tilde a}$, two adjoint fields $X, \, Y$ and the superpotential
\begin{align}
W_A^{mon}= \mathrm{Tr} X^{n+1}+\mathrm{Tr} X Y^2+V_{0,0}^++V_{0,0}^-
\end{align}
where $V_{0,0}^\pm$ are a pair of monopole operators of Theory A with topological symmetry charge $\pm1$.
\item Theory B is the 3d $\mathcal N = 2$ $U(3 n N_f-N_c-4 n-2)$ gauge theory with $N_f$ pairs of fundamental fields $q_{\tilde a}$ and anti-fundemental fields $\tilde q_a$, two adjoint fields $\hat X, \, \hat Y$, and $3 n {N_f}^2$ gauge singlet fields $M_{s,t}{}^{\tilde a a}$ for $a,\tilde a = 1,\dots,N_f$, $s = 0,\dots,n-1$ and $t = 0,1,2$.
The superpotential is given by
\begin{align}
W_B^{mon} = \mathrm{Tr} \hat X^{n+1}+\mathrm{Tr} \hat X \hat Y^2+\sum_{s = 0}^{n-1} \sum_{t = 0}^2 M_{s,t} \tilde q \hat X^{n-1-s} \hat Y^{2-t} q+\hat V_{0,0}^+ + \hat V_{0,0}^- \,.
\end{align}
where $\hat V_{0,0}^\pm$ are a pair of monopole operators of Theory B with topological symmetry charge $\pm1$.
\end{itemize}
On the other hand, the duality deformed by $W_0^\pm$ is given as follows.
\begin{itemize}
\item Theory A is the 3d $\mathcal N = 2$ $U(N_c)$ gauge theory with $N_f$ pairs of fundamental fields $Q^a$ and anti-fundamental fields $\tilde Q^{\tilde a}$, two adjoint fields $X, \, Y$ and the superpotential
\begin{align}
W_A^{mon}= \mathrm{Tr} X^{n+1}+\mathrm{Tr} X Y^2+W_{0}^++W_{0}^-
\end{align}
where $W_{0}^\pm$ are a pair of monopole operators of Theory A with topological symmetry charge $\pm2$.
\item Theory B is the 3d $\mathcal N = 2$ $U(3 n N_f-N_c-2 n+2)$ gauge theory with $N_f$ pairs of fundamental fields $q_{\tilde a}$ and anti-fundemental fields $\tilde q_a$, two adjoint fields $\hat X, \, \hat Y$, and $3 n {N_f}^2$ gauge singlet fields $M_{s,t}{}^{\tilde a a}$ for $a,\tilde a = 1,\dots,N_f$, $s = 0,\dots,n-1$ and $t = 0,1,2$.
The superpotential is given by
\begin{align}
\label{eq:dual mon sup}
W_B^{mon} = \mathrm{Tr} \hat X^{n+1}+\mathrm{Tr} \hat X \hat Y^2+\sum_{s = 0}^{n-1} \sum_{t = 0}^2 M_{s,t} \tilde q \hat X^{n-1-s} \hat Y^{2-t} q+\hat W_{0}^+ + \hat W_{0}^- \,.
\end{align}
where $\hat W_{0}^\pm$ are a pair of monopole operators of Theory B with topological symmetry charge $\pm2$.
\\
\end{itemize}

\subsection{RG flows of double adjoint matters and the conformal manifold}
\label{sec:RG}

We should stress that although the generic forms of the dualities are given as above, our analysis in sections \ref{sec:one adjoint} and \ref{sec:two adjoints} are valid when the monopole superpotential triggers an RG flow to a new fixed point distinct from the original ones without the monopole superpotentials; otherwise, the IR fixed point would have extra symmetries other than the one we assume, which are not taken into account in our index computation. Indeed, the \emph{relevant} monopole deformation is not always the case and has to be checked using, e.g., the $F$-maximization \cite{Jafferis:2010un}. The $F$-maximization determines the superconformal $R$-charges, and hence the conformal dimensions, of the chiral fields in the IR. A relevant operators then has the IR dimension less than 2. For example, in section 4, we conducted the $F$-maximization for a particular set of examples with one adjoint matter and showed their monopole deformations are relevant and lead to new IR fixed points. For the theory with two adjoints, on the other hand, the $F$-maximization was performed for $n = 3$ in \cite{Hwang:2018uyj}, whose result for $\Delta_Q$ is reported here in Table \ref{tab:Delta}. This result also determines the $R$-charges of the monopole operators $V_{0,0}^\pm$ and $W_0^\pm$; see Table \ref{tab:Delta_V} and Table \ref{tab:Delta_W}, which can be used to determine the relevance of the monopole superpotentials.
\begin{table}[tbp]
\centering
\begin{tabular}{|c|ccccc|}
\hline
$\Delta_Q$ & $N_f = 1$ & $N_f = 2$ & $N_f = 3$ & $N_f = 4$ & $N_f = 5$ \\
\hline
$U(2)$ & $0.212$ & 0.348 & - & - & - \\
$U(3)$ & $0.080$ & 0.293 & 0.352 & 0.384 & 0.404 \\
\hline
\end{tabular}
\caption{\label{tab:Delta} The IR $R$-charge of the fundamental field $Q$ of the double adjoint theory without the monopole superpotential \cite{Hwang:2018uyj}. The values for $U(2)$ with $N_f \geq 3$ are absent because $\mathrm{Tr} X^{4}$ is irrelevant in those cases, which were thus excluded for the original HKP duality \cite{Hwang:2018uyj} for the double adjoint theory without the monopole superpotential.}
\end{table}
\begin{table}[tbp]
\centering
\begin{tabular}{|c|ccccc|}
\hline
$\Delta_{V_{0,0}^\pm}$ & $N_f = 1$ & $N_f = 2$ & $N_f = 3$ & $N_f = 4$ & $N_f = 5$ \\
\hline
$U(2)$ & $0.538$ & 1.05 & - & - & - \\
$U(3)$ & $0.42$ & 0.914 & 1.44 & 1.96 & 2.48 \\
\hline
\end{tabular}
\caption{\label{tab:Delta_V} The IR $R$-charge of the monopole operator $V_{0,0}^\pm$ of the double adjoint theory without the monopole superpotential \cite{Hwang:2018uyj}. Again the values for $U(2)$ with $N_f \geq 3$ are absent because $\mathrm{Tr} X^{4}$ is irrelevant.}
\end{table}
\begin{table}[tbp]
\centering
\begin{tabular}{|c|ccccc|}
\hline
$\Delta_{W_0^\pm}$ & $N_f = 1$ & $N_f = 2$ & $N_f = 3$ & $N_f = 4$ & $N_f = 5$ \\
\hline
$U(2)$ & $1.58$ & 2.61 & - & - & - \\
$U(3)$ & $1.34$ & 2.33 & 3.39 & 4.43 & 5.46 \\
\hline
\end{tabular}
\caption{\label{tab:Delta_W} The IR $R$-charge of the monopole operator $W_{0}^\pm$ of the double adjoint theory without the monopole superpotential. Again the values for $U(2)$ with $N_f \geq 3$ are absent because $\mathrm{Tr} X^{4}$ is irrelevant.}
\end{table}
Hence, we would like to conclude the paper making some comments on the RG-flows of those examples with two adjoint matters in the presence of monopole superpotentials.
\\

Let us first consider the deformation by $W_0^\pm$, whose IR $R$-charge is shown in Table \ref{tab:Delta_W} for $N_c = 2, 3$. In both cases, $W_0^\pm$ is relevant for $N_f = 1$ because its IR $R$-charge is less than 2. Hence, we expect that the deformation by $W_0^\pm$ triggers an RG flow to a different IR fixed point and gives rise to a new duality with the monopole superpotential as we proposed. Indeed, we have computed the indices for $N_f = 1$, which show a perfect match under the proposed duality. See appendix \ref{sec:index results}.

On the other hand, for $N_f \geq 2$, $W_0^\pm$ has the IR $R$-charge greater than 2 and therefore is an irrelevant deformation. Nevertheless, interestingly, we have found another RG flow, followed by a marginal deformation, that may lead us to the expected fixed point enjoying the proposed duality. For instance, if we consider the $U(2)$ theory with $N_f = 2$, we notice that $\det (\tilde Q Q)$ is a relevant deformation of the theory because its $R$-charge is $4 \Delta_Q = 1.392 < 2$. See Table \ref{tab:Delta}. Once we turn on $\det (\tilde Q Q)$ in the superpotential, the theory is supposed to flow a new fixed point without $U(1)_A$, which is broken by $\det (\tilde Q Q)$. The corresponding index is given by
\begin{align}
\label{eq:w/ w}
I &= 1+\left(1+w+w^{-1}\right) x^\frac34+\left(1+\mathbf 2_t \mathbf 2_u\right) x+\left(w+w^{-1}\right) x^\frac54 \nonumber \\
&\quad +\left(3+w^2+2 w+2 w^{-1}+w^{-2}+\mathbf 2_t \mathbf 2_u\right) x^\frac32+\left(\mathbf 2_t \mathbf 2_u \left(w+w^{-1}\right)+2 \, \mathbf 2_t \mathbf 2_u+w+w^{-1}\right) x^\frac74 \nonumber \\
&\quad +\left(-\mathbf 3_t-\mathbf 3_u-1+\mathbf 3_t \mathbf 3_u+\mathbf 2_t+\mathbf 2_u+2 w^2+2 w^{-2}\right) x^2+\dots \,,
\end{align}
for which we have flipped the decoupled operator $\mathrm{Tr} X$, whose $R$-charge is 1/2. For the $x^2$ term, it is known that only the conserved currents and the marginal operators contribute to this order, with negative and positive signs respectively \cite{Razamat:2016gzx}. Indeed, the theory deformed by $\det (\tilde Q Q)$ preserves $SU(2)_t \times SU(2)_u \times U(1)_T$, whose currents should contribute $-\left(\mathbf 3_t+\mathbf 3_u+1\right) x^2$ to the index, which is exactly the case shown in \eqref{eq:w/ w}. In addition, this case also has marginal operators, whose couplings would parametrize the conformal manifold. Especially, let us focus on part of the conformal manifold parametrized by \emph{monopole} operators. There are four marginal monopole operators: $(V_0^\pm)^2$ and $W_0^\pm$, contributing $2 \left(w^2+w^{-2}\right) x^2$ to the index. Since they are charged under $U(1)_T$, their nonzero couplings break $U(1)_T$, whose conserved current then combines with one of the marginal operators and becomes a long multiplet \cite{Green:2010da}. Therefore, only the other three remain exactly marginal and describe part of the conformal manifold. Note that $U(1)_T$ is broken on a generic point of this three-dimensional manifold, where the index is given by
\begin{align}
\label{eq:w/o w}
I &= 1+3 x^\frac34+\left(1+\mathbf 2_t \mathbf 2_u\right) x+2 x^\frac54+\left(9+\mathbf 2_t \mathbf 2_u\right) x^\frac32+\left(4 \, \mathbf 2_t \mathbf 2_u+2\right) x^\frac74 \nonumber \\
&\quad +\left(-\mathbf 3_t-\mathbf 3_u+\mathbf 3_t \mathbf 3_u+\mathbf 2_t+\mathbf 2_u+3\right) x^2+\dots \,.
\end{align}
Notice that the negative contribution to the $x^2$ term is $-\left(\mathbf 3_t+\mathbf 3_u\right) x^2$, which is consistent with the fact that the global symmetry is now $SU(2)_t \times SU(2)_u$ without any abelian symmetry. Also one can see the positive contribution $3 x^2$ of the three exactly marginal monopole operators. Their couplings parametrize a three-dimensional conformal manifold where the theory on a generic point preserves $SU(2)_t \times SU(2)_u$, while there is a special point with the additional $U(1)_T$ symmetry preserved, whose index is given by \eqref{eq:w/ w}.

Our duality proposal then implies that the dual $U(12)$ theory deformed by $W_0^\pm$ flows to a certain point on this three-dimensional conformal manifold where $SU(2)_t \times SU(2)_u$ is preserved, while $U(1)_T$ is broken. Unfortunately, we weren't able to study the $F$-maximization of the dual $U(12)$ theory due to the limit of the computation power. Nevertheless, assuming the monopole deformation is relevant for the dual theory or at least there is an RG flow to a conformal manifold containing the expected monopole-deformed fixed point, we have checked the index of the dual $U(12)$ theory is exactly the same as \eqref{eq:w/o w}. Furthermore, one can also move along the conformal manifold to the special point with the extra $U(1)_T$ symmetry. On the original side, this can be done by turning on $\det (\tilde Q Q)$ but turning off all the monopole terms in the superpotential. This suggests another duality between the proposed dual pair with different superpotentials preserving not only $SU(2)_t \times SU(2)_u$ but also $U(1)_T$. Indeed, we have computed the indices of the dual pair keeping the $U(1)_T$ fugacity $w$ up to $x^5$, which show perfect agreement. See the first few terms given in \eqref{eq:w/ w}. This is strong evidence of the existence of the duality preserving $U(1)_T$ for the $U(2)$ theory with $N_f = 2$. It would also be interesting to investigate if there are other cases exhibiting a similar RG flow to the conformal manifold and a special point thereof with the extra $U(1)_T$ symmetry, which would lead to another duality preserving $U(1)_T$.

The above example demonstrates that, even if the monopole operator is irrelevant, there can be some other relevant deformation triggering an RG flow to the conformal manifold, connected to the monopole-deformed fixed point by a marginal deformation. On the other hand, one can attempt to engineer a sequence of RG flows directly flowing to the monopole-deformed fixed point by coupling the theory to another interacting CFT \cite{Benini:2017dud}. Imposing a suitable interaction between the original theory and the other CFT, one can make the monopole operator relevant and initiate an RG flow to a new fixed point by turning on such relevant monopole deformation. The new fixed point would have extra fields coming from the coupled CFT, which however can be made massive and integrated out. Thus, the resulting IR theory is the expected monopole-deformed fixed point enjoying the monopole duality. It would be interesting to study if such RG flows to the monopole-deformed fixed points enjoying our dualities can be engineered for cases with irrelevant monopole operators.
\\

Next, we consider the deformation by $V_{0,0}^\pm$. Again we focus on $N_c = 2, \, 3$. The IR $R$-charge of $V_{0,0}^\pm$ is shown in Table \ref{tab:Delta_V}. The proposed duality requires $\tilde N_c = 3 n N_f-N_c-4 n-2 \geq 0$, which is only satisfied for $N_f \geq 2$ since we take $n = 3$. One can see that $V_{0,0}^\pm$ is a relevant deformation for $2 \leq N_f \leq 4$, while it is irrelevant for $N_f \geq 5$.

We have computed the indices for those relevant cases, which show perfect matches under the proposed duality. In particular, the index for the $U(2)$ theory with $N_f = 3$ and those for the $U(3)$ theories with $N_f = 3, 4$ are given by\footnote{Note that we have omitted the $SU(N_f)_t \times SU(N_f)_u$ fugacities for the computational simplicity.}
\begin{align}
I^{(N_c,N_f) = (2,3)} &= 1+10 \sqrt{x}+x^{3/4}+65 x+19 x^{5/4}+311 x^{3/2}+145 x^{7/4}+1203 x^2+\dots \,, \\
I^{(N_c,N_f) = (3,3)} &= 1 + 9 x^{1/3} + \sqrt{x} + 45 x^{2/3} + x^{3/4} + 18 x^{5/6} + 167 x + 18 x^{13/12} + 126 x^{7/6} \nonumber \\
		&\quad + x^{5/4} + 531 x^{4/3} + 126 x^{17/12} + 573 x^{3/2} + 36 x^{19/12} + 1575 x^{5/3} + 571 x^{7/4} \nonumber \\
		&\quad + 2043 x^{11/6} + 360 x^{23/12} + 4453 x^2 + \dots \,, \\
I^{(N_c,N_f) = (3,4)} &= 1 + \sqrt{x} + 17 x^{3/4} + 2 x + 33 x^{5/4} + 172 x^{3/2} + \dots \,,
\end{align}
which agree with the dual indices $I^{(\tilde N_c,N_f) = (11,3)}, \, I^{(\tilde N_c,N_f) = (10,3)}$ and $I^{(\tilde N_c,N_f) = (19,4)}$, respectively. Thus, we expect that the deformation by $V_{0,0}^\pm$ triggers an RG flow to a new interacting fixed point, to which both dual theories flow.

On the other hand, for $N_f = 2$, the computed indices still agree under the duality but do not have the standard form of the superconformal index, which must start with 1 because the identity operator of the SCFT on $\mathbb R^3$ always contributes 1 to the index. However, the indices for $N_f = 2$ are given by\footnote{In this case, the index should only be interpreted as the supersymmetric partition function on $S^2 \times S^1$ rather than the superconformal index.}
\begin{align}
I^{(N_c,N_f) = (2,2)} &= -x^\frac14-5 x^\frac34-14 x^\frac54-29 x^\frac74+\dots \,, \label{eq:neg ind} \\
I^{(N_c,N_f) = (3,2)} &= 0 \,.
\end{align}
Especially, the index for $N_c = 3$ vanishes, which signals that the theory has no supersymmetric vacuum.
For $N_c = 2$, the index doesn't vanish but starts with a nonzero power of $x$ rather than 1. We thus conclude that the $U(2)$ theory also doesn't flow to a conformal fixed point in the IR although the interpretation of the nonzero negative terms in \eqref{eq:neg ind} is unclear. It would be interesting to examine their interpretation and more detailed IR dynamics of this theory.
\\

\acknowledgments

We would like to thank K.~Avner, S.~Pasquetti, and M.~Sacchi for valuable discussions. This research is supported by NRF-2021R1A6A1A10042944 (JP) and NRF-2021R1A2C1012440 (JP, SK). CH is partially supported by the STFC consolidated grant ST/T000694/1.

\newpage

\appendix

\section{More results of the superconformal index computation}
\label{sec:index results}

In this appendix, we provide the list of superconformal indices for the three monopole dualities we propose. Each dual pair show the perfect match of the indices, which is strong evidence of the proposed dualities.

\subsection{Single adjoint with $\Delta W_A = V_{\alpha}^{+}+V_{\alpha}^{-}$}
\begin{table}[H]
	\centering
	\begin{tabular}{|c|c|c|c|}
		\hline
		n	&	$\Delta W_A$	&	$(N_f , N_c , \tilde{N}_c)$	&	SCI	\\ 
		\hline
		&				&	$(2,0,0)$				&	$\begin{array}{c}
		1
		\end{array}$	\\
		\cline{3-4}
		&				&	$(3,0,1)$				&	$\begin{array}{c}
		1
		\end{array}$	\\
		\cline{3-4}
		&				&	$(3,1,0)$				&	$\begin{array}{c}
		1 \!+\! 9 x^{2/3} \!+\! 36 x^{4/3} \!+\! 84 x^{2} \!+\! 135 x^{8/3} \!+\! 198 x^{10/3} \!+\! 327 x^{4} \!+\! O(x^{13/3})
		\end{array}$	\\
		\cline{3-4}
		&				&	$(4,0,2)$				&	$\begin{array}{c}
		1
		\end{array}$	\\
		\cline{3-4}
		1&$V_0^{+}\!+\!V_0^{-}\!\!$&	$(4,1,1)$				&	$\begin{array}{c}
		1 + 16 x + 70 x^2 + 128 x^3 + 259 x^4 + 464 x^5 + 326 x^6 + O(x^7)
		\end{array}$	\\
		\cline{3-4}
		&				&	$(4,2,0)$				&	$\begin{array}{c}
		1 \!+\! 16 \sqrt{x} \!+\! 136 x \!+\! 800 x^{3/2} \!+\! 3620 x^2 \!+\! 13344 x^{5/2} \!+\! O(x^{3})
		\end{array}$	\\
		\cline{3-4}
		&				&	$(5,0,3)$				&	$\begin{array}{c}
		1
		\end{array}$	\\
		\cline{3-4}
		&				&	$(5,1,2)$				&	$\begin{array}{c}
		1+25 x^{6/5}-48 x^2+225 x^{12/5}+25 x^{14/5}-700 x^{16/5}+O\left(x^{18/5}\right)
		\end{array}$	\\
		\cline{3-4}
		&				&	$(5,2,1)$				&	$\begin{array}{c}
		1+25 x^{4/5}+325 x^{8/5}-48 x^2+2825 x^{12/5}-1150 x^{14/5}+O\left(x^{16/5}\right)
		\end{array}$	\\
		\cline{3-4}
		&				&	$(5,3,0)$				&	$\begin{array}{c}
		1\!+\!25 x^{2/5}\!+\!325 x^{4/5}\!+\!2925 x^{6/5}\!+\!20450 x^{8/5}\!+\!118130 x^2\!+\!O\left(x^{11/5}\right)
		\end{array}$	\\
		\hline  
		&				&	$(2,0,0)$				&	$\begin{array}{c}
		1
		\end{array}$	\\
		\cline{3-4}
		&				&	$(3,0,2)$				&	$\begin{array}{c}
		1
		\end{array}$	\\
		\cline{3-4}
		&$V_0^{+}\!+\!V_0^{-}\!\!$&	$(3,1,1)$				&	$\begin{array}{c}
		1 \!+\! 10 x^{2/3} \!+\! 45 x^{4/3} \!+\! 120 x^{2} \!+\! 220 x^{8/3} \!+\! 342 x^{10/3} \!+\! 560 x^{4} \!+\! O(x^{13/3})
		\end{array}$	\\
		\cline{3-4}
		&				&	$(3,2,0)$				&	$\begin{array}{c}
		1+9 x^{2/9}+45 x^{4/9}+165 x^{2/3}+504 x^{8/9}+1359 x^{10/9}+3327 x^{4/3}\\
		+7515 x^{14/9}+15876 x^{16/9}+31681 x^2+O\left(x^{19/9}\right)
		\end{array}$	\\
		\cline{2-4}
		2&				&	$(1,0,0)$				&	$\begin{array}{c}
		1
		\end{array}$	\\
		\cline{3-4}
		&				&	$(2,1,1)$				&	$\begin{array}{c}
		1 \!+\! 5 x^{2/3} \!+\! 15 x^{4/3} \!+\! 19 x^{2} \!+\! 20 x^{8/3} \!+\! 31 x^{10/3} \!+\! 59 x^4 \!+\! O(x^{14/3})
		\end{array}$	\\
		\cline{3-4}
		&$V_1^{+}\!+\!V_1^{-}\!\!$&	$(3,1,3)$				&	$\begin{array}{c}
		1 \!+\! x^{2/3} \!+\! 9 x^{10/9} \!+\! 2 x^{4/3} \!+\! 9 x^{16/9} \!-\! 16 x^{2} \!+\! 36 x^{20/9} \!+\! O(x^{8/3})
		\end{array}$	\\
		\cline{3-4}
		&				&	$(3,2,2)$				&	$\begin{array}{c}
		1 \!+\! 10 x^{2/3} \!+\! 66 x^{4/3} \!+\! 294 x^{2} \!+\! 980 x^{8/3} \!+\! O(x^{10/3})
		\end{array}$	\\
		\cline{3-4}
		&				&	$(3,3,1)$				&	$\begin{array}{c}
		1+9 x^{2/9}+45 x^{4/9}+166 x^{2/3}+513 x^{8/9}+1413 x^{10/9}+3575 x^{4/3}\\
		+8442 x^{14/9}+18819 x^{16/9}+39939 x^2+O\left(x^{19/9}\right)
		\end{array}$	\\
		\hline
		&				&	$(2,0,0)$				&	$\begin{array}{c}
		1
		\end{array}$	\\
		\cline{3-4}
		&$V_0^{+}\!+\!V_0^{-}\!\!$&	$(3,1,2)$				&	$\begin{array}{c}
		1 \!+\! \sqrt{x} \!+\! 9 x^{2/3} \!+\! x \!+\! 9 x^{7/6} \!+\! 36 x^{4/3} \!+\! 9 x^{5/3} \!+\! 36 x^{11/6} \!+\! O(x^{2})
		\end{array}$	\\
		\cline{3-4}
		&				&	$(3,2,1)$				&	$\begin{array}{c}
		1+9 x^{1/3}+\sqrt{x}+45 x^{2/3}+18 x^{5/6}+166 x+117 x^{7/6}+513 x^{4/3}\\
		+489 x^{3/2}+1440 x^{5/3}+1584 x^{11/6}+3735 x^2+O\left(x^{13/6}\right)
		\end{array}$	\\
		\cline{2-4}
		3&				&	$(2,1,1)$				&	$\begin{array}{c}
		1 \!+\! 5 \sqrt{x} \!+\! 14 x \!+\! 31 x^{3/2} \!+\! 44 x^{2} \!+\! 56 x^{5/2} \!+\! 69 x^{3} \!+\! 96 x^{7/2} \!+\! O(x^{4})
		\end{array}$	\\
		\cline{3-4}
		&$V_1^{+}\!+\!V_1^{-}\!\!$&	$(3,2,3)$				&	$\begin{array}{c}
		1+\sqrt{x}+9 x^{2/3}+2 x+18 x^{7/6}+45 x^{4/3}+3 x^{3/2}+27 x^{5/3}\\
		+117 x^{11/6}+151 x^2+36 x^{13/6}+198 x^{7/3}+O\left(x^{5/2}\right)
		\end{array}$	\\
		\cline{3-4}
		&				&	$(3,3,2)$				&	$\begin{array}{c}
		1+9 x^{1/3}+\sqrt{x}+45 x^{2/3}+18 x^{5/6}+167 x+126 x^{7/6}+531 x^{4/3}\\
		+573 x^{3/2}+1575 x^{5/3}+2025 x^{11/6}+4453 x^2+O\left(x^{13/6}\right)
		\end{array}$	\\
		\hline
	\end{tabular}
\end{table}

\begin{table}[H]
	\centering
	\begin{tabular}{|c|c|c|c|}
		\hline
		n	&	$\Delta W_A$	&	$(N_f , N_c , \tilde{N}_c)$	&	SCI	\\ 
		\hline
		&				&	$(2,2,2)$				&	$\begin{array}{c}
		1 \!+\! 5 \sqrt{x} \!+\! 22 x \!+\! 70 x^{3/2} \!+\! 179 x^{2} \!+\! 374 x^{5/2} \!+\! 661 x^{3} + O(x^{7/2})
		\end{array}$	\\
		\cline{3-4}
		3&$V_2^{+}\!+\!V_2^{-}\!\!$&	$(3,3,4)$				&	$\begin{array}{c}
		1+\sqrt{x}+9 x^{2/3}+4 x+18 x^{7/6}+45 x^{4/3}+6 x^{3/2}+54 x^{5/3}\\
		+126 x^{11/6}+160 x^2+90 x^{13/6}+378 x^{7/3}+550 x^{5/2}+O\left(x^{8/3}\right)
		\end{array}$	\\
		\cline{3-4}
		&				&	$(3,4,3)$				&	$\begin{array}{c}
		1+9 x^{1/3}+\sqrt{x}+45 x^{2/3}+18 x^{5/6}+169 x+126 x^{7/6}+549 x^{4/3}\\
		+576 x^{3/2}+1674 x^{5/3}+2079 x^{11/6}+4873 x^2+O\left(x^{13/6}\right)
		\end{array}$	\\
		\hline
		&				&	$(3,1,3)$				&	$\begin{array}{c}
		1+x^{2/5}+9 x^{2/3}+x^{4/5}+9 x^{16/15}+x^{6/5}+36 x^{4/3}\\
		+9 x^{22/15}+36 x^{26/15}+9 x^{28/15}+84 x^2+O\left(x^{31/15}\right)
		\end{array}$	\\
		\cline{3-4}
		&$V_0^{+}\!+\!V_0^{-}\!\!$&	$(3,2,2)$				&	$\begin{array}{c}
		1+10 x^{2/5}+65 x^{4/5}+310 x^{6/5}+1210 x^{8/5}+4002 x^2\\
		+11605 x^{12/5}+30020 x^{14/5}+O\left(x^{16/5}\right)
		\end{array}$	\\
		\cline{3-4}
		&				&	$(3,3,1)$				&	$\begin{array}{c}
		1+9 x^{2/15}+45 x^{4/15}+166 x^{2/5}+513 x^{8/15}+1413 x^{2/3}\\
		+3574 x^{4/5}+8442 x^{14/15}+18855 x^{16/15}+40196 x^{6/5}\\
		+82332 x^{4/3}+162819 x^{22/15}+312130 x^{8/5}+581958 x^{26/15}\\
		+1058085 x^{28/15}+1880074 x^2+O\left(x^{31/15}\right)
		\end{array}$	\\
		\cline{2-4}
		&				&	$(2,1,1)$				&	$\begin{array}{c}
		1 \!+\! 5 x^{2/5} \!+\! 14 x^{4/5} \!+\! 30 x^{6/5} \!+\! 56 x^{8/5} \!+\! 80 x^{2} \!+\! 105 x^{12/5} \!+\! \mathcal{O}(x^{14/5})
		\end{array}$	\\
		\cline{3-4}
		&				&	$(3,2,4)$				&	$\begin{array}{c}
		1+x^{2/5}+9 x^{2/3}+2 x^{4/5}+18 x^{16/15}+2 x^{6/5}+45 x^{4/3}\\
		+27 x^{22/15}+4 x^{8/5}\!+\!117 x^{26/15}\!+\!36 x^{28/15}\!+\!151 x^2\!+\!O\left(x^{31/15}\right)
		\end{array}$	\\
		\cline{3-4}
		&$V_1^{+}\!+\!V_1^{-}\!\!$&	$(3,3,3)$				&	$\begin{array}{c}
		1+10 x^{2/5}+65 x^{4/5}+330 x^{6/5}+1411 x^{8/5}\\
		+5276 x^2+17685 x^{12/5}+53940 x^{14/5}+O\left(x^{16/5}\right)
		\end{array}$	\\
		\cline{3-4}
		4&				&	$(3,4,2)$				&	$\begin{array}{c}
		1+9 x^{2/15}+45 x^{4/15}+166 x^{2/5}+513 x^{8/15}+1413 x^{2/3}\\
		+3575 x^{4/5}+8451 x^{14/15}+18909 x^{16/15}+40443 x^{6/5}\\
		+83259 x^{4/3}+165807 x^{22/15}+320729 x^{8/5}+604629 x^{26/15}\\
		+1113786 x^{28/15}+2009121 x^2+O\left(x^{31/15}\right)
		\end{array}$	\\
		\cline{2-4}
		&				&	$(2,2,2)$				&	$\begin{array}{c}
		1+5 x^{2/5}+20 x^{4/5}+61 x^{6/5}+165 x^{8/5}+373 x^2+756 x^{12/5}\\
		+1361 x^{14/5}+O\left(x^{16/5}\right)
		\end{array}$	\\
		\cline{3-4}
		&				&	$(3,3,5)$				&	$\begin{array}{c}
		1+x^{2/5}+9 x^{2/3}+2 x^{4/5}+18 x^{16/15}+5 x^{6/5}+45 x^{4/3}\\
		+36 x^{22/15}\!+\!7 x^{8/5}\!+\!126 x^{26/15}\!+\!72 x^{28/15}\!+\!158 x^2\!+\!O\left(x^{31/15}\right)
		\end{array}$	\\
		\cline{3-4}
		&$V_2^{+}\!+\!V_2^{-}\!\!$&	$(3,4,4)$				&	$\begin{array}{c}
		1\!+\!10 x^{2/5}\!+\!65 x^{4/5}\!+\!332 x^{6/5}\!+\!1451 x^{8/5}\!+\!5606 x^2\!+\!O\left(x^{11/5}\right)
		\end{array}$	\\
		\cline{3-4}
		&				&	$(3,5,3)$				&	$\begin{array}{c}
		1+9 x^{2/15}+45 x^{4/15}+166 x^{2/5}+513 x^{8/15}+1413 x^{2/3}\\
		+3575 x^{4/5}+8451 x^{14/15}+18909 x^{16/15}+40446 x^{6/5}\\
		+83286 x^{4/3}+165951 x^{22/15}+321308 x^{8/5}+606582 x^{26/15}\\
		+1119600 x^{28/15}+2024869 x^2+O\left(x^{31/15}\right)
		\end{array}$	\\
		\cline{2-4}
		&$V_3^{+}\!+\!V_3^{-}\!\!$&	$(2,3,3)$				&	$\begin{array}{c}
		1+5 x^{2/5}+22 x^{4/5}+77 x^{6/5}+238 x^{8/5}+641 x^2\\
		+1575 x^{12/5}+3495 x^{14/5}+O\left(x^{16/5}\right)
		\end{array}$	\\
		\cline{3-4}
		&				&	$(3,5,5)$				&	$\begin{array}{c}
		1+10 x^{2/5}+67 x^{4/5}+352 x^{6/5}+1584 x^{8/5}+6320 x^2\\
		+22946 x^{12/5}+76934 x^{14/5}+O\left(x^{16/5}\right)
		\end{array}$	\\
		\hline
	\end{tabular}
	\caption{ The superconformal index results for the single adjoint theories with $\Delta W_A = V_\alpha^++V_\alpha^-$. For simplicity, the $SU(N_f)_t \times SU(N_f)_u$ flavor fugacities are all omitted. This table includes all possible cases up to $n\leq 4$, $N_f \leq 3$, and $N_c, \tilde{N}_c\leq5$ with conditions $\Delta_Q,\Delta_q>0$. Especially, we exhibit more results with $N_f \leq 5$ for $n = 1$, in which case the adjoint field becomes massive such that the corresponding duality is nothing but the Benini--Benvenuti--Pasquetti duality \cite{Benini:2017dud}. In addition, the case with $\Delta W_A = V_0^++V_0^-$ corresponds to the Amariti--Cassia duality \cite{Amariti:2018wht}. Therefore, our index results provide nontrivial evidence for those dualities.}
\end{table}

\subsection{Double adjoints with $\Delta W_A = V_{0,0}^{+}+V_{0,0}^{-}$}
\begin{table}[H]
	\centering
	\begin{tabular}{|c|c|c|}
		\hline
		n	&	$(N_f , N_c , \tilde{N}_c)$	&	SCI	\\ 
		\hline
		&	$(3,13,0)$				&	$\begin{array}{c}
		1 + 9 x^{1/6} + 45 x^{1/3} + 9 x^{5/12} + 165 \sqrt{x} + 81 x^{7/12} + 504 x^{2/3} + 405 x^{3/4} + 1404 x^{5/6} \\
		+ 1485 x^{11/12} + 3732 x + 4536 x^{13/12} + 9549 x^{7/6} + 12396 x^{5/4} + 23382 x^{4/3}  \\
		+ 31428 x^{17/12} + 54766 x^{3/2} + 75132 x^{19/12} + 123264 x^{5/3} + 170757 x^{7/4} + O(x^{11/6})
		\end{array}$	\\
		\cline{2-3}
		&	$(3,12,1)$				&	$\begin{array}{c}
		1 + 9 x^{1/12}+ 45 x^{1/6} + 165 x^{1/4} + 504 x^{1/3} + 1368 x^{5/12} + 3409 \sqrt{x} + 7938 x^{7/12}\\
		+ 17496 x^{2/3} + 36869 x^{3/4} + 74817 x^{5/6} + 146952 x^{11/12} + 280531 x + 522279 x^{13/12}\\
		+ 950868 x^{7/6} + 1696594 x^{5/4} + 2972232 x^{4/3} + 5120685 x^{17/12}  + O(x^{3/2})
		\end{array}$	\\
		\cline{2-3}
		&	$(3,11,2)$				&	$\begin{array}{c}
		1+9 x^{1/4}+55 \sqrt{x}+265 x^{3/4}+1100 x+4072 x^{5/4}+13793 x^{3/2}+43396 x^{7/4}\\
		+128283 x^2+O\left(x^{9/4}\right)
		\end{array}$	\\
		\cline{2-3}
		&	$(3,10,3)$				&	$\begin{array}{c}
		1 + 9 x^{1/6} + 45 x^{1/3} + 9 x^{5/12} + 166 \sqrt{x} + 81 x^{7/12} + 513 x^{2/3} + 406 x^{3/4} + 1458 x^{5/6} \\
		+ 1512 x^{11/12} + 3980 x + 4743 x^{13/12} + 10485 x^{7/6} + 13453 x^{5/4} + 26541 x^{4/3}\\
		 + 35667 x^{17/12} + 64663 x^{3/2} + 89559 x^{19/12} + 152487 x^{5/3} + 214672 x^{7/4} + O(x^{11/6})
		\end{array}$	\\
		\cline{2-3}
		&	$(3,9,4)$				&	$\begin{array}{c}
		1 + 9 x^{1/12} + 45 x^{1/6} + 165 x^{1/4} + 504 x^{1/3} + 1368 x^{5/12} + 3409 \sqrt{x} + 7938 x^{7/12}\\
		+ 17496 x^{2/3} + 36869 x^{3/4} + 74817 x^{5/6} + 146952 x^{11/12} + 280532 x + 522297 x^{13/12}\\
		+ 951003 x^{7/6} + 1697246 x^{5/4} + 2974644 x^{4/3} + 5128209 x^{17/12} + O\left(x^{3/2}\right)
		\end{array}$	\\
		\cline{2-3}
		&	$(3,8,5)$				&	$\begin{array}{c}
		1 + 9 x^{1/4} + 55 \sqrt{x} + 265 x^{3/4} + 1100 x + 4081 x^{5/4} + 13893 x^{3/2} + 44080 x^{7/4} \\
		+ 131902 x^2 + O\left(x^{9/4}\right)
		\end{array}$	\\
		\cline{2-3}		
	3	&	$(3,7,6)$				&	$\begin{array}{c}
		1 + 9 x^{1/6} + 45 x^{1/3} + 9 x^{5/12} + 166 \sqrt{x} + 81 x^{7/12} + 513 x^{2/3} + 406 x^{3/4} + 1458 x^{5/6} \\
		+ 1512 x^{11/12} + 3980 x + 4743 x^{13/12} + 10485 x^{7/6} + 13453 x^{5/4} + 26541 x^{4/3} \\
		+ 35676 x^{17/12} + 64663 x^{3/2} + 89640 x^{19/12} + 152496 x^{5/3} + 215077 x^{7/4} + O\left(x^{11/6}\right)
		\end{array}$	\\
		\cline{2-3}
		&	$(3,6,7)$				&	$\begin{array}{c}
		1 + 9 x^{1/3} + \sqrt{x} + 9 x^{7/12} + 45 x^{2/3} + x^{3/4} + 18 x^{5/6} + 81 x^{11/12} + 167 x + 27 x^{13/12}\\
		+ 171 x^{7/6} + 406 x^{5/4} + 540 x^{4/3} + 288 x^{17/12} + 978 x^{3/2} + 1539 x^{19/12}\\
		+ 1791 x^{5/3} + 1870 x^{7/4} + 4086 x^{11/6} + 5193 x^{23/12} + 6478 x^2 + O\left(x^{25/12}\right)
		\end{array}$	\\
		\cline{2-3}
		&	$(3,5,8)$				&	$\begin{array}{c}
		1+10 \sqrt{x}+10 x^{3/4}+65 x+109 x^{5/4}+384 x^{3/2}+748 x^{7/4}+2041 x^2+O\left(x^{9/4}\right)
		\end{array}$	\\
		\cline{2-3}
		&	$(3,4,9)$				&	$\begin{array}{c}
		1 + 9 x^{1/6} + 45 x^{1/3} + 166 \sqrt{x} + 513 x^{2/3} + x^{3/4} + 1413 x^{5/6} + 18 x^{11/12} + 3575 x\\
		+ 126 x^{13/12} + 8451 x^{7/6} + 571 x^{5/4} + 18909 x^{4/3} + 2016 x^{17/12} + 40444 x^{3/2} \\
		+ 6111 x^{19/12} + 83277 x^{5/3} + 16696 x^{7/4} + 165978 x^{11/6} + 42120 x^{23/12} + O\left(x^{2}\right)
		\end{array}$	\\
		\cline{2-3}
		&	$(3,3,10)$				&	$\begin{array}{c}
		1 + 9 x^{1/3} + \sqrt{x} + 45 x^{2/3} + x^{3/4} + 18 x^{5/6} + 167 x + 18 x^{13/12} + 126 x^{7/6}\\
		+ x^{5/4} + 531 x^{4/3} + 126 x^{17/12} + 573 x^{3/2} + 36 x^{19/12} + 1575 x^{5/3} + 571 x^{7/4}\\
		+ 2043 x^{11/6} + 360 x^{23/12} + 4453 x^2 + O\left(x^{25/12}\right)
		\end{array}$	\\
		\cline{2-3}
		&	$(3,2,11)$				&	$\begin{array}{c}
		1+10 \sqrt{x}+x^{3/4}+65 x+19 x^{5/4}+311 x^{3/2}+145 x^{7/4}+1203 x^2+O\left(x^{9/4}\right)
		\end{array}$	\\
		\cline{2-3}
		&	$(3,1,12)$				&	$\begin{array}{c}
		1 + \sqrt{x} + 9 x^{2/3} + x^{3/4} + x + 9 x^{7/6} + 36 x^{4/3} + 9 x^{17/12} + x^{3/2} + 9 x^{5/3} \\
		+ 36 x^{11/6} + 84 x^2 + O\left(x^{25/12}\right)
		\end{array}$	\\
		\hline  
	\end{tabular}
	\caption{The superconformal index results for the double adjoint theories with $\Delta W_A = V_{0,0}^+ + V_{0,0}^-$. Here we list the cases with $(n,N_f) = (3,3)$ and $1 \leq N_c \leq 13$. In each case, the gauge invariant operators with negative conformal dimension are flipped for the expansion of the index. The $SU(N_f)_t \times SU(N_f)_u$ flavor fugacities are all omitted for simplicity.}
\end{table}

\subsection{Double adjoints with $\Delta W_A = W_{0}^{+}+W_{0}^{-}$}

\begin{table}[H]
	\centering
	\begin{tabular}{|c|c|c|}
		\hline
		n	&	$(N_f , N_c , \tilde{N}_c)$	&	SCI	\\ 
		\hline
		&	$(1,5,0)$				&	$\begin{array}{c}
		1+x^{1/4}+2 \sqrt{x}+2 x^{3/4}+3 x+3 x^{5/4}+3 x^{3/2}+2 x^{7/4}+2 x^2+2 x^{9/4}\\
		+3 x^{5/2}+3 x^{11/4}+4 x^3+O\left(x^{13/4}\right)
		\end{array}$	\\
		\cline{2-3}
		&	$(1,4,1)$				&	$\begin{array}{c}
		1+x^{1/4}+3 \sqrt{x}+7 x^{3/4}+10 x+14 x^{5/4}+26 x^{3/2}+29 x^{7/4}+34 x^2\\
		+49 x^{9/4}+51 x^{5/2}+50 x^{11/4}+73 x^3+O\left(x^{13/4}\right)
		\end{array}$	\\
		\cline{2-3}
		&	$(1,3,2)$				&	$\begin{array}{c}
		1+x^{1/4}+3 \sqrt{x}+7 x^{3/4}+11 x+19 x^{5/4}+33 x^{3/2}+44 x^{7/4}+61 x^2\\
		+84 x^{9/4}+100 x^{5/2}+118 x^{11/4}+145 x^3+O\left(x^{13/4}\right)
		\end{array}$	\\
		\cline{2-3}
		&	$(1,2,3)$				&	$\begin{array}{c}
		1+2 \sqrt{x}+4 x^{3/4}+4 x+8 x^{5/4}+14 x^{3/2}+12 x^{7/4}+17 x^2+24 x^{9/4}\\
		+20 x^{5/2}+20 x^{11/4}+31 x^3+O\left(x^{13/4}\right)
		\end{array}$	\\
		\cline{2-3}
		&	$(1,1,4)$				&	$\begin{array}{c}
		1+2 \sqrt{x}+3 x^{3/4}+3 x+3 x^{5/4}+8 x^{3/2}+3 x^{7/4}+4 x^2+7 x^{9/4}\\
		+5 x^{5/2}+2 x^{11/4}+10 x^3+O\left(x^{13/4}\right)
		\end{array}$	\\
		\cline{2-3}
		&	$(2,14,0)$				&	$\begin{array}{c}
		1+4 \sqrt{x}+10 x+16 x^{3/2}+19 x^2+20 x^{5/2}+26 x^3+O\left(x^{7/2}\right)
		\end{array}$	\\
		\cline{2-3}
		&	$(2,13,1)$				&	$\begin{array}{c}
		1+4 x^{1/4}+11 \sqrt{x}+31 x^{3/4}+74 x+156 x^{5/4}+320 x^{3/2}+611 x^{7/4}\\
		+1078 x^2+1846 x^{9/4}+3016 x^{5/2}+4660 x^{11/4}+7013 x^3+O\left(x^{13/4}\right)
		\end{array}$	\\
		\cline{2-3}
		&	$(2,12,2)$				&	$\begin{array}{c}
		1+4 x^{1/4}+15 \sqrt{x}+43 x^{3/4}+117 x+283 x^{5/4}+655 x^{3/2}+1416 x^{7/4}\\
		+2944 x^2+5840 x^{9/4}+11190 x^{5/2}+20656 x^{11/4}+36955 x^3+O\left(x^{13/4}\right)
		\end{array}$	\\
		\cline{2-3}
		&	$(2,11,3)$				&	$\begin{array}{c}
		1+4 x^{1/4}+15 \sqrt{x}+47 x^{3/4}+129 x+331 x^{5/4}+794 x^{3/2}+1803 x^{7/4}\\
		+3923 x^2+8213 x^{9/4}+16607 x^{5/2}+32591 x^{11/4}+62180 x^3+O\left(x^{13/4}\right)
		\end{array}$	\\
		\cline{2-3}
		3&	$(2,10,4)$				&	$\begin{array}{c}
		1+4 x^{1/4}+15 \sqrt{x}+47 x^{3/4}+133 x+343 x^{5/4}+842 x^{3/2}+1947 x^{7/4}\\
		+4322 x^2+9244 x^{9/4}+19138 x^{5/2}+38460 x^{11/4}+75365 x^3+O\left(x^{13/4}\right)
		\end{array}$	\\
		\cline{2-3}
		&	$(2,9,5)$				&	$\begin{array}{c}
		1+4 x^{1/4}+15 \sqrt{x}+47 x^{3/4}+133 x+347 x^{5/4}+854 x^{3/2}+1995 x^{7/4}\\
		+4466 x^2+9648 x^{9/4}+20177 x^{5/2}+41031 x^{11/4}+81348 x^3+O\left(x^{13/4}\right)
		\end{array}$	\\
		\cline{2-3}
		&	$(2,8,6)$				&	$\begin{array}{c}
		1+4 x^{1/4}+15 \sqrt{x}+47 x^{3/4}+133 x+347 x^{5/4}+858 x^{3/2}+2007 x^{7/4}\\
		+4514 x^2+9788 x^{9/4}+20569 x^{5/2}+42031 x^{11/4}+83799 x^3+O\left(x^{13/4}\right)
		\end{array}$	\\
		\cline{2-3}		
		&	$(2,7,7)$				&	$\begin{array}{c}
		1+4 x^{1/4}+15 \sqrt{x}+47 x^{3/4}+133 x+347 x^{5/4}+858 x^{3/2}+2011 x^{7/4}\\
		+4522 x^2+9824 x^{9/4}+20665 x^{5/2}+42291 x^{11/4}+84443 x^3+O\left(x^{13/4}\right)
		\end{array}$	\\
		\cline{2-3}
		&	$(2,6,8)$				&	$\begin{array}{c}
		1+5 \sqrt{x}+7 x^{3/4}+20 x+41 x^{5/4}+95 x^{3/2}+176 x^{7/4}+379 x^2\\
		+705 x^{9/4}+1359 x^{5/2}+2494 x^{11/4}+4587 x^3+O\left(x^{13/4}\right)
		\end{array}$	\\
		\cline{2-3}
		&	$(2,5,9)$				&	$\begin{array}{c}
		1+4 x^{1/4}+11 \sqrt{x}+31 x^{3/4}+79 x+183 x^{5/4}+412 x^{3/2}+889 x^{7/4}\\
		+1833 x^2+3690 x^{9/4}+7232 x^{5/2}+13784 x^{11/4}+25751 x^3+O\left(x^{13/4}\right)
		\end{array}$	\\
		\cline{2-3}
		&	$(2,4,10)$				&	$\begin{array}{c}
		1+5 \sqrt{x}+3 x^{3/4}+20 x+21 x^{5/4}+73 x^{3/2}+96 x^{7/4}+241 x^2\\
		+367 x^{9/4}+750 x^{5/2}+1193 x^{11/4}+2200 x^3+O\left(x^{13/4}\right)
		\end{array}$	\\
		\cline{2-3}
		&	$(2,3,11)$				&	$\begin{array}{c}
		1+\sqrt{x}+7 x^{3/4}+2 x+13 x^{5/4}+37 x^{3/2}+26 x^{7/4}+77 x^2\\
		+164 x^{9/4}+154 x^{5/2}+339 x^{11/4}+603 x^3+O\left(x^{13/4}\right)
		\end{array}$	\\
		\cline{2-3}
		&	$(2,2,12)$				&	$\begin{array}{c}
		1+\sqrt{x}+3 x^{3/4}+6 x+5 x^{5/4}+18 x^{3/2}+23 x^{7/4}+28 x^2\\
		+49 x^{9/4}+70 x^{5/2}+73 x^{11/4}+103 x^3+O\left(x^{13/4}\right)
		\end{array}$	\\
		\cline{2-3}
		&	$(2,1,13)$				&	$\begin{array}{c}
		1+\sqrt{x}+3 x^{3/4}+x+6 x^{5/4}+5 x^{3/2}+6 x^{7/4}-2 x^2\\
		+10 x^{9/4}+4 x^{5/2}+3 x^{11/4}+8 x^3+O\left(x^{13/4}\right)
		\end{array}$	\\
		\hline
	\end{tabular}
\end{table}

\begin{table}[H]
	\centering
	\begin{tabular}{|c|c|c|}
		\hline
		n	&	$(N_f , N_c , \tilde{N}_c)$	&	SCI	\\ 
		\hline
		&	$(2,22,0)$				&	$\begin{array}{c}
		1+4 x^{1/3}+14 x^{2/3}+36 x+81 x^{4/3}+156 x^{5/3}+272 x^2 \\
		+428 x^{7/3}+628 x^{8/3}+O\left(x^{17/6}\right)
		\end{array}$	\\
		\cline{2-3}
		&	$(2,21,1)$				&	$\begin{array}{c}
		1+4 x^{1/6}+11 x^{1/3}+28 \sqrt{x}+62 x^{2/3}+131 x^{5/6}+264 x+500 x^{7/6} \\
		+917 x^{4/3}+1619 x^{3/2}+2771 x^{5/3}+4630 x^{11/6}+7510 x^2 \\
		+11915 x^{13/6}+18502 x^{7/3}+28116 x^{5/2}+41987 x^{8/3}+O\left(x^{17/6}\right)
		\end{array}$	\\
		\cline{2-3}
		&	$(2,20,2)$				&	$\begin{array}{c}
		1+4 x^{1/6}+15 x^{1/3}+40 \sqrt{x}+105 x^{2/3}+239 x^{5/6}+535 x+1103 x^{7/6} \\
		+2233 x^{4/3}+4284 x^{3/2}+8075 x^{5/3}+14652 x^{11/6}+26146 x^2 \\
		+45332 x^{13/6}+77380 x^{7/3}+129092 x^{5/2}+212259 x^{8/3}+O\left(x^{17/6}\right)
		\end{array}$	\\
		\cline{2-3}
		&	$(2,19,3)$				&	$\begin{array}{c}
		1+4 x^{1/6}+15 x^{1/3}+44 \sqrt{x}+117 x^{2/3}+287 x^{5/6}+658 x+1439 x^{7/6} \\
		+3008 x^{4/3}+6071 x^{3/2}+11870 x^{5/3}+22569 x^{11/6}+41879 x^2 \\
		+75983 x^{13/6}+135121 x^{7/3}+235897 x^{5/2}+404861 x^{8/3}+O\left(x^{17/6}\right)
		\end{array}$	\\
		\cline{2-3}		
		&	$(2,18,4)$				&	$\begin{array}{c}
		1+4 x^{1/6}+15 x^{1/3}+44 \sqrt{x}+121 x^{2/3}+299 x^{5/6}+706 x+1567 x^{7/6} \\
		+3359 x^{4/3}+6911 x^{3/2}+13829 x^{5/3}+26856 x^{11/6}+50982 x^2 \\
		+94560 x^{13/6}+172085 x^{7/3}+307324 x^{5/2}+540035 x^{8/3}+O\left(x^{17/6}\right)
		\end{array}$	\\
		\cline{2-3}
		&	$(2,17,5)$				&	$\begin{array}{c}
		1+4 x^{1/6}+15 x^{1/3}+44 \sqrt{x}+121 x^{2/3}+303 x^{5/6}+718 x+1615 x^{7/6} \\
		+3487 x^{4/3}+7267 x^{3/2}+14684 x^{5/3}+28880 x^{11/6}+55441 x^2 \\
		+104155 x^{13/6}+191864 x^{7/3}+347173 x^{5/2}+617955 x^{8/3}+O\left(x^{17/6}\right)
		\end{array}$	\\
		\cline{2-3}
		5&	$(2,16,6)$				&	$\begin{array}{c}
		1 + 4 x^{1/6} + 15 x^{1/3} + 44 \sqrt{x} + 121 x^{2/3} + 303 x^{5/6} + 722 x + 1627 x^{7/6}\\
		+ 3535 x^{4/3} + 7395 x^{3/2} + 15040 x^{5/3} + 29740 x^{11/6} + 57480 x^2 \\
		+ 108679 x^{13/6} + 201631 x^{7/3} + 367444 x^{5/2} + 659006 x^{8/3} + O\left(x^{19/6}\right)
		\end{array}$	\\
		\cline{2-3}
		&	$(2,15,7)$				&	$\begin{array}{c}
		1+4 x^{1/6}+15 x^{1/3}+44 \sqrt{x}+121 x^{2/3}+303 x^{5/6}+722 x+1631 x^{7/6}\\
		+3547 x^{4/3}+7443 x^{3/2}+15168 x^{5/3}+30096 x^{11/6}+58340 x^2\\
		+110723 x^{13/6}+206170 x^{7/3}+377276 x^{5/2}+679445 x^{8/3}+O\left(x^{19/6}\right)
		\end{array}$	\\
		\cline{2-3}
		&	$(2,14,8)$				&	$\begin{array}{c}
		1+4 x^{1/6}+15 x^{1/3}+44 \sqrt{x}+121 x^{2/3}+303 x^{5/6}+722 x+1631 x^{7/6}\\
		+3551 x^{4/3}+7455 x^{3/2}+15216 x^{5/3}+30224 x^{11/6}+58696 x^2\\
		+111583 x^{13/6}+208214 x^{7/3}+381816 x^{5/2}+689280 x^{8/3}+O\left(x^{19/6}\right)
		\end{array}$	\\
		\cline{2-3}
		&	$(2,13,9)$				&	$\begin{array}{c}
		1+4 x^{1/6}+15 x^{1/3}+44 \sqrt{x}+121 x^{2/3}+303 x^{5/6}+722 x+1631 x^{7/6}\\
		+3551 x^{4/3}+7459 x^{3/2}+15228 x^{5/3}+30272 x^{11/6}+58824 x^2\\
		+111939 x^{13/6}+209070 x^{7/3}+383848 x^{5/2}+693776 x^{8/3}+O\left(x^{19/6}\right)
		\end{array}$	\\
		\cline{2-3}
		&	$(2,12,10)$				&	$\begin{array}{c}
		1+4 x^{1/6}+15 x^{1/3}+44 \sqrt{x}+121 x^{2/3}+303 x^{5/6}+722 x+1631 x^{7/6}\\
		+3551 x^{4/3}+7459 x^{3/2}+15232 x^{5/3}+30284 x^{11/6}+58872 x^2\\
		+112063 x^{13/6}+209414 x^{7/3}+384660 x^{5/2}+695692 x^{8/3}+O\left(x^{19/6}\right)
		\end{array}$	\\
		\cline{2-3}
		&	$(2,11,11)$				&	$\begin{array}{c}
		1+4 x^{1/6}+15 x^{1/3}+44 \sqrt{x}+121 x^{2/3}+303 x^{5/6}+722 x+1631 x^{7/6}\\
		+3551 x^{4/3}+7459 x^{3/2}+15232 x^{5/3}+30288 x^{11/6}+58880 x^2\\
		+112099 x^{13/6}+209494 x^{7/3}+384888 x^{5/2}+696196 x^{8/3}+O\left(x^{19/6}\right)
		\end{array}$	\\
		\cline{2-3}
		&	$(2,10,12)$				&	$\begin{array}{c}
		1+5 x^{1/3}+4 \sqrt{x}+20 x^{2/3}+27 x^{5/6}+75 x+121 x^{7/6}+268 x^{4/3}\\
		+456 x^{3/2}+900 x^{5/3}+1541 x^{11/6}+2843 x^2+4826 x^{13/6}\\
		+8506 x^{7/3}+14238 x^{5/2}+24248 x^{8/3}+O\left(x^{19/6}\right)
		\end{array}$	\\
		\cline{2-3}
		&	$(2,9,13)$				&	$\begin{array}{c}
		1+4 x^{1/6}+11 x^{1/3}+28 \sqrt{x}+67 x^{2/3}+151 x^{5/6}+324 x+667 x^{7/6}\\
		+1330 x^{4/3}+2581 x^{3/2}+4891 x^{5/3}+9070 x^{11/6}+16495 x^2\\
		+29496 x^{13/6}+51924 x^{7/3}+90107 x^{5/2}+154331 x^{8/3}+O\left(x^{19/6}\right)
		\end{array}$	\\
		\hline
	\end{tabular}
\end{table}

\begin{table}[H]
	\centering
	\begin{tabular}{|c|c|c|}
		\hline
		n	&	$(N_f , N_c , \tilde{N}_c)$	&	SCI	\\ 
		\hline
		&	$(2,8,14)$				&	$\begin{array}{c}
		1+5 x^{1/3}+20 x^{2/3}+7 x^{5/6}+65 x+41 x^{7/6}+190 x^{4/3}+176 x^{3/2}\\
		+536 x^{5/3}+611 x^{11/6}+1454 x^2+1876 x^{13/6}+3843 x^{7/3}\\
		+5322 x^{5/2}+9856 x^{8/3}+O\left(x^{19/6}\right)
		\end{array}$	\\
		\cline{2-3}		
		&	$(2,7,15)$				&	$\begin{array}{c}
		1+4 x^{1/6}+11 x^{1/3}+28 \sqrt{x}+63 x^{2/3}+135 x^{5/6}+280 x+555 x^{7/6}\\
		+1068 x^{4/3}+2001 x^{3/2}+3661 x^{5/3}+6570 x^{11/6}+11573 x^2\\
		+20058 x^{13/6}+34248 x^{7/3}+57685 x^{5/2}+95975 x^{8/3}+O\left(x^{19/6}\right)
		\end{array}$	\\
		\cline{2-3}
		&	$(2,6,16)$				&	$\begin{array}{c}
		1+5 x^{1/3}+20 x^{2/3}+3 x^{5/6}+65 x+21 x^{7/6}+190 x^{4/3}+96 x^{3/2}\\
		+514 x^{5/3}+351 x^{11/6}+1316 x^2+1116 x^{13/6}+3234 x^{7/3}\\
		+3220 x^{5/2}+7696 x^{8/3}+O\left(x^{19/6}\right)
		\end{array}$	\\
		\cline{2-3}
		5&	$(2,5,17)$				&	$\begin{array}{c}
		1+x^{1/3}+4 \sqrt{x}+2 x^{2/3}+11 x^{5/6}+13 x+21 x^{7/6}+47 x^{4/3}\\
		+58 x^{3/2}+117 x^{5/3}+171 x^{11/6}+260 x^2+451 x^{13/6}+635 x^{7/3}\\
		+1032 x^{5/2}+1533 x^{8/3}+O\left(x^{19/6}\right)
		\end{array}$	\\
		\cline{2-3}		
		&	$(2,4,18)$				&	$\begin{array}{c}
		1+x^{1/3}+6 x^{2/3}+3 x^{5/6}+11 x+5 x^{7/6}+31 x^{4/3}+26 x^{3/2}\\
		+68 x^{5/3}+57 x^{11/6}+142 x^2+160 x^{13/6}+314 x^{7/3}+355 x^{5/2}\\
		+627 x^{8/3}+O\left(x^{19/6}\right)
		\end{array}$	\\
		\cline{2-3}
		&	$(2,3,19)$				&	$\begin{array}{c}
		1+x^{1/3}+2 x^{2/3}+7 x^{5/6}+3 x+13 x^{7/6}+4 x^{4/3}+26 x^{3/2}\\
		+39 x^{5/3}+38 x^{11/6}+80 x^2+57 x^{13/6}+157 x^{7/3}+201 x^{5/2}\\
		+237 x^{8/3}+O\left(x^{19/6}\right)
		\end{array}$	\\
		\cline{2-3}
		&	$(2,2,20)$				&	$\begin{array}{c}
		1+x^{1/3}+2 x^{2/3}+3 x^{5/6}+6 x+5 x^{7/6}+11 x^{4/3}+7 x^{3/2}\\
		+23 x^{5/3}+25 x^{11/6}+33 x^2+39 x^{13/6}+47 x^{7/3}+61 x^{5/2}\\
		+96 x^{8/3}+O\left(x^{19/6}\right)
		\end{array}$	\\
		\cline{2-3}
		&	$(2,1,21)$				&	$\begin{array}{c}
		1+x^{1/3}+x^{2/3}+3 x^{5/6}+x+6 x^{7/6}+x^{4/3}+6 x^{3/2}\\
		+5 x^{5/3}+6 x^{11/6}-2 x^2+6 x^{13/6}+4 x^{7/3}+10 x^{5/2}\\
		+4 x^{8/3}+O\left(x^{19/6}\right)
		\end{array}$	\\
		\hline  
	\end{tabular}
	\caption{The superconformal index results for the double adjoint theories with $\Delta W_A = W_{0}^+ + W_{0}^-$. Here we list the cases with $(n,N_f) = (3,1), \, (3,2)$ and $(n,N_f) = (5,2)$. In each case, the gauge invariant operators with negative conformal dimension are flipped for the expansion of the index. The $SU(N_f)_t \times SU(N_f)_u$ flavor fugacities are all omitted for simplicity.}
\end{table}


\bibliography{mybib}
\bibliographystyle{JHEP}
\end{document}